\documentclass{emulateapj}
\usepackage{amsbsy}
\usepackage{graphicx}

\usepackage{amssymb,amsmath}

\def\gsim{\;\rlap{\lower 2.5pt
\hbox{$\sim$}}\raise 1.5pt\hbox{$>$}\;}
\def\lsim{\;\rlap{\lower 2.5pt
\hbox{$\sim$}}\raise 1.5pt\hbox{$<$}\;}

\def\eg{{\it e.g.\ }} 
\def\etal{{et al.\thinspace}}

\newcommand{\rmnum}[1]{\romannumeral #1}
\newcommand{\Rmnum}[1]{\expandafter\@slowromancap\romannumeral #1@}

\begin{document}

\title{Modeling the Pollution of Pristine Gas in the Early Universe}


\author{Liubin Pan\altaffilmark{1},  Evan Scannapieco\altaffilmark{2}, \& John Scalo\altaffilmark{3}}
\altaffiltext{1}{Harvard-Smithsonian Center for Astrophysics,
60 Garden St., Cambridge, MA 02138;  {\tt lpan@cfa.harvard.edu}}
\altaffiltext{2} {School of Earth and Space Exploration, Arizona State University, P.O. Box 871404, Tempe, AZ, 85287-1494; {\tt evan.scannapieco@asu.edu}}
\altaffiltext{3} {Department of Astronomy, University of Texas, Austin, TX 78712; \tt{parrot@astro.as.utexas.edu}}

\begin{abstract}

We conduct a comprehensive theoretical and numerical investigation of the pollution of pristine
gas in turbulent flows, designed to provide useful new tools for modeling the evolution of the first 
generation of stars.
The properties of such Population III (Pop III) stars are 
thought to be very different than those of later stellar generations, because cooling is dramatically 
different in gas with a metallicity below a critical value $Z_{\rm c},$ which lies between $\sim 10^{-6}$ and $\sim 10^{-3} Z_\odot$. 
The critical value is much smaller than the typical overall average metallicity, 
$\left< Z \right>$, and therefore the mixing efficiency of the pristine gas in the interstellar medium plays a crucial  
role in determining the transition from Pop III to normal star formation.  The small critical value, $Z_{\rm c}$, 
corresponds to the far left tail of the probability distribution function (PDF) of the 
metal abundance. Based on closure models for the PDF formulation of turbulent mixing, 
we derive evolution equations for the fraction of gas, $P$, lying below $Z_{\rm c},$ in 
statistically-homogeneous compressible turbulence. Our simulation data shows that the evolution of the pristine fraction 
$P$ can be well approximated by a generalized ``self-convolution'' 
model, which predicts 
that $\dot P = - \frac{n}{\tau_{\rm con}} P (1-P^{1/n}),$ where $n$ is a measure of the locality of the mixing 
or PDF convolution events and the convolution  timescale $\tau_{\rm con}$ is determined by the 
rate at which turbulence stretches the pollutants.   Carrying out a suite of numerical simulations with 
turbulent Mach numbers ranging from $M = 0.9$ to $6.2,$ we are able to provide accurate fits to $n$ and $\tau_{\rm con}$ 
as a function of $M$, $Z_{\rm c}/\left<Z\right>,$ and the length scale, $L_{\rm p}$, at which pollutants are added to the 
flow.  For pristine fractions above $P= 0.9,$ mixing occurs only in the regions surrounding blobs of 
pollutants,  such that $n=1$.  For smaller values  of $P,$  $n$ is larger as the mixing process becomes more global.
We show how these results can be used to construct one-zone models for the evolution 
of Pop III stars in a single high-redshift galaxy, as well as subgrid models for tracking the 
evolution of the first stars in large cosmological numerical simulations.

\end{abstract}

\section{Introduction}

All stable elements heavier than lithium were forged in stars. Big bang nucleosynthesis produced helium efficiently, 
but it was halted by the expansion of the universe before it could go much further  (Walker et al.\ 1991).  On the other hand, 
all  stars observed to date have substantial mass fractions of carbon, silicon, iron and other elements that are the products 
of the final stages of stellar evolution.    In fact, even the most pristine stars observed (Cayrel et al.\ 2004; Frebel et al.\ 2008; 
Caffau et al.\ 2011) have been polluted by this  material.  The earliest stellar generation, referred to as Population III (Pop III), 
is missing.

While this absence could be due 
to the formation of an extremely small number of metal-free stars,  detailed theoretical studies suggest that it 
is more likely that these stars were too massive to survive to the present day
(Scannapieco \etal 2006; Brook \etal 2007). In fact, the absence of heavy elements drastically 
decreases the cooling rates in collapsing star-forming gas, such that that primordial gas clouds 
would have been much less susceptible to fragmentation, perhaps forming $10^3$-10$^4$ 
solar mass star forming clumps (Hutchins 1976; Abel, Bryan, \& Norman 2000; Bromm, Coppi, \& Larson 2002; Bromm \& Loeb 2003).    
Furthermore, the strong accretion rates onto the central protostellar cores in 
such pristine clumps cannot be arrested by radiation pressure, bipolar outflows, or rotation (Ferrara 2001), 
meaning that  these regions may have formed stars with masses hundreds of times greater than the sun.  On the other hand, 
recent work suggests that physical processes such as enhanced HD cooling in shock-compressed primordial 
gas (Johnson \& Bromm 2006), a lack of magnetic fields in primordial turbulent clouds 
(Padoan \etal 2007), photoionization of turbulent primordial clouds (Clark \etal 2011a), 
fragmentation of the protostellar accretion disks (\eg Stacy \etal 2010, Clark et al.\ 2011b), 
early termination of accretion (McKee \& Tan 2008, Hosokawa  \etal 2011), and 
gravitational torques (Greif \etal 2012), may have led to  primordial stars with 
masses  $\approx 10 M_\odot$  in single or perhaps binary systems (Turk, Abel, \& O'Shea 2009).

Even at these comparatively low masses, direct detections of primordial stars would be possible 
only  through observations of the high-redshift universe, relying on what is likely to have 
been an extended transition between metal free  and pre-enriched (Population II/I) star 
formation (Scannapieco \etal 2003; Jimenez \& Haiman 2006; Trenti \& Stiavelli 2007; 2009; Maio \etal 2010), 
as well as the unusual observable signatures of metal free stars. During their lifetimes,  for example, 
the lack of heavy elements in Pop III stars drastically reduces their opacities, resulting 
in much higher surface temperatures and strong ultraviolet spectroscopic features that 
would distinguish them from current stellar populations (Schaerer 2002; Nagao \etal 2008). 
Alternatively, if they were extremely massive, 
Pop III stars could be detectable as they ended their lives as tremendously powerful 
pair-production supernovae (Bond \etal 1984; Heger \& Woosley 2002; Scannapieco \etal 2005; Whalen \etal 2012).   
Furthermore, the reliance on H$_2$ + HD cooling in primordial gas may have effects beyond 
masses of stars, such as affecting the phase nature of the interstellar medium and the star formation rate (Norman \& Spaans 1997).

Whatever the detection method, when and where metal free gas condensed into stars is a question of 
fundamental importance in planning
searches for this remarkable early generation of stars.   On cosmological scales, the key issue is the time it takes for heavy 
elements to propagate from one galaxy to another.  As shown in Scannapieco et al.\ (2003), the distances between 
these oases  of early star-formation are so vast that for several 100 million years, the universe was divided into 
two regions: One in which galaxies formed out of material that was already polluted with heavy elements well above the 
minimum ``critical mass fraction," $Z_{\rm c},$ at which normal stellar evolution occurs 
(Schneider et al.\ 2003; Bromm \& Loeb 2003; Omukai et al.\ 2005), and one in which galaxies were formed from initially pristine material.  

The evolution of  initially-pristine galaxies is especially interesting, as it depends on two important theoretical issues.  The first of these is the uncertain value of $Z_{\rm c},$  which is  expected to lie in the range from $\sim 10^{-8}$ to $\sim 10^{-5}$,  (or from $10^{-6}$ to $10^{-3}$ times  solar metallicity), depending on whether  the cooling is dominated by dust  grains (Omukai et al.\ 2005) or by the fine structure lines of carbon and oxygen (Bromm and Loeb 2003).   The second important issue is the rate at which the gas within the galaxy can be polluted above this critical value by the turbulent mixing of heavy elements (Pan \& Scalo 2007). Within a given galaxy, the key  quantity to characterize the transition is the  fraction, $P(Z_{\rm c}, t)$, of the interstellar gas  with metal concentration $Z$ below $Z_{\rm c}$ as a function of time. The temporal  
behavior of this fraction depends  not only on the rate at which new sources of metals  are released to the interstellar gas, but, more importantly, 
on the transport and mixing 
of the metals in the interstellar gas. For example, a high mixing efficiency would result in a rapid decrease in $P(Z_{\rm c}, t)$, 
and hence a sharp transition as the average concentration exceeds the threshold $Z_{\rm c}$. On the other hand, 
a low mixing efficiency would lead to a  gradual transition. 
The interstellar medium is known to  be turbulent and highly compressible, and the turbulent motions are 
likely to be supersonic. Therefore, understanding  mixing  in supersonic turbulence is crucial to understanding the  evolution 
of primordial gas in early galaxies.   The evolution of the pristine fraction was considered by Oey (2000, 2003) in 
the context of the sequential enrichment model, which, however, does not  reflect or correctly 
capture the physics of mixing in interstellar turbulence.


In Pan, Scannapieco, \& Scalo (2012; hereafter PSS), we developed a theoretical 
approach to model the evolution of the pristine fraction in statistically homogeneous turbulence.  
The starting point of our theoretical model 
was the PDF method for turbulent mixing, since the pristine fraction  $P(Z_{\rm c}, t)$ 
corresponds to the far left tail of the metallicity PDF.  
The PDF equation for passive scalars cannot be solved exactly,  
and we adopted several closure models from the literature and derived predictions 
for the evolution of the pristine fraction.  
Using numerical simulations, we showed 
that a class of PDF closure models, called self-convolution models, 
provided successful fitting functions to the evolution of 
$P(Z_{\rm c}, t)$ for a limited range of flow Mach number and pollution properties.  
These models are based on the physical picture of
turbulence stretching pollutants and causing a cascade 
of concentration structures toward small scales (Pan \& Scalo 2007), a picture that
is generally valid in turbulent flows at all Mach numbers (Pan \& Scannapieco 2010; hereafter PS10).
Mixing occurs as the  scale of the structures becomes sufficiently small for 
molecular diffusivity to efficiently operate, and the homogenization between 
neighboring structures can be described as a convolution of the concentration PDF.  
As discussed in more 
detail below, 
the models are dependent on 
two major parameters: $ \tau_{\rm con},$  
which sets
the characteristic timescale for convolution of the metal abundance PDF through 
turbulent stretching of concentration structures, and 
$n$, which 
quantifies the degree of spatial locality of the PDF convolution process. 

Here we use a suite of numerical simulations to expand 
these results and show that the generalized self-convolution model
provides good fits for all the turbulence and pollutant conditions relevant for 
primordial star formation. 
Note that, besides affecting the temperature directly, H$_2$ + HD cooling without heavy elements makes 
neutral primordial gas less compressible (Scalo \& Biswas 2002; Spaans \& Silk 2005), emphasizing 
the importance of studying the Mach number dependence of the pollution processes.
By comparing  this model to simulations with Mach numbers
$M=0.9,$ 2.1, 3.5, \& 6.2, in which pollutants are added at different length scales, and 4 different initial 
values of $P,$ we are able to obtain detailed fits of $ \tau_{\rm con},$ and $n$ over all Mach numbers, 
initial pollution fraction, and pollution scales of interest.  
We tabulate and obtain empirical fits to our results, and show how they can be used both to make 
simple one-zone estimates of the evolution of $P$ within a single high-redhshift galaxy, as well as 
to construct a  subgrid model for numerical simulations, 
which tracks the evolution of the primordial fraction below the resolution scale. The model is 
expected to improve  the prediction for the evolution of the primordial gas fraction in early galaxies, and should also be 
applicable to any physical problem in which the unresolved, unmixed fraction needs to be tracked throughout a simulation. 

A realistic simulation for the pollution of the pristine gas in early galaxies needs to 
properly specify the driving mechanism of the interstellar turbulence.  A variety of physical processes 
may contribute to turbulent motions in the interstellar medium. Turbulence can be produced at galactic 
scales, e.g, by the gas infall from the halo or by merger events during the assembly of the galaxy. 
Supernova explosions by the first stars also contribute to the turbulent energy. In the present work, 
we adopt a solenoidal driving force in the simulations,  which may be a good approximation 
if the interstellar turbulence is mainly driven by large-scale motions or instabilities, 
associated with galaxy formation, mergers or interactions.  On the other hand, if the primary 
source for turbulent energy is stellar winds and supernova explosions, the driving 
force may be compressive rather than solenoidal.
 In that case, several questions need to be addressed by future 
studies. First, as shown by Federrath et al.\ (2010),  a compressively driven 
supersonic turbulent flow shows significantly different statistics than the solenoidal
case. The amplitude of the density fluctuations is much larger due to stronger 
compressions, and the intermittency of the velocity field is significantly higher.  
We will speculate how these features may qualitatively affect the parameters in 
our convolution models. A quantitative understanding may be obtained by a 
suite of  simulations using driving forces at different degrees of compressibility. 
Second, if supernovae are the primary energy source of turbulence,  
the injection of heavy elements is highly correlated with the turbulent driving force. This 
is not accounted for in our simulations either. These issues are complicated and
require a systematic investigation. This is, however, beyond of the scope of the current 
paper, which is essentially  an initial step that provides the theoretical framework and useful 
subgrid methodology  for the  modeling of 
the pristine gas pollution in early galaxies.

The structure of this work is as follows.  In \S 2 we introduce the PDF formulation for turbulent 
mixing, and discuss the fundamental mixing physics. In \S 3, we describe the 
self- convolution closure models, and show how they can be used to predict the evolution of
the pristine fraction in statistically homogeneous turbulence.  In \S4 we describe 
our suite of numerical simulations, and in \S5 we use them to test and constrain 
the self-convolution models over the full range of turbulence
and pollution conditions necessary to model primordial star formation.
Having fixed the parameters in the theoretical models with our simulation results
we show how they can then be applied to one zone models of high-redshift galaxies in \S6,
and in \S7 we show how they can be used to construct subgrid models for the unresolved
primordial fraction in large, numerical simulations. A summary of our conclusions 
is given in \S 8.  

\section{The PDF Formulation for Turbulent Mixing}

The mixing of heavy elements in the interstellar medium can be studied by tracking the evolution of 
the concentration field, $C({\bf x}, t)$, defined as the ratio of the 
local density of these elements to the total gas density. The concentration field 
obeys the advection-diffusion equation, 
\begin{equation}
\frac{\partial  C} {\partial t}+    v_i \frac{\partial C}{\partial x_i}  =  \frac{1}{\rho} \frac{\partial}{\partial x_i} \left(\rho \gamma \frac{\partial C}{\partial x_i}\right) +  S({\bf x},t),
\label{advection}
\end{equation}   
where $\rho({\bf x}, t)$ and  ${\bf v}({\bf x}, t)$ denote the density and velocity fields 
in interstellar turbulence, $\gamma$ is the molecular diffusivity, and the term $S({\bf x},t)$ 
represents continuing sources of heavy elements or pollutants. The concentration 
field, $C({\bf x}, t)$, could 
also
represent the local abundance of a specific element, 
but here we are interested in the mass fraction of all metals at a given location. 
The pristine mass fraction in a flow corresponds to 
the low tail of the probability distribution function (PDF) of the pollutant 
concentration, so we adopt a PDF approach for turbulent mixing.   

This approach was first established for the turbulent velocity field (Monin 1967 
and Lundgren1967) and was later extended to mixing of passive or reactive 
species in turbulent flows (Ievlev 1973; Dopazo and O'Brien 1974;  Pope 1976; 
O'Brien 1980; Pope 1985; Kollemann 1990; Dopazo et al.\ 1997). 
It has been particularly 
successful in the field of reacting turbulent flows (e.g., Haworth 2010). However,  most 
of the
work on the PDF modeling of turbulent mixing has been dedicated to incompressible 
or weakly compressible turbulence. In order to apply the method to 
the interstellar media of galaxies, where strong density fluctuations exist, PSS generalized 
the PDF formulation to mixing in highly compressible turbulent flows at large Mach numbers, emphasizing the importance 
of using a density weighting scheme.

As in PSS, herewe use a statistical ensemble to define a density-weighted concentration PDF, 
$\Phi(Z; {\bf x}, t) \equiv \langle \rho \phi(Z; {\bf x}, t) \rangle/ \langle \rho({\bf x}, t) \rangle$, 
where $\phi(Z; {\bf x}, t) \equiv \delta(Z-C({\bf x}, t) )$ is the fine-grained PDF in a 
single realization, $\langle \cdot \cdot \cdot\rangle$ denotes the ensemble average, and $Z$ 
is the sampling variable.  
The probability distribution defined here at a given position and time is an average over 
many independent realizations in the statistical ensemble. The ensemble average density, $\langle \rho({\bf x}, t) \rangle$, 
is in general a function of ${\bf x}$ and $t$. In PSS, $\langle \rho({\bf x}, t) \rangle$ 
was implicitly assumed to be constant. Below we consider the more general case that accounts for the spatial and 
temporal variations of $\langle \rho({\bf x}, t) \rangle$. 
An important motivation for using a density-weighting factor is that, when studying mixing 
of primordial gas in early galaxies, it is appropriate to consider the mass fraction, rather than the volume fraction, 
of the interstellar gas with $Z \le Z_{\rm c}$.

An equation for  $\Phi(Z; {\bf x}, t)$ can be derived using the advection-diffusion equation 
(\ref{advection}), for $C({\bf x}, t),$ and the continuity equation, for $\rho({\bf x}, t)$. Applying the 
same method as in Appendix A of PSS and accounting for the spatial and temporal dependence 
of $\langle \rho \rangle$, we find, 
\begin{gather}
\frac {\partial (\langle \rho \rangle \Phi)}{\partial t} +  \frac{\partial} {\partial x_i} \left( \langle \rho \rangle \Phi {\langle v_i|C=Z \rangle}_{\rho} \right)
=  \hspace {5cm}\notag \\  \hspace {2cm}- \frac {\partial}{\partial Z} \left( \langle \rho \rangle \Phi {\left \langle \frac{1}{\rho}  \frac{\partial} {\partial x_i} \left(\rho \gamma \frac{\partial C}{\partial x_i} \right) \left\vert  \vphantom{\frac{1}{1}} \right. C=Z \right \rangle}_{\rho} \right) \notag \\
\hspace {-0.6cm}-\frac {\partial}{\partial Z} \left(  \langle \rho \rangle \Phi {\langle S|C=Z\rangle}_{\rho} \right), 
\label{pdfeq}
\end{gather}
where $\langle ...|C=Z \rangle_{\rho}$ denotes the conditional ensemble average with 
density weighting.  For any physical quantity, $A({\bf x}, t)$, this ensemble average is  
defined as,
\begin{equation}
\langle A|C=Z \rangle_{\rho} \equiv \frac{\langle \rho A|C=Z \rangle }{ \langle \rho|C=Z \rangle}, 
\label{densweightconditionalave}
\end{equation}
where the conditional average $\langle ...|C=Z \rangle$ without density-weighting is evaluated by 
selecting and counting only those realizations satisfying the constraint that the concentration $C({\bf x}, t)$
at ${\bf x}$ and  $t$ is equal to $Z$. Setting $\langle \rho({\bf x}, t) \rangle$ to be constant reduces 
eq. (\ref{pdfeq}) to eq.\ (2.2) in PSS.  
Derivations of analogous PDF equations for passive 
and reacting scalar turbulence in the incompressible case can be found in e.g., Pope (2000)  and Fox (2003). 
The PDF equation is essentially a Liouville equation for the conservation 
of the concentration probability in phase space. In analogy to the Liouville equation 
in kinetic theory, the concentration field corresponds to the particle 
momentum and the advection-diffusion equation 
corresponds to the particle equation of motion (PSS). 

Although the PDF equation is derived in the context of a 
statistical ensemble, it is also useful for the study of scalar statistics in a 
single realization, i.e., the real flow. If the flow and the scalar 
statistics are spatially homogeneous, $\langle \rho ({\bf x},t)\rangle$ and 
$\Phi(Z; {\bf x},t)$, are independent of ${\bf x}$, and the ergodic theorem indicates 
that the statistics over an ensemble is equivalent to that over the 
spatial domain of a single realization. This means that $\Phi(Z; {\bf x},t)$ is 
equal to the PDF, $\Phi(Z; t) =  \int_V \rho \delta (Z-C({\bf x}, t)) dx^3 / \int_V \rho dx^3$, 
computed from the density and concentration fluctuations over the entire volume, $V$, 
of the real flow domain. At galactic length scales, the assumption of 
statistical homogeneity is likely invalid. For example, coherent mean flows, such as galactic 
rotation, infall or outflow, may exist at large scales 
in the interstellar medium. 
Also a 
large-scale 
metallicity gradient may develop 
if the star formation rate has a radial dependence,
 and
the metallicity statistics can vary  substantially from region to region.

However, the ensemble-defined PDF $\Phi (Z; {\bf x},t)$ may 
still be used to study this spatial dependence if it is understood as 
corresponding to the concentration PDF for local 
fluctuations in a region of considerable size in the real 
flow. If the size is selected to be large enough to allow 
sufficient statistics, but small enough for local statistical homogeneity to be 
restored (i.e., considerably smaller than the characteristic scale for 
the mean flow or mean concentration gradient),  then the ergodic 
theorem will apply locally, and $\Phi(Z; {\bf x},t)$ will represent the 
PDF in the region around ${\bf x}$. In fact, in an attempt to build
a subgrid model for the pollution of pristine interstellar gas, 
a concentrationPDF characterizing the fluctuations in local regions is 
defined by applying a spatial filter to the real flow (see \S 7 and Appendix A), 
and the equation derived for the filtered PDF is identical to eq.\ (\ref{pdfeq}) for 
$\Phi (Z; {\bf x},t)$ over an ensemble. This confirms the equivalence 
of $\Phi (Z; {\bf x},t)$ to the local concentration PDF in the real flow 
and the applicability of eq.\ (\ref{pdfeq}) to the study of mixing in a statistically inhomogeneous setting.   
 
The last three terms in eq.\ (\ref{pdfeq}) correspond to 
turbulent advection, molecular diffusivity and source terms in the 
advection-diffusion equation. We give a brief general 
discussion for each term below, and refer the 
interested reader to PSS for more details. 

\subsection{The Advection Term}

The second term in eq.\ (\ref{pdfeq}) is the advection term.  As it takes a divergence form,
it conserves  the global PDF with density-weighting, i.e., the integral 
of $\langle \rho \rangle \Phi(Z; {\bf x}, t)$ over the entire flow domain. 
This corresponds to a fundamental issue in mixing physics: the turbulent velocity field does not homogenize at 
all by itself. Intuitively, a velocity field moves, stretches and 
redistributes the concentration field, but it does not change the {\it mass} fraction 
of the fluid elements at a given concentration level, and thus does not truly mix.  
On the other hand, without density-weighting, the advection 
term in the PDF equation would not be a divergence term if the 
flow is compressible, but instead be a term representing the effect of expansions and compressions 
on the volume fraction of fluid elements at a given concentration. This 
effect is different from mixing, and makes the PDF modeling more complicated. 
Therefore, in addition to the practical reasons mentioned above, adopting a 
density-weighting scheme is also strongly motivated on a theoretical basis.  

The advection term vanishes and need not be considered in a flow that is 
statistically homogeneous. In case of statistical inhomogeneity, the 
term corresponds to the transport of the concentration PDF by the 
velocity field:
Turbulent advection causes changes in the local PDF as
it moves the fluid elements around. If one is 
interested in the concentration fluctuations in a local region, 
this transport effect must be accounted for carefully.  
A similar advection term exists in the equation for the 
filtered PDF of the local concentration fluctuations, 
which is used to build a subgrid model for the pollution of 
pristine gas in early galaxies  in \S 7. In that case, a proper treatment of the advection term is essential, 
as it is responsible for the flux of pristine mass fraction into 
or out of a computation cell due to the velocity field (see \S 7). 
However, an exact treatment of the advection term is impossible due the usual 
difficulty in turbulence theory known as the closure problem 
(see, e.g., Pope 2000). A similar problem exists for the diffusivity term, which is
discussed in more details in \S 2.2. 
We will adopt  the commonly-used eddy-diffusivity approximation to model 
the advection term in \S 7.
     
\subsection{The Diffusivity Term}      

As shown in Pan \& Scalo (2007) and PSS, the molecular diffusivity 
term in eq.\ (\ref{pdfeq}),
i.e., the first term on the right hand side, is the only term responsible for the
homogenization of the concentration fluctuations.  
The term
can be rewritten as (see Appendix A),  
\begin{gather}
- \frac {\partial}{\partial Z} \left( \langle \rho \rangle \Phi {\left \langle \frac{1}{\rho}  \frac{\partial}{\partial x_i} 
\left(\rho \gamma \frac{\partial C} {\partial x_i} \right)  \left\vert  \vphantom{\frac{1}{1}} \right. C=Z \right\rangle}_{\rho} \right) =   
\frac{\partial}{\partial x_i}  \left\langle \rho \gamma \frac{\partial \phi}{ \partial x_i} \right\rangle \notag\\ 
\hspace{1.5cm}- \frac {\partial^2}{\partial Z^2} \left( \langle \rho \rangle \Phi \left\langle \gamma \left(\frac{\partial C} {\partial x_i} \right)^2  \left\vert  \vphantom{\frac{1}{1}} \right. C=Z\right \rangle_{\rho} \right), 
\label{ensemblediffusivityterm}
\end{gather}
where the first term on the right hand side is  a spatial diffusion of the fine-grained 
concentration PDF (defined in \S2) by the molecular diffusivity. As a divergence term,  it conserves 
the global PDF and does not contribute to true homogenization.  The last term can be 
thought of as an anti-diffusion process in concentration space, as the coefficient is negative definite. 
It continuously narrows the PDF toward the mean value. Unfortunately, this diffusivity 
term does not have an exact or closed form. As it involves concentration gradients, it is 
nonlocal and dependent on the two-point concentration PDF. 
Deriving an equation for the two-point PDF gives rise to terms that require 
knowledge of three-point statistics, and so on, leading to a chain of multipoint
PDF equations similar to the BBGKY hierarchy in kinetic theory (Lundgren 1967; Dopazo and O'Brein 1974).  
Assumptions must be made to truncate the hierarchy to obtain a closed set of equations. 
This is the so-called closure problem. 
In PSS, we considered a number of closure models and showed that a class of models 
based on the convolution of the concentration PDF are particularly successful in fitting the simulation 
results for the pollution of pristine material in turbulent flows. One of these convolution 
models for closure of the diffusivity term was used in our earlier modeling of 
the primordial fraction (Pan \& Scalo 2007). We will summarize these models in \S 3.

Although the diffusivity term lacks an apparent dependence on the velocity field, 
an efficient homogenization of the concentration field does rely on the existence 
of turbulent motions. The action of the diffusivity term is very slow at large 
length scales where the pollutants are injected, because the molecular diffusivity 
$\gamma$ is usually tiny in most natural environments including the interstellar medium.  
For example, Pan \& Scalo (2007) estimated $\gamma$ for the Galaxy's ISM, 
weighted by residence time in different phases of the neutral gas, of about $10^{20}$ cm$^{2}$ 
s$^{-1}$, with a corresponding diffusivity scale of $\simeq 0.06(L_{100}/v_{10})^{1/2}$ pc, where $L_{100}$ 
is the galactic turbulence integral scale in units of 100 pc, and $v_{10}$ is the turbulent rms 
velocity in units of 10 km s$^{-1}$.  This is the scale to which turbulence must 
stretch the pollutants in order for mixing to occur. Therefore, mixing by molecular 
diffusivity is negligible in the absence of a velocity field. A turbulent velocity can act 
as a catalyst and significantly accelerate the mixing process. This implicit role of turbulence 
on scalar homogenization is through the dependence of the diffusivity term on the 
concentration gradients (see eq.\ (\ref{pdfeq})). By continuously stretching the 
pollutants, turbulence produces structures at smaller and smaller scales, 
resulting in an enormous increase in the concentration gradients. Once the structures 
reach a small scale called the diffusion scale, where molecular diffusivity 
operates faster than turbulent stretching, they are homogenized efficiently 
by the diffusivity term. This suggests that the mixing timescale is essentially 
determined by turbulent stretching, even though the velocity itself does not truly mix. 
It is the cooperation of molecular diffusivity and turbulent motions that gives a
 significant mixing efficiency.   
 
 \subsection{The Source Term}    
 
The last term in eq.\ (\ref{pdfeq}) is the source term, corresponding to the injection of new pollutants 
into the turbulent flow. In general, pollutants are any source materials with a 
composition pattern different from that in the existing flow. 
Thus for mixing of heavy elements in the interstellar media of galaxies,  
the source term would include both ejecta from supernova and stellar winds and, if it exists,
infall of low-metallicity or primordial gas. To evaluate the source term, it is actually not 
necessary to compute its conditional average form in eq.\ (\ref{pdfeq}). Instead, 
it can be estimated by directly considering the rate at which the pollutants are 
injected and how they affect the concentration PDF in the flow. For example, assuming that the 
supernova ejecta are nearly pure metals, the source term for the supernova contribution
 would take a delta function form at $Z=1$ with a coefficient depending on the 
 supernova rate, ejecta mass etc  (see \S 6 and Pan and Scalo 2007).  On the other hand,  
 a primordial infall would give a delta function at $Z=0$. Therefore, the effect of continuous 
sources of  primordial  gas  and new metals is to force spikes in the concentration 
PDF at small and large  concentration values, respectively.

\section{Modeling the Diffusivity Term}

The primary goal of this study is to investigate how the diffusivity term 
in the PDF equation, representing the homogenization by molecular diffusivity 
catalyzed by turbulent motions, reduces the fraction of pristine 
material in a turbulent flow. To understand the fundamental physics, we 
consider an idealized problem: mixing of decaying scalars (i.e., $S({\bf x}, t)$ =0) 
in statistically stationary and homogeneous turbulence. The initial scalar field is 
also assumed to be statistically homogeneous. Clearly, this idealized 
problem is much simpler than in a realistic galactic environment.
However, the simplified setting is extremely useful for understanding the 
underlying physics. 
As discussed in \S 2, under the assumption of  statistical homogeneity, 
the advection term vanishes  and the PDF, $\Phi(Z; {\bf x}, t)$, 
is independent of ${\bf x}$ and is equivalent to that computed from the spatial 
fluctuations in a single realization. With these simplifications, the PDF equation becomes,  
\begin{equation}
\frac {\partial \Phi (Z; t)}{\partial t} = - \frac {\partial^2}{\partial Z^2} \left(\Phi \left\langle \gamma \left(\frac{\partial C} {\partial x_i} \right)^2  \left\vert  \vphantom{\frac{1}{1}} \right. C=Z\right \rangle_{\rho} \right)  
\label{simplepdfeq}
\end{equation}
where we used eq.\ (\ref{ensemblediffusivityterm}).
The diffusivity term is the only term in the simplified PDF equation.

In analogy with the mixing of primordial gas in early galaxies, we set 
the initial condition of the decaying scalar to be bimodal, consisting 
of pure pollutants ($Z=1$) and completely unpolluted flow ($Z=0$). This 
corresponds to a double-delta function form for the initial concentration PDF, 
\begin{equation}  
\Phi(Z; 0) = P_0 \delta(Z) +  H_0 \delta(Z-1)  
\label{initialpdf}
\end{equation}
where $P_0$ and $H_0$ are the initial probabilities/fractions of pristine gas 
and pollutants, respectively, and $P_0 +H_0 =1$ from normalization.

Before introducing closure models for the diffusivity term, we discuss the 
evolution of the concentration variance, which helps reveal the general 
physics of turbulent mixing. In terms of the density-weighted PDF, the average 
concentration with density weighting is written as $\langle Z \rangle \equiv \int Z \Phi(Z; t) dZ $,  
which is equal to $\langle {\rho} C \rangle/ \langle {\rho} \rangle$.   Similarly, 
the density-weighted variance is expressed as 
$\langle (\delta Z)^2 \rangle \equiv \int (Z- \langle Z \rangle)^2 \Phi(Z; t) dZ $, 
which is equivalent to $\langle \rho (\delta C)^2 \rangle/ \langle \rho \rangle$ with 
$\delta C  =  C - \langle \rho C \rangle/\langle \rho \rangle$ being 
the fluctuating part of the concentration field.  Taking the 2nd-order moment of 
eq.\ (\ref{simplepdfeq}) yields $\partial_t \langle {\rho} (\delta C)^2 \rangle = - 2 \langle {\rho} \gamma 
( \partial_i C) ^2  \rangle$, which can also be derived directly from the 
advection diffusion equation using the assumption of statistically homogeneity (see PS10). 
We therefore have, 
\begin{equation} 
\frac{d \langle (\delta Z)^2 \rangle}{dt} = - \frac{\langle (\delta Z)^2 \rangle}{\tau_{\rm m}}.  
\label{ensemblevariance} 
\end{equation}
The mixing timescale, $\tau_{\rm m}$, is  the ratio of the concentration variance to its dissipation rate, 
\begin{equation}
\tau_{\rm m} = \langle {\rho} (\delta C)^2 \rangle/(2 \langle {\rho} \gamma (\partial_i C)^2  \rangle). 
\end{equation}
Clearly,  $\tau_{m}$ is the timescale for the variance decay, and thus also characterizes the rate 
at which the diffusivity term reduces the PDF width.

As discussed in \S 2, the mixing timescale, $\tau_{\rm m}$, depends on the 
rate of turbulent stretching, which produces concentration structures at small 
scales and feeds molecular diffusivity with large concentration gradients. 
In the classical phenomenology for turbulent mixing, the continuous production of 
small-scale structures is described as a cascade, in the sense that the 
process proceeds progressively faster toward smaller scales. The picture is similar 
to the cascade of kinetic energy.  It predicts that the mixing timescale is 
determined mainly by the eddy turnover time at the scale where the 
pollutants are injected, but insensitive to the small diffusion scale, where 
the molecular diffusivity acts to homogenize. The prediction has been confirmed 
by  PP10 using simulated supersonic turbulent flows with solenoidal driving force.  
They found that, at all Mach numbers explored, the mixing time was close 
to the eddy turnover time at the pollutant injection scale, suggesting that the 
cascade picture, originally proposed for mixing in incompressible flows, is valid also 
for highly compressible turbulence.

PS10 also found that compressible modes in solenoidally-driven supersonic turbulence 
do not make a significant contribution to the cascade of concentration structures to small scales. 
Compressible modes consist of  both expansions and compressions. As passive scalars simply follow the flow motions, 
the compression events would decrease the length scale of  the 
concentration structures, or equivalently increase the concentration gradients. 
This makes a contribution to enhance the mixing rate. On the other hand, the expansion 
events  would cause  the mixing process to slow down. The two opposite effects tend to 
counteract each other. However, they do not exactly cancel out, and the effect of 
compressions appear to win slightly.  Using the density fluctuations as a 
measure for the strength of the compression events, PP10 found that, in solenoidally driven flows, the net 
contribution of compressible modes to the enhancement of the concentration gradients is 
much smaller than the solenoidal modes (see \S 5 of PP10). 
A limitation in the effect of compressible modes on mixing is that  the squeezing effect by 
compressions is not continuous due to the gas pressure. It is likely a compressed region 
would expand before being squeezed by a second compression event.

Such limitation does not exist in the stretching by solenoidal modes, which operates 
continuous and unlimited by the gas pressure.  The stretching effect by incompressible 
modes appears to be the primary ``mixer" in the simulated flows with solenoidal 
driving at all Mach numbers.  As a consequence, a useful measure for the mixing 
efficiency would be the fraction of energy contained in solenoidal modes in the inertial 
range of the flow, which is responsible for the cascade of passive scalars toward the 
diffusion scale. A statistical analysis of the simulated velocity fields by PP10 
showed that the solenoidal energy fraction in the inertial range decreases 
with $M$ for $M\lsim 3$ and then saturates at an equipartition value of 2/3 at $M\gsim 3$. 
This provides a satisfactory explanation for the behavior of  the mixing timescale 
normalized to the flow dynamical time as a function of $M$. The normalized mixing 
timescale increases with $M$ for $M\lsim 3$ and saturates at larger $M$.  
This finding supports our argument above that  compressible modes are less 
efficient at enhancing mixing  in solenoidally driven flows, and, with a larger 
fraction of compressible energy  in the inertial range, the mixing is slower. 

It remains to be checked if  the normalized timescale as a function of $M$ 
has the same behavior in supersonic turbulence with completely 
compressive driving. One issue is that, quantitatively, the energy fraction of 
solenoidal or compressible modes at inertial-range scales as a function of 
$M$ in fully-developed flows with  compressive driving 
may be different from the solenoidal case.  Second, as shown in Federrath et al.\ (2010), at 
the same $M$, the density fluctuations in a compressively driven flow are much stronger. 
This implies that the net effect of compressible modes on amplifying the concentration gradients 
would be  more efficient than in the solenoidal driving case.  
If the contribution from compressible modes to the mixing efficiency in highly supersonic 
compressively driven flows is comparable to or even faster than the solenoidal modes,  
the behavior of $\tau_{\rm m}$ as function of $M$ may be qualitatively different, 
with the normalized timescale decreasing with $M$ at sufficiently high $M$. 
On the other hand,  if compressible modes in supersonic flows with compressive driving 
are still less efficient at enhancing mixing than solenoidal modes, one may expect a similar 
behavior for the normalized mixing timescale. We will investigate these possibilities in a future work.

\subsection{Self-convolution PDF Models}

A variety of closure models have been developed for the 
diffusivity term in the PDF equation. PSS considered several 
existing models from the literature, including the mapping closure 
model (Chen et al.\ 1989), based on an approximation for the exact 
but unclosed form of the diffusivity term, and a class of models, referred 
to as self-convolution models by PSS, based largely on a physical 
picture of the turbulent mixing process (Curl 1963, Dopazo 1979, Janicka et al.\ 1979, Venaille and Sommeria 2007, 
Villermaux and Duplat 2003, Duplat and Villermaux 2008). One of the 
self-convolution models was used in the initial study of pollution of 
pristine gas by Pan and Scalo (2007).
By a detailed comparison 
with numerical simulations of turbulent mixing in two compressible flows 
at Mach 0.9 and 6.2, PSS showed that the convolution models 
provide both clear physical insights and successful fitting 
functions for the decay of the pristine mass fraction. Here we 
give a brief introduction of the convolution models, and refer the interested reader to PSS for details.

There has been  compelling evidence that 
the dominant  scalar structures at small scales are 2D sheets or edges 
(e.g., Pan and Scannapieco 2011 and references therein). The rate 
at which the scalar sheets are produced is determined mainly by 
the turbulent stretching rate at large length scales. With time, the sheets become thinner, 
and once the thickness 
of the sheets is 
sufficiently small for molecular diffusivity to efficiently operate, 
the neighboring sheets  
are homogenized, leading to a reduction 
in the PDF width.

The physical picture outlined above can be approximately 
described by an integral equation for the concentration PDF, 
\begin{gather}
\frac{\partial \Phi(Z; t)}{\partial t} = s(t) \bigg\{\int\limits_{0}^{1} \Phi(Z_{1};t) 
\int\limits_{0}^{1} \Phi(Z_{2}; t)  \times \hspace {5cm}\notag \\ \hspace {2cm} \delta \left( Z-\frac{Z_{1} +Z_{2}}{2} \right) dZ_{1} dZ_{2}  
- \Phi(Z; t) \bigg\}, 
\label{curl}
\end{gather}
where $Z_1$ and $Z_2$ denote the concentrations in two 
nearby sheets prior to the mixing by molecular diffusivity, 
and the delta function in the integrand arises from the assumption that 
a perfect homogenization occurs instantaneously once two 
scalar sheets are sufficiently stretched for molecular diffusivity to 
take effect. Here $s(t)$ is the turbulent stretching rate
that controls the rate at which the PDF  convolution proceeds.
The last term in eq.\ (\ref{curl}) is the ``destruction" of the previous PDF due to 
the mixing event.  Using the properties of delta functions, eq.\ (\ref{curl}) can be 
written as $\partial_t \Phi (Z; t) =  s(t) [2\int\limits_{0}^{1} \Phi(Z'; t) \Phi(2Z -Z'; t) dZ' - \Phi(Z; t)]$, 
which shows that turbulent mixing is essentially assumed to be a self-convolution process. 

For a reason to be clarified soon, 
eq.\ (\ref{curl}) 
was referred to as the discrete 
convolution model in PSS. 
It was first proposed by Curl (1963) in a study of droplet interactions 
in a two-liquid system, and was later extended to model mixing in turbulent 
flows (Dopazo 1979, Janicka et al.\ 1979). Several variants and generalizations of eq.\ (\ref{curl})
have been proposed to solve the problems of the model for turbulent mixing. 
One problem of Curl's model is that, for a double delta initial PDF (eq.\ (\ref{initialpdf})), 
it produces unphysical spikes in between the initial delta functions. 
In order to avoid this, Dopazo (1979) and Janicka et al.\ (1979) suggested 
replacing the delta function in eq.\ (\ref{curl}) by a general function, $J(Z; Z_1, Z_2)$, that 
is smooth in between $Z_1$ and $Z_2$. 
PSS showed that, with this modification, the model gives essentially the same
prediction for the evolution of the pristine mass fraction.  Another weakness 
of the convolution PDF models with double integral equations is that, for 
mixing in incompressible flows, they substantially overestimate the PDF tails 
at late times (Kolleman 1990). However, the model offers an insightful 
picture for the mixing of pristine gas and provides useful fitting 
functions to the pristine fraction decay in certain physical regimes.    

More recently, Venaille and Sommeria (2007) developed a ``continuous" version 
of the self-convolution model, based on an extension of the Curl (1963) model in 
Laplace space. We first define the Laplace transform, ${\Psi}(\zeta; t)$, of the 
concentration PDF as $\Psi (\zeta; t) = \int_0^{\infty} \Phi(Z; t) \exp(-Z \zeta) dZ$. 
Using the convolution theorem, the Laplace transform of eq.\ (\ref{curl}) reads,
\begin{equation}
\frac{\partial \Psi (\zeta; t)}{\partial t} = s (t) \left[\Psi (\zeta/2; t)^2 - \Psi (\zeta; t)\right].  
\label{convolution}
\end{equation}
Rewriting eq.\ (\ref{convolution}) in a difference form, we have $\Psi (\zeta; t+ \delta t) = \epsilon \Psi (\zeta/2; t)^2 +(1-\epsilon) \Psi (\zeta; t)$
where $\epsilon = s(t) \delta t$ with $\delta t$ an infinitesimal time step. The difference 
equation has the following interpretation: during a timestep $\delta t$, mixing occurs in an 
infinitesimal fraction, $\epsilon$, of the flow, and in this fraction of the flow the scalar PDF undergoes a 
convolution. This suggests that in Curl's model the PDF convolution occurs locally 
in space.  Also note that, whenever a mixing event occurs, it appears as a single 
complete convolution in the model, and in this sense the convolution process is ``discrete". 

The continuous convolution model essentially assumes that the convolution occurs 
everywhere in the flow at any given time, but in an infinitesimal time the number of convolutions is infinitesimal 
and equal to $\epsilon$ (Duplat and Villermaux 2008). 
The assumption can be represented by $\Psi (\zeta; t+ \delta t) = \Psi (\zeta/(1+\epsilon); t)^{(1+\epsilon)}$. 
The Taylor expansion of this equation gives $\Psi (\zeta/(1+\epsilon); t)^{(1+\epsilon)} 
\simeq \Psi (\zeta; t) +  \epsilon [\Psi(\zeta; t) \ln (\Psi(\zeta; t) ) -\zeta \partial \Psi(\zeta; t)/\partial \zeta]$.  
Taking the limit $\delta t \to 0$, we obtain, 
\begin{equation}
\frac{\partial \Psi (\zeta; t)}{\partial t} = s(t) \left[ \Psi \ln (\Psi) -  \zeta \frac {\partial \Psi}{\partial \zeta} \right]. 
\label{cconvolution}
\end{equation} 
The equation was first derived by Venaille and Sommeria (2007), who showed that the 
predicted PDF evolves toward Gaussian in the long time limit. In the continuous version,  
the PDF convolution occurs globally in space. The model prediction has been tested 
against experimental results by Venaille and Sommeria (2008). 
Similar to Curl's model,  the continuous model cannot be applied to predict the 
evolution of the entire PDF right at the beginning if the initial PDF is a double-delta function (Venaille and Sommeria 2007).  
Fortunately, for the problem of pristine gas pollution, the model provides a useful 
prediction that works immediately from the initial time (PSS). 

A more general extension of the self-convolution model in Laplace space 
was given in Duplat and Villermaux (2008),
\begin{equation}
\frac{\partial \Psi (\zeta; t)}{\partial t} = s(t) n \left[ \Psi \left(\frac{\zeta}{1+1/n}; t\right)^{(1+1/n)} - \Psi (\zeta; t) \right]. 
\label{nconvolution}
\end{equation} 
With $n=1$ and in the limit $n \to \infty$, the equation becomes 
eq.\ (\ref{convolution}) for Curl's original model and eq.\ (\ref{cconvolution}) for the 
model of Venaille and Sommeria 
(2007), respectively.  
In deriving eq.\ (\ref{nconvolution}), it was assumed that a fraction, $n \epsilon$, of the 
flow experiences mixing/convolution events during a time interval $\delta t$, and the 
number of convolutions in this fraction of the flow is $1/n$. From the discussion 
above for Curl's model and its continuous version, $n$ characterizes the degree of spatial 
locality of the PDF convolution.  Larger values of $n$ correspond to more ``global" 
convolutions in the spatial space,  and the parameter $n$ may be a function of time in general.  

Eq.\ (\ref{nconvolution}) was referred to as the generalized convolution model in PSS, 
where we found that with increasing $n$ the tails of the predicted PDFs 
become narrower. For example, the discrete model with $n=1$ predicts exceedingly fat 
PDF tails, while in the continuous model ($n \to \infty$) the PDF approaches Gaussian at late times.
In other words,  more ``global" PDF convolutions produce narrower PDF tails.

Finally, we point out that  self-convolution models 
were not originally intended for mixing in highly 
compressible flows and they do not directly account for how compressible 
modes and the density fluctuations in supersonic turbulence may 
affect the concentration PDF. The diffusivity term in the PDF equation (see eq.\ (\ref{simplepdfeq})) has 
a dependence on the density field, suggesting that the flow compressibility 
may have potentially important effects on the PDF evolution. To our knowledge, the effect of compressibility has 
not been investigated in existing PDF models for turbulent mixing. 
Here we take the following approach: We compare the predictions of the 
convolution models for the primordial fraction against simulation results, 
and examine whether, by adjusting their parameters, they 
can be applied to study the pristine gas pollution in supersonic  turbulence. 
Indeed, we find that, by varying the parameter $n$, the self-convolution 
models give satisfactory predictions for the the pollution of pristine gas in turbulent flows at different degrees of compressibility.  
Nevertheless, new closure models are strongly motivated to 
directly and explicitly address the effects of shocks and flow 
compressibility on the scalar PDF in supersonic turbulence. 

\subsection{Mass Fraction of Pristine Gas}

The pristine fraction, defined as the mass fraction of the 
interstellar gas with metallicity smaller than the critical value, $Z_{\rm c}$, 
can be evaluated from the concentration PDF by $P( Z_{\rm c}, t) = \int\limits_0^{Z_{\rm c}} \Phi(Z', t) dZ'$. 
The fraction can be calculated easily if the PDF evolution is known.  The threshold 
metallicity, $Z_{\rm c}$, for the transition to Pop II 
star formation is small but finite, in the range from $10^{-8}$ to $10^{-5}$ by mass
(see Bromm \& Yoshida 2011, Schneider et al.\ 2012 and references therein). 
We also consider the fraction, $P(t)$, in the limit of an infinitesimal threshold, 
i.e., $ P(t) = \lim_{ Z_{\rm c} \to 0} P( Z_{\rm c}, t)$, which corresponds to the 
mass fraction of exactly metal-free gas. Clearly, the fraction $P(t)$ is zero 
unless the concentration PDF, $\Phi(Z; t)$, has a delta function component at $Z=0$.  
Equations of $P(t)$ can be exactly derived from the self-convolution models in \S 3.1.   

There is a subtle issue about the decay of the exactly metal-free fraction, 
$P(t)$, and the pristine fraction, $P( Z_{\rm c}, t)$, with a finite threshold. PSS 
pointed out that the nonlocal nature of the Laplacian operator in the molecular diffusivity 
term leads to an essentially instantaneous decrease of $P(t)$. Physically, a tiny but 
finite fraction of the pollutant atoms can have extremely fast thermal speed, 
corresponding to the high tail of the Maxwellian distribution, and may reach and 
pollute the pristine gas at large distances in a short time. Even though the degree of 
pollution by these atoms at large distances is negligibly tiny, they do reduce the mass 
of gas that is {\it exactly} metal-free, and this occurs at a timescale much 
shorter than the sound crossing time. Therefore, with the molecular diffusivity 
alone, $P(t)$ would decrease to zero almost instantaneously, regardless of the 
amplitude of  the molecular diffusivity $\gamma$.  On the other hand, it takes a finite 
time for the molecular diffusivity to enrich the entire flow up to a finite threshold $Z_{\rm c}$. 
In fact, the decay of $P(Z_{\rm c}, t)$ with $Z_{\rm c}$, say, $\simeq 10^{-8}$ by molecular diffusivity alone 
is very slow because $\gamma$ is typically tiny, $\simeq 10^{20}$ cm$^{2}$ s$^{-1}$ in Galactic neutral ISM 
(see Pan and Scalo 2007). 
An efficient mixing rate relies on the presence of  turbulent motions. 

An ideal model for the pollution of pristine gas should 
accurately capture 
both the rapid decay of $P(t)$ and the evolution behavior of $P(Z_{\rm c}, t)$. However, 
none of the models considered in PSS satisfy both constraints. For example, 
the mapping closure model by Chen et al.\ (1989) does predict an instantaneous decay of $P(t)$, 
but a comparison with simulation results shows that its prediction for $P(Z_{\rm c}, t)$ 
is poor in general, especially in highly supersonic flows. On the 
other hand, the convolution PDF models introduced in \S 3.1 do not 
reduce $P(t)$ to zero immediately, instead the delta function component at $Z=0$
remains finite at any finite time. This is inconsistent with the expectation of 
an instantaneous reduction of $P(t)$, and the reason is that the Laplacian operator 
in the molecular diffusivity term was not directly incorporated in these models. 
Despite this inconsistency, PSS found a very interesting result: the evolution
equations of $P(t)$ derived from the convolution models provide excellent fitting 
functions for the simulation results for the decay of the pristine 
fraction $P(Z_{\rm c}, t)$ with a small but finite threshold.

Here we take the same approach as PSS and carry out a more systematic 
parameter study required to accurately span the range of astrophysical 
environments of interest. We will use the $P(t)$ equations from the 
convolution models to fit the simulation results for $P( Z_{\rm c}, t)$ 
with different thresholds, $Z_{\rm c}$, for scalars with different initial conditions 
evolving in a number of turbulent flows.
This systematic procedure gives best-fit parameters 
in the convolution models as functions of $Z_{\rm c}$, the initial pollutant 
conditions and the flow Mach number. The numerically-tested $P(t)$ 
equations with the best-fit parameters then provide a new 
tool to model the pollution of the primordial gas and the 
transition from Pop III to Pop II star formation in early galaxies.  

We derive the equations of $P(t)$ from the 
convolution models using the PDF equations in Laplace space.
Since the delta function at $Z=0$ persists in these models, 
we decompose the concentration PDF into two terms,  
\begin{equation}
\Phi(Z;t) =P(t) \delta (Z) + \Phi_{\rm e}(Z; t) 
\label{decomp}
\end{equation}   
where $P(t)$ is the fraction of exactly metal-free gas, 
and$\Phi_{\rm e} (Z; t)$ is the concentration PDF in the 
enriched part of the flow, which satisfies the condition 
$\lim_{Z\to 0} \int_0^Z \Phi_{\rm e} (Z';t) dZ' =0$. 
The Laplace transform of eq.\ (\ref{decomp}) gives, 
\begin{equation}
\Psi(\zeta; t) = P(t) + \Psi_{\rm e}(\zeta; t). 
\label{lapdecomp}
\end{equation}
where $\Psi_{\rm e}(\zeta; t)$ is the Laplace transform of 
$\Phi_{\rm e}(Z;t)$.  From the condition $\lim_{Z\to 0} \int_0^Z \Phi_{\rm e}(Z';t) dZ' =0$, 
we have $\Psi_{\rm e}(\zeta; t) \to 0$ in the limit $\zeta \to +\infty$. 

Inserting eq.\ (\ref{lapdecomp}) to the PDF equation (\ref{nconvolution}) 
for the generalized convolution model and taking the 
limit $\zeta \to +\infty$ yields: 
\begin{equation}
\frac{dP}{dt} = -\frac{n}{\tau_{\rm con}} P(1-P^{1/n}).  
\label{pfnconv}
\end{equation}
For later convenience, we have replaced the turbulent stretching rate, $s$, 
by a ``convolution" timescale $\tau_{\rm con} \equiv s(t) ^{-1}$.    

Setting $n=1$ in eq.\ (\ref{pfnconv}), we obtain the equation of $P(t)$ for Curl's model, 
\begin{equation}
\frac{dP(t)}{dt}= -\frac{1}{\tau_{\rm con}}P(1-P), 
\label{pfcurl}
\end{equation}  
which was first given in Pan \& Scalo (2007). An alternative derivation of this 
equation from the PDF equation in the double integral form is presented in PSS. 
From eq.\ (\ref{pfcurl}), we see an interesting and simple physical 
picture for the pollution of the pristine gas by turbulent mixing: The primordial 
fraction is reduced when the fluid elements that are exactly metal-free 
and the rest of the flow that has been polluted by sources or previous 
mixing events, are brought close enough by turbulent stretching for the molecular diffusivity to homogenize. 
Taking $n \to \infty$, eq.\ (\ref{pfnconv}) becomes, 
\begin{equation}
\frac{d P(t)}{dt} =  \frac{P \ln(P)}{\tau_{\rm con}},
\label{pfcconv}
\end{equation}
which is the prediction of the continuous convolution model of Venaille and Sommeria (2007) 
for the pristine faction evolution. 

Assuming both $n$  and $\tau_{\rm con}$ are constant with time,  
equation (\ref{pfnconv}) has an analytic solution,  
\begin{equation}
P(t) = \frac{P_0}{\left[P_0^{1/n} + (1-P_0^{1/n} ) \exp\left( t /\tau_{\rm con} \right)  \right]^n}, 
\label{pfnconvsolution}
\end{equation}
where $P_0$ is the initial pristine fraction. 
This
equation becomes
\begin{equation}
P(t) = \frac{P_0}{P_0 + (1-P_0) \exp \left( \frac{t}{\tau_{\rm con}} \right)},  
\label{pfintegralsolution}
\end{equation}
for Curl's ``discrete" model  with $n=1$ and 
\begin{equation}
P(t)  = P_0^ {\exp(t/\tau_{\rm con})},
\label{pfcconvsolution} 
\end{equation}
for  the Venaille and Sommeria (2007)  model
with $n \to \infty$.

These convolution models predict that the pollution of primordial gas in 
turbulent flows proceeds at a timescale $\tau_{\rm con} \simeq s^{-1}$, 
which is essentially the timescale of turbulent stretching at large 
scales,  anticipated by  the cascade picture for turbulent 
mixing.  Also note that the pollution timescale is essentially 
independent of the molecular diffusivity $\gamma$. Again, this is because 
the mixing rate is largely controlled by how fast the velocity field 
produces and feeds fine structures to the molecular diffusivity, 
but insensitive to the diffusion scale at which molecular diffusivity operates.

\section{Numerical Simulations}

To calibrate $n$ and $\tau_{\rm con}$ as function of the flow and pollutant 
properties, we carried out numerical simulations for mixing in compressible turbulence 
using the FLASH code (version 3.2), a multidimensional hydrodynamic code 
(Fryxell et al.\ 2000) that solves the Riemann problem on a Cartesian grid using 
a directionally-split Piecewise-Parabolic Method 
(Colella \& Woodward 1984; Colella \& Glaz 1985; Fryxell, M\" uller, \& Arnett 1989). 
The hydrodynamic equations were evolved in a periodic box of unit size with 512$^3$ 
grid points. Simulation runs at a lower resolution (256$^3$) were also conducted 
to check the potential effect of numerical diffusion. An isothermal equation of 
state with unit sound speed was adopted in all our simulations. The 
turbulent flows were driven and maintained at a steady state by a large-scale 
solenoidal external force, which was set be a Gaussian stochastic vector that 
decorrelates exponentially with a timescale equal to a quarter of the sound crossing time. 
The driving force was generated in Fourier space, and it included all independent modes 
with wave numbers in the range from $2\pi$ and $6\pi$. Each mode was given the same 
amount of power. We defined a characteristic driving length scale 
$L_{\rm f} \equiv \int  \frac{2 \pi} {k} \mathcal{P}_{\rm f}(k)d{\bf k}/\int \mathcal{P}_{\rm f}(k)d{\bf k}$, 
where $ \mathcal{P}_{\rm f}(k)$ is the power spectrum of the driving force. 
Calculating $L_{\rm f} $ from our forcing spectrum, we found $L_{\rm f} = 0.46$ in units in which 
the box size is unity. By adjusting the amplitude of the driving force, we simulated four 
flows with different (density-weighted) rms velocities, $v_{\rm rms}$. 
For each flow,  we defined a dynamical timescale, $\tau_{\rm dyn} \equiv L_{\rm f}/v_{\rm rms}$, 
and all the simulation runs lasted for about 5 $\tau_{\rm dyn}$. 
We computed the mean rms velocity by a temporal average after each flow reached a steady state, 
and the rms Mach numbers, $M$, i.e., the ratio of the rms 
velocity to the sound speed, in the four flows were 
$M=0.9$, 2.1, 3.6, and 6.2, respectively. 
The simulation setup for the turbulent velocity field is the same as in 
PS10 and PSS, to which we refer the interested reader for details.  

To study turbulent mixing, we evolved a number of decaying scalar fields in the 
four simulated flows. In each flow, we solved the advection equations of all scalar 
fields starting at the same time after the flow had already become fully developed 
and statistically stationary. The initial condition of the scalar fields was taken to be 
bimodal, consisting of pure pollutants and completely unpolluted material only. 
Such a bimodal field was obtained by setting the pollutant concentration, $C$, to 
unity in selected regions, representing pure pollutants, and to zero in the rest of the simulation 
box, corresponding to the unpolluted flow. 
The rate at which the pollution of the pristine material proceeds in our simulations
depends not only on the flow properties but also 
on the initial configuration of the pollutants. 
Two parameters in the initial condition are of particular interest. The first one is the initial pollutant fraction, $H_0$, 
defined as the ratio of the heavy element mass to the total flow mass 
in the simulation box. The fraction is related to the initial primordial 
fraction, $P_0$, by $P_0 + H_0 =1$. Obviously, with more pollutants 
in the flow, i.e., a larger value of $H_0$,  one would expect 
a faster pollution of the pristine gas. 

The mixing/pollution timescale also depends on how the pollutants 
are spatially distributed in the flow. For illustration, let us consider two 
different distribution patterns for the same amount of pollutants. In the 
first pattern, the pollutants are released in the form of a single blob, while 
in the second the pollutants are divided into many blobs of similar sizes 
evenly distributed in the flow. Intuitively, the pollution process would be 
considerably faster in the latter case. In that case, the 
pollution injection scale, $L_{\rm p}$, which is essentially the average distance 
between the pollutant locations, is smaller, and the mixing timescale should 
be shorter since it is determined by the eddy turnover time at $L_{\rm p}$ (PS10).  We thus
expect that a smaller $L_{\rm p}$ 
would result in a faster decay of the pristine mass fraction,
and  we will quantitatively examine the dependence of the pollution of the pristine flow 
on this parameter. In the context of the mixing of heavy elements in the interstellar media of 
early galaxies, the pollutant fraction, $H_0$, is related to the number or the rate 
of the supernova events, and the pollutant injection scale corresponds to the 
average distance between the explosion locations.

\begin{table*}
\begin{center}
\caption{Initial configuration of passive scalar fields evolved in our simulated turbulent flows}
\label{tbl-1}
\vspace{3mm}
\scalebox{1.1}{
\begin{tabular}{c@{\hspace{10mm}}c@{\hspace{10mm}}c@{\hspace{10mm}}c@{\hspace{10mm}}c@{\hspace{10mm}}c}
\hline\hline
Category & Pollutant configuration & $ H_0 =0.5 $ & $H_0 =0.1$ & $H_0 =10^{-2}$ & $H_0=10^{-3}$  \\
\hline
\rmnum{1} &  ~1 cube           &  \rmnum{1}A &   \rmnum{1}B   & \rmnum{1}C & \rmnum{1}D    \\
\rmnum{2} &  1 ball             &   \rmnum{2}A &   \rmnum{2}B   & \rmnum{2}C  &  \rmnum{2}D    \\
\rmnum{3} &  2$^3$ balls   &  \rmnum{3} A &  \rmnum{3} B  & \rmnum{3} C &  \rmnum{3} D   \\
\rmnum{4} &  4$^3$ balls   &  \rmnum{4}A &  \rmnum{4}B   & \rmnum{4}C   & \rmnum{4}  D   \\
\rmnum{5} &  8$^3$ balls   &  \rmnum{5}A &   \rmnum{5}B   & \rmnum{5}C  &  \rmnum{5}D    \\
\hline
\end{tabular}}
\end{center}
\end{table*}
 
In order to conduct a systematic study of the parameter dependence of $P$, we 
included in each simulated flow a total of 20 scalar fields with different initial conditions. Table 1 
summarizes the fields, which are divided into 5 categories based on the 
geometry and the spatial distribution of the pollutants. For categories 
\rmnum{1}\ and \rmnum{2}, the initial pollutant configuration is a single blob 
located right at the center of the simulation box, and the geometrical shape of the blob was 
set to be a cube and a spherical ball, respectively.
Clearly, for a single pollutant blob, the pollutant separation and hence the 
injection length scale, $L_{\rm p}$, is given by the box size.  
Considering that the flow driving scale $L_{\rm f}$ in our simulations is 0.46 
of the  box size, we have $L_{\rm p} \simeq 2 L_{\rm f}$ for 
categories \rmnum{1}\  and \rmnum{2}. In the other three categories, 
the pollutants are divided into 
identical spherical blobs, equally spaced in the simulation box. 
The number of blobs is 8, 64 and 512, respectively, for category \rmnum{3}, \rmnum{4}, and \rmnum{5}. 
For scalar fields in these three categories, $L_{\rm p}$ corresponds 
to $\frac{1}{2}$, $\frac{1}{4}$ and $\frac{1}{8}$ box size (or equivalently $\simeq 1, 0.5$ and $0.25 L_{\rm f}$), 
respectively.  For reference, the scale at which  the energy cascade and 
hence the inertial range starts in our simulated flows is about $\frac{1}{4}$ box size.  
The injection scale ($\frac{1}{8}$ box size) of category \rmnum{5}\ scalar fields is  well 
within the inertial range. 

There are four scalar fields in each category, which differ in the initial pollutant mass fraction, 
$H_0$. The four scalar fields in category \rmnum{1}, named \rmnum{1}A, \rmnum{1}B, \rmnum{1}C, 
and \rmnum{1}D, have $H_0 = 0.5$,  0.1, 0.01, and 0.001, respectively. Scalar fields in other 
categories are named in the same way. These exact $H_0$ values were achieved by tuning 
the size of the pollutant blob(s).
In category \rmnum{1}, the length of the pollutant cube 
is set to 0.79, 0.47, 0.22 and 0.1 in units of the box size, or 1.7, 1.0, 0.48 and 0.22 in 
units of the driving scale $L_{\rm f}$, for scalars A, B, C and D,  respectively,  in the Mach 0.9 flow.
For the four scalars in category \rmnum{2}, the radius, $r_{\rm p}$, of the spherical 
ball is 0.49, 0.29, 0.14, and 0.063 box size. In units of $L_{\rm f}$,  $r_{\rm p} = 1.1$, 
0.63, 0.30, and 0.14 $L_{\rm f}$, respectively. The radius of each pollutant ball for
the corresponding scalar in categories \rmnum{3}, \rmnum{4}, and \rmnum{5} is smaller 
by a factor of 2, 4, and 8, respectively, than the $r_{\rm p}$ value for category \rmnum{2}. 
This is because the numbers of balls in those categories are 
larger than that in category \rmnum{2}\ by factors of $8$, $64$ and $512$. The radii $r_{\rm p}$ 
given here are the values used in the $M=0.9$ flow. At larger $M$, $r_{\rm p}$ for 
each corresponding scalar is slightly different. Due to significant 
density fluctuations in flows at higher $M$, using pollutant blobs 
of the same size at the same locations leads to different values of $H_0$. 
We thus tuned the pollutant size to guarantee the initial pollutant {\it mass} 
fraction, $H_0$, is exactly 0.5, 0.1, 0.01, and $0.001$ for the 
scalars in each flow.
 
We also made an attempt to investigate a smaller value ($10^{-4}$) of $H_0$, 
which would also be interesting for mixing in the interstellar medium of early galaxies. 
However,  a tiny $H_0$ corresponds to a small pollutant size, and 
due to the limited numerical resolution, the pollutant size for $H_0 = 10^{-4}$ 
is too close to the resolution scale of our simulations. In that case, numerical diffusion took 
effect and significantly polluted the 
surrounding flow from the beginning, 
leading to a different evolution behavior for the pristine fraction 
at early times than the other cases with $H_0 \ge 10^{-3}$.  In the interstellar medium, 
the pollutant size is essentially a supernova remnant stall diameter, $\simeq150$ pc, 
with little dependence on parameters (see Thornton et al.\ 1998, Hanayama \& Tomisaka 2006). 
This is expected to lie within the inertial range of interstellar turbulence. 
Therefore, the real homogenization of fresh metals 
from supernovae by molecular diffusivity must wait for turbulent stretching to 
bring the concentration structures to the diffusion scale, 
which is tiny in comparison to the remnant size.  It is thus appropriate to consider pollutants with 
initial sizes significantly larger than the diffusion/resolution scale of the 
turbulent flow, and in the present work we do not explore scalar cases with $H_0 \le 10^{-4}$.  We 
point out that the first three scalar fields in category \rmnum{1}, i.e., \rmnum{1}A, \rmnum{1}B,  and \rmnum{1}C, 
in Mach 0.9 and 6.2 flows have been studied in details in PSS. 
In this paper, we perform a more systematic study covering a much larger parameter space.  
 
Neither the viscous term in the hydrodynamic equations nor the diffusivity term in the 
advection-diffusion equation were explicitly included in our code. Therefore, both kinetic 
energy dissipation and scalar homogenization (or dissipation) 
are through numerical diffusion in our simulations. The diffusion scale where the scalar 
homogenization occurs is close to the resolution scale, and so is the energy 
dissipation or the Kolmorgorov scale. To examine whether our results for primordial 
gas mixing depend on the amplitude of numerical diffusion, we also performed simulations 
at the resolution of $256^3$, and the results at the two resolutions are compared 
in \S 5.4.5. Otherwise, unless explicitly stated, the results reported below are from  the $512^3$ simulations.     

\section{Results}

\subsection{The Concentration Field}

\begin{figure*}
\centerline{\includegraphics[width=2\columnwidth]{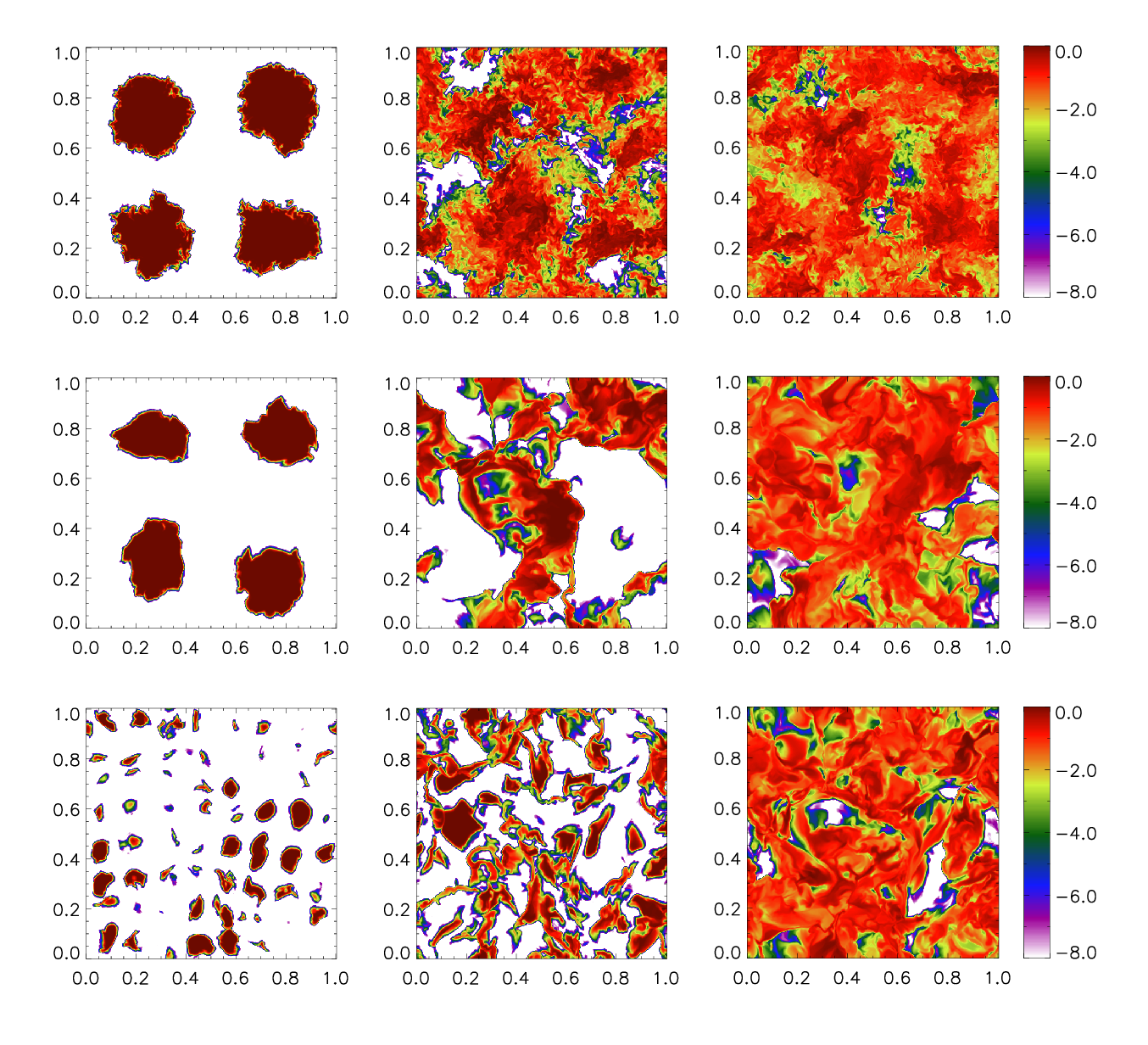}}
\caption{Evolution of the concentration fields of three scalars on a slice 
of the simulation grid. The color scale for the concentration field is logarithmic 
with white regions representing the unpolluted flow with $C \le 10^{-8}$. From left to right, the 
top three panels show snapshots of scalar field \rmnum{3}B in the Mach 0.9 
flow at $t=0.11$, 0.68 and 1.06 $\tau_{\rm dyn}$.  The density-weighted scalar 
variances at these times are 0.086, 0.038, and 0.015, respectively. 
The initial condition of this scalar field is eight equally-spaced 
blobs with a total pollutant fraction $H_0 = 0.1$. The central three panels 
plot a scalar field with the same initial condition but in the highly supersonic flow 
with $M=6.2$. The three snapshots correspond to $t=0.11$ (left), 0.76 (mid), and 1.41 $\tau_{\rm dyn}$ (right), 
with the scalar variance being 0.078, 0.023, and 0.006, respectively. 
The bottom panels show case \rmnum{5}B in the Mach 6.2 flow.  
This scalar field initially consists of 512 pollutant blobs, and the total 
pollutant fraction $H_0$ is also equal to 0.1. At $t=0$, 64 blobs lie on the selected 
slice, and the three snapshots are taken at  $t= 0.11 $, 0.33, $0.76$ $\tau_{\rm dyn}$.
At these snapshots, the scalar variance is 0.07,  0.038,  and 0.011, respectively.}
\label{images} 
\end{figure*}

In Fig.\ 1, we show the concentration on a slice (on the $y$-$z$ plane at $x =0.25$) 
of the simulation grid for three scalar fields (rows) at three different times (columns).  
The concentration is 
plotted 
on a logarithmic scale with the white color representing 
concentration levels below $10^{-8}$. The top three panels, from left to right,  correspond 
to case \rmnum{3}B in the Mach 0.9 flow at $t=0.11, 0.68$, and $1.06 \tau_{\rm dyn}$, 
respectively. At $t=0$, four spherical pollutant blobs lie on the $x=0.25$ 
plane. With time, turbulence stretches and spreads out the pollutants, and 
structures at small scales are continuously produced. In particular, 
we observe prominent ``cliff" structures with sharp concentration 
gradients. These sheet-like structures are typical of passive 
scalars in incompressible turbulence (e.g., Watanabe and Gotoh 2004). 
As the length scale of the scalar structures reaches the (numerical) diffusion scale, mixing 
occurs between the pollutants/polluted flow and the pristine regions. 
The mixing process reduces the volume fraction of the pristine flow (white regions), 
and at $1.06\,  \tau_{\rm dyn}$ almost the entire flow is polluted. The density-weighted 
concentration variance decreases from the initial value of 0.09 to 0.086, 0.038, and 0.015, 
respectively,  for the three snapshots from the left to the right. 

The three panels in the second row show snapshots of the same  
scalar case (\rmnum{3}B), but in the Mach 6.2 flow, at $t=0.11$ (left), 0.76 (mid), 
and 1.41 $\tau_{\rm dyn}$ (right), respectively.  Comparing with the top panels,  
we see that, even at later times (in units of $\tau_{\rm dyn}$), 
the surviving pristine volume is larger than in the $M=0.9$ flow, 
suggesting that the pollution of the pristine gas is slower in turbulent flows at 
higher $M$.  The concentration field appears to be smoother than in 
the $M=0.9$ flow. As explained in detail in PS10, this is because in highly 
supersonic turbulence the visual impression of the scalar field is 
dominated by expansion events, which occupy most volume of the flow 
domain. Since a passive scalar simply follows the flow velocity, an expanding 
region tends to produce coherent and smooth structures 
at large scales. We note that the scalar in the supersonic 
flow has a smoother appearance also at small scales, and the likely reason is that
the code used in our simulations applies a larger effective numerical diffusion
to stabilize stronger stocks in flows with larger $M$ .  
Although compressible modes play a key role in shaping the large-scale geometry 
of the scalar field, the primary mixing agent is still stretching by solenoidal modes 
even in our simulated flows at very high $M$ (PS10). 
The concentration variances at the three snapshots shown here are 0.078, 0.023, and 0.006, 
respectively. Note that,  even though the scalar variances in the right 
two panels are smaller than the corresponding snapshots in 
the $M=0.9$ flow, the remaining pristine fraction in the $M=6.2$ flow appears to be
larger. This is because in turbulent flows with higher $M$ the scalar 
PDF tails are broader, leading to a larger pristine 
fraction at the same concentration variance. 
A more detailed discussion on this issue is given in \S 5.2. 

The bottom three panels in Fig.\ 1 plot the evolution of the scalar field \rmnum{5}B in 
the Mach 6.2 flow, which also has $H_0 = 0.1$. Unlike case \rmnum{3}B shown in the top and central panels,  
this field initially consists of 512 small blobs, and has a smaller injection scale, $L_{\rm p}$.  
At $t=0$, 64 blobs lie on the slice shown here. At early times, 
some blobs appear to be small dots or filaments because they are being advected out of the 
selected slice. The three panels correspond to $t= 0.11 $, 0.33, $0.76$ $\tau_{\rm dyn}$. 
The mixing/pollution process proceeds much faster than case \rmnum{3}B 
in the same flow. One reason is that, for a smaller pollutant size,  turbulent 
stretching of the pollutant is faster, and thus mixing of each individual blob 
with the surrounding flow is more efficient. Also, since the separation between the pollutant blobs 
is small, the polluted/mixed regions by the individual pollutant blobs start to overlap quickly, 
resulting in a much faster erasure of the pristine flow material. As a reference, 
the scalar variances at the three snapshots from the left to the right are 0.07,  0.038,  and 0.011, respectively.

\subsection{The PDF Evolution}

\begin{figure*}
\centerline{\includegraphics[width=2\columnwidth]{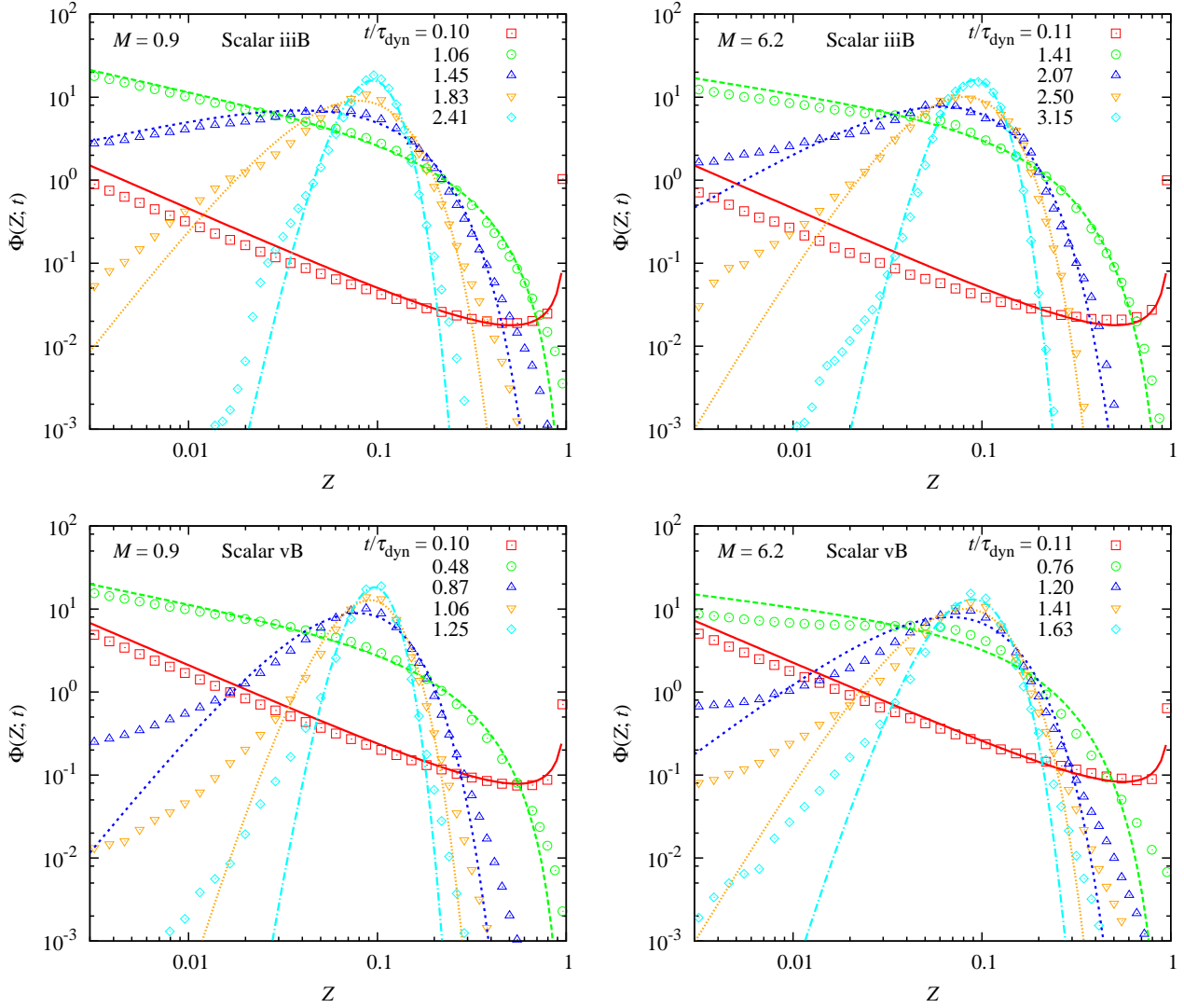}}
\caption{The density-weighted concentration PDFs 
of four scalar fields as a function of time. The left and 
right panels correspond to scalar fields advected in the Mach 0.9 and Mach 
6.2 flows, respectively. Top and bottom panels show results for scalar fields 
in two different categories, \rmnum{3}\ and \rmnum{5}.
The initial pollutant separation for the two scalars is 1/2 and 1/8 box size, respectively.  All scalar fields 
shown here have the same initial pollutant fraction $P= 0.1$, 
and thus their PDFs have the same mean (0.1). 
The lines are beta distribution functions with the same mean 
and variances as the corresponding PDFs measured from the 
simulations (data points).   
 }
\label{pdf} 
\end{figure*}

In this subsection, we discuss simulation results for the evolution 
of the concentration PDF.  Fig.\ (\ref{pdf}) plots the PDF as a function 
of time for four scalar fields. 
For all scalars, the heights of the two spikes at $Z=0$ and $Z=1$ 
decrease at  early times, and mixing causes a probability flux toward the central 
part, which gradually fills the concentration space between the two spikes.
Both spikes are eventually removed, and for an initial PDF with 
negative skewness, or $P_0 > H_0$, the left spike lasts longer than the right one.  At later times, 
a central peak forms around the mean concentration, and 
the PDF becomes unimodal. After that, the PDF 
continuously narrows toward the mean value, a process 
described in \S 2.2 as anti-diffusion in concentration space.

In PSS, we tested the predictions of various models for the PDF evolution against 
simulation data for scalar case \rmnum{1}B  in Mach 0.9 
and 6.2 flows. The initial condition of this scalar is a single cubic pollutant with $H_0 =0.1$. 
It was found to be very challenging for PDF models to accurately predict the scalar PDF tails, 
especially for scalar fields in highly supersonic flows. Here we do not attempt to obtain successful 
model fits to the measured  PDFs,  as the main goal of this work is to understand the evolution of the pristine 
fraction, rather than the full details of the entire PDF. However, 
including a model prediction for the PDF evolution in our figure is 
useful, because it provides 
a guideline to compare the fatness of the PDF tails for different scalar fields
in different flows. For this purpose, we consider the beta 
distribution function as a PDF model for passive scalar mixing, which has 
been shown to provide a good approximation for the PDF shape of 
decaying scalars with a double-delta initial condition in incompressible turbulence (e.g., Girimaji 1991). 
The beta distribution function is defined as,  
\begin{equation}
\Phi_{\beta} (Z) = \frac{\Gamma(\beta_1 +\beta_2)}{\Gamma(\beta_1) \Gamma(\beta_2)} Z^ {\beta_1-1}(1-Z)^{\beta_2-1}
\label{beta}
\end{equation} 
where $\Gamma$ is the Gamma function. To compare 
the beta distribution with the simulation results,  one can determine the two 
parameters, $\beta_1$ and $\beta_2$, in eq.\ (\ref{beta}) by equating the 
mean and variance of the beta PDF to those measured from the 
simulation data. For each measured PDF (data points) shown in Fig.\ (\ref{pdf}), 
we plot  a beta distribution (line), where the beta parameters 
are fixed using the concentration mean and variance at the corresponding time.

The left top panel in Fig.\ (\ref{pdf}) shows the result for case \rmnum{3}B 
in the Mach 0.9 flow. The initial condition of this case is eight equally-spaced 
spherical blobs with $L_{\rm p}$ equal to 1/2 box size. The total pollutant 
fraction, $H_0$, of this scalar field is 0.1. The PDFs are measured at five different 
times as indicated in the legend. For this scalar, the fitting quality of the beta distribution 
functions is generally good except at far tails.  
In PSS, we showed that, at later evolution times, the PDF of scalar \rmnum{1}B 
in the $M=0.9$ flow is well fit by a Gamma distribution, as predicted by the 
PDF model of Villermaux and Duplat  (2003).  The initial condition of  scalar \rmnum{1}B 
is a single cubic pollutant, and it has a twice larger $L_{\rm p}$ than \rmnum{3}B 
shown here. 
The performance of the Villermaux and Duplat (2003) model is less satisfactory for 
scalars with smaller  $L_{\rm p}$, e.g., it significantly underestimates the PDF tails for the 
scalars in Fig.\ (\ref{pdf}). Also that model is is invalid at the early evolution stage 
(see PSS).  On the other hand, the beta distribution does provide acceptable fits 
to the measured PDFs at early times, as seen in Fig.\ (\ref{pdf}).

The top right panel shows the PDF of the same scalar field in the Mach 6.2 flow. Using the 
beta distributions as a reference, we see that at late times, the left PDF tails are broader than in 
the Mach 0.9 case. The bottom two panels plot the results for case \rmnum{5}B in the same two flows.
This scalar field also has $H_0= 0.1$, but the injection 
scale, $L_{\rm p}$, is significantly smaller, $\simeq \frac{1}{8}$ box size (see Table 1). 
A comparison of the time series indicated 
by the legends in the top and bottom panels 
shows that the PDF variance decays much faster for 
scalar fields with smaller injection scale
(see \S 5.3),
but at similar values of the  variance,  the PDF tails are 
broader for scalar fields
with smaller $L_{\rm p}$. 
These observations are consistent with the findings of PS10, who studied the dependence of 
the PDF shape on 
$M$ and $L_{\rm p}$ in details, and found that the PDF tails become broader with increasing 
$M$ or decreasing $L_{\rm p}$. The physical origin of this behavior is probably 
related to the phenomenon of turbulent intermittency, i.e., the existence of strong 
non-Gaussian velocity structures at small scales. Supporting this interpretation 
is the fact that the degree of non-Gaussianity of the velocity field 
increases as $M$ increases or as $L_{\rm p}$ decreases, which coincides with the trend 
of the PDF tails of passive scalars.   As discussed in \S 3.1, the convolution PDF models 
with smaller $n$ predict fatter tails, meaning that  $n$ would decrease with 
increasing $M$ or decreasing $L_{\rm p}$ if one attempts to fit the measured PDFs 
with the predictions of the convolution models.  
Extending the intermittency argument here to mixing in supersonic flows with totally compressive 
driving, we expect that, at the same $M$, the passive scalar PDF would have fatter tails than in 
our flows with solenoidal driving. This is because the compressively driven supersonic flows are significantly  
more intermittent (Federrath et al.\ 2010).

We point out that the fatness of the PDF tails as a function of $M$ and $L_{\rm p}$ 
for decaying scalars in the current study is less clear-cut than for the forced scalars 
examined in PS10. The general trend is sometimes not clearly obeyed in our 
simulations here, especially for scalar fields in high Mach-number flows ($M=3.5$ and 6.2). 
For example, as seen in Fig.\ (\ref{pdf}),  it appears that the PDF tail 
of scalar \rmnum{5}B in the Mach 6.2 flow is less broad than in the Mach 0.9 flow.  
A possible reason is that the measurement of the PDFs of decaying scalars is 
less precise than in the case of forced scalars, for which the PDFs can be 
computed by averaging over many snapshots.  
Simulations with higher resolutions may help us establish a robust trend for the 
PDF tail of decaying scalars in turbulent flows at large $M$, as they provide 
better statistics and better resolution of complexities, such as strong 
density fluctuations, in highly supersonic turbulence.  

Finally, we stress that the pristine fraction corresponds to the 
probability contained in the far left tail of the PDF with $Z<10^{-8}-10^{-5}$, which is beyond the range 
of $Z$ 
values  shown in Fig.\ (\ref{pdf}). 
Nevertheless, the PDF tails shown 
in Fig.\ (\ref{pdf}) can be used to infer the trends of the pristine 
fraction with varying 
$M$ and $L_p.$ 
Since the PDF tail broadens with $M$, we would expect that, 
with the same concentration variance, the pristine fraction contained in the PDF 
would be higher for a scalar field evolving  
in a flow with higher Mach number
or smaller pollutant injection scale.  In fact,
the dependence  of  the PDF tails 
on $M$ and $L_{\rm p}$  induces interesting effects on the pristine fraction 
as a function of time, which will be discussed in detail in \S 5.4.  

\subsection{The Variance Decay}

\begin{figure}
\centerline{\includegraphics[width=1\columnwidth]{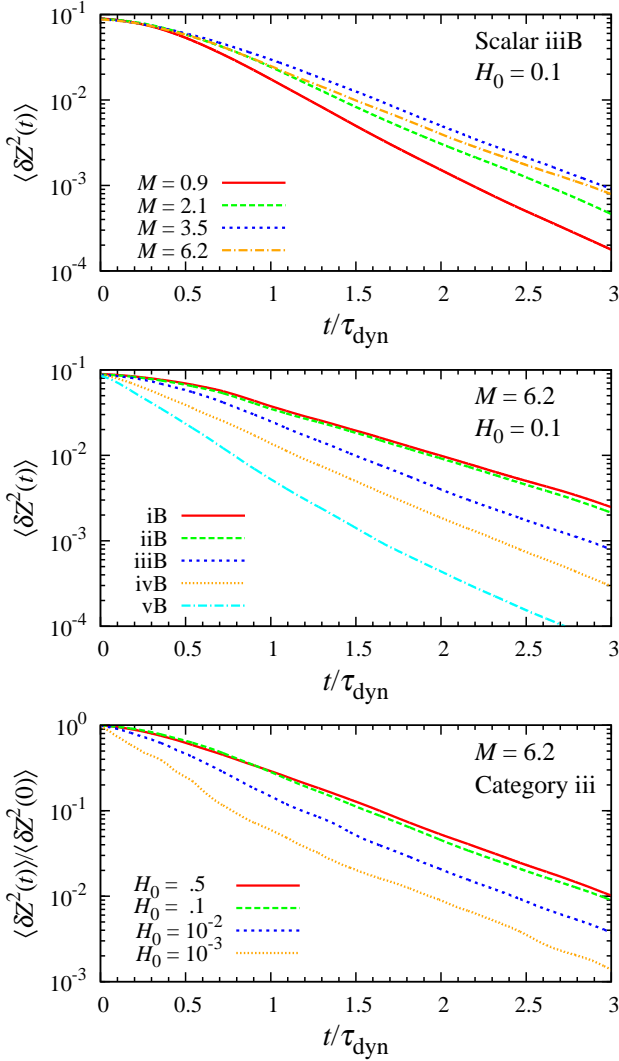}}
\caption{The density-weighted concentration variance as a function 
of time. Top panel: scalar field \rmnum{3}B in four simulated flows with 
$M= 0.9$, 2.1, 3.5 and 6.2. The scalar field has a total pollutant fraction, $H_0$, of 0.1 
and the injection length scale, $L_{\rm p}$, is  $\simeq 1/2$ box size.  
Mid panel: B scalar fields ($H_0 =0.1$) in the $M =6.2$ flow. The five curves 
correspond to five categories in Table 1 with different pollutant shape and injection length scales. 
Bottom panel: scalar fields from category \rmnum{3}\ in the $M =6.2$ flow. Each curve is 
for a different value of the initial pollutant fraction, and the variance 
of each scalar field is normalized to its initial value.}
\label{var} 
\end{figure}

Fig.\ (\ref{var}) plots the variance decay of a number of scalar fields. 
The top panel shows the results for scalar fields \rmnum{3}B in the four simulated flows at different $M$. 
For these scalar fields, $L_{\rm p} \simeq L_{\rm f}$, $H_0 =0.1$, and the initial 
variance is 0.09. The variance decay first slows down with increasing Mach 
number, then becomes slightly faster as $M$ increases from $3.5$ to 6. The same behavior has been found in PS10, where a 
physical explanation was given (see also \S 3). In our simulated flows, 
compressible modes are inefficient in producing small-scale structures. 
Therefore, the mixing efficiency decreases as the fraction of kinetic 
energy contained in compressible modes at inertial-range scales  
increases. At $M\gsim 3$, this fraction saturates at an equipartition value of 1/3, 
and the mixing timescale becomes essentially constant. 
The slightly faster mixing as $M$ increases to $6.2$ is because 
of the effect of strong compression in our simulated flow with $M =6.2$. 
Due to the limited numerical resolution, the strongest compression events in this flow can directly squeeze the 
scalar structures to the diffusion scale and  provide some contribution to the mixing efficiency. 
As discussed in \S 3, the effect of compressible modes on mixing could be 
stronger in a highly supersonic flow with compressive driving, and the 
behavior of the normalized mixing timescale as a function of $M$ in 
compressively driven flows will be studied in a future work. 
In Fig.\ (\ref{var}), we see that the variance decrease is approximately exponential. 
The mixing timescale $\tau_{\rm m}$ is measured to be 0.45, 0.48, 0.58, 0.57 $\tau_{\rm dyn}$ for $M=0.9$, 2.1, 3.5, and 6.2, respectively.  
The 20\% increase in $\tau_{\rm m}$ as $M$ goes from 0.9 to 6.2 is 
consistent with the results of PS10. 

The middle panel of  Fig.\ (\ref{var}) shows the variance of five B scalars
in the $M=6.2$ flow. Each case is from one of the five categories listed 
in Table 1, and they all have $H_0 =0.1$. The curves for 
scalar fields \rmnum{1}B and  \rmnum{2}B are very close to each other. 
The initial pollutant distributions of these two scalar fields 
are a single cube and spherical ball, respectively, and the 
similarity of their variance decay suggests  that the mixing timescale 
is essentially independent of the geometrical shape of the pollutants. 
On the other hand, the mixing timescale decreases steadily as the average 
pollutant separation becomes smaller. The injection scale, $L_{\rm p}$, for 
scalar fields \rmnum{3}B, \rmnum{4}B, and \rmnum{5}B is $\frac{1}{2}$, 
$\frac{1}{4}$, and $\frac{1}{8}$ box size, respectively. We attempted to 
measure $\tau_{\rm m}$ by fitting the five curves  with exponentials in the 
time interval from $0$ to $\simeq 2 \tau_{\rm dyn}$,  which is the time range 
of primary interest for the pristine gas pollution (see \S 5.4).  
The measured values of $\tau_{\rm m}$ are 0.72, 0.71, 0.57, 0.48, and 0.34 $\tau_{\rm dyn}$, 
respectively, for the five curves from top to bottom. The mixing timescale is 
determined by the eddy turnover time at the pollutant injection scale, 
and thus decreases with decreasing $L_{\rm p}$. This physical picture 
also provides an explanation for the scale dependence of the mixing timescale 
found by de Avillez and Mac Low (2002) in a suite of numerical simulations of
mixing in supernova-driven interstellar turbulence.   
  
In the bottom panel, we plot the variance decay of four scalar fields 
from category \rmnum{3}\ in the Mach 6.2 flow. Different curves 
correspond to different values of $H_{0}$.  Unlike the top two panels, 
here we normalize the concentration variance to its initial value, $\langle \delta Z^2(0) \rangle,$
which makes it easier to compare the variance 
decay timescale of different scalars.  For a double-delta PDF (eq.\ (\ref{initialpdf})), 
the initial variance $\langle \delta Z^2(0) \rangle$ is equal to $P_0 H_0 = P_0 (1-P_0) = H_0(1-H_0)$. 
For $H_0 \le 0.5$,  $\langle \delta Z^2(0) \rangle$ decreases with decreasing $H_0$. 
This suggests that, for scalar PDFs 
close to a double delta shape, the variance is 
not a good indicator of the 
pristine fraction, as 
the smaller variance for scalar fields with smaller $H_0$ in bottom 
panel of Fig.\ (\ref{var}) actually corresponds to a larger pristine fraction.
While a better indictor would be the variance normalized 
to the average concentration squared, which measures the 
rms of the fluctuations relative to the mean,     
the  variance plot normalized to the 
initial value nevertheless provides useful information for the timescale of the mixing  process. 

With decreasing $H_0$, the radius of each individual pollutant 
blob becomes smaller, decreasing from about 0.5 box size ($H_0=0.5$) to 
only 0.06 box size ($H_0=10^{-3}$). 
The top two curves for $H_0 =0.5$ and 0.1 are close to each other, 
and the reason is that, for these two scalar fields, both the 
pollutant size and the pollutant separation are close to the flow driving 
scale, $L_{\rm f}$, and the scale (1/4 box size) at which the inertial range of the 
flow starts. The mixing timescales for these two scalar cases are 
thus  given by the turnover time of large eddies of similar sizes. 
The situation is different for the rest two scalar fields. 
For $P \le 0.01$, the size of each individual blob is significantly smaller than 
$L_{\rm f}$. It is also smaller than the average separation, $L_{\rm p}$ ($\simeq L_{\rm f}$), 
between the pollutant blobs. In this case, the mixing process around each blob is 
not synchronized with that over the entire flow. This divides the 
variance evolution into two phases. The early phase 
occurs faster and
is controlled by the turbulent stretching 
rate at smaller scales (the pollutant size). 
This explains the faster variance decay 
for smaller $H_0$ at early times.  After each blob is stretched, spread and 
mixed to a size close to the average pollutant separation, the mixing 
process starts to proceed at a single pace, and the timescale is determined by the 
turnover time of eddies of size $L_{\rm p}$. As seen in the bottom panel of Fig.\ (\ref{var}), 
the variance decay is exponential for all cases at late times with 
essentially the same timescale ($0.6 \tau_{\rm dyn}$). The existence 
of two phases for scalar fields with small $H_0$ also leaves a signature 
in the evolution of the pristine  gas fraction.

\begin{figure*}
\centerline{\includegraphics[width=2\columnwidth]{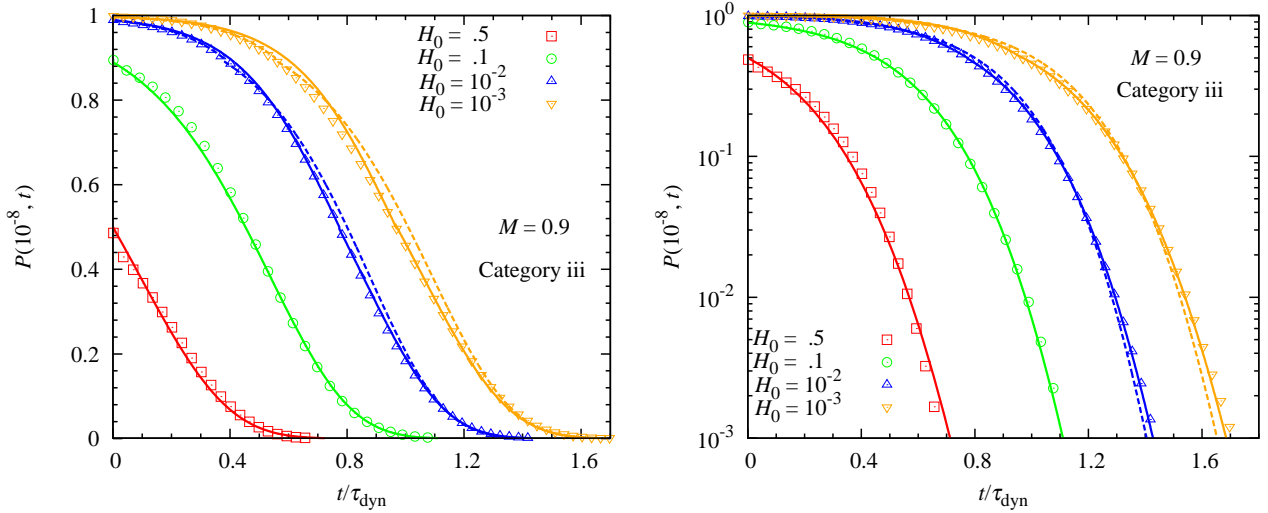}}
\caption{The pristine fraction, $P(10^{-8}, t)$, as a function of time 
for the four scalar fields from category \rmnum{3}\  in the  Mach 0.9 flow.  
The left and right panels show the same plot but on  a linear-linear and 
a linear-log scale, respectively. Lines are fitting functions based on the prediction 
of the self-convolution PDF models.  For scalar fields \rmnum{3}C ($H_0=0.01$)
and \rmnum{3}D ($H_0=0.001$), the dashed lines are obtained by 
a two-phase fitting that connects at $P(10^{-8}, t) =0.9$, while the solid lines 
connect the two phases at $P(10^{-8}, t) =0.5$.  The fitting parameters are given in the text.} 
\label{Mach0.9Four} 
\end{figure*}

\begin{figure*}
\centerline{\includegraphics[width=2\columnwidth]{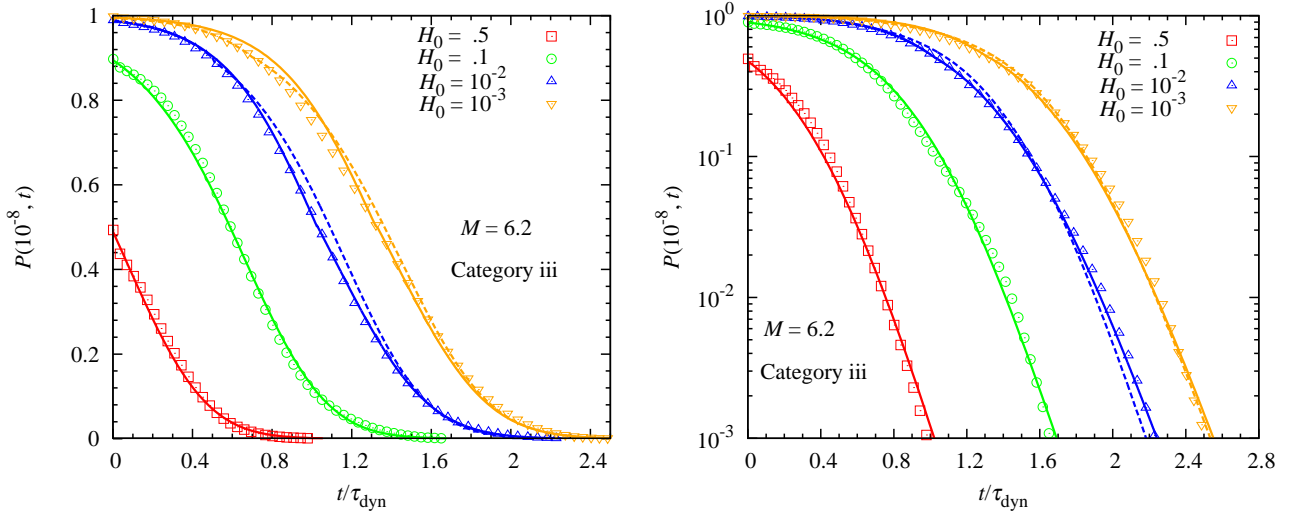}}
\caption{Same as Fig.\ (\ref{Mach0.9Four}), but for scalar fields in the Mach 6.2 flow. 
See text for details.} 
\label{Mach6.2Four} 
\end{figure*}  
 
\subsection{The Pristine Fraction}

\subsubsection{General Results}

We now present results for the decay of the pristine mass 
fraction in our simulated flows.  In PSS, we have shown 
results for three scalar fields,  \rmnum{1}A,  \rmnum{1}B  and \rmnum{1}C 
from category \rmnum{1}\ (with $H_0 =0.5$, 0.1 and 0.01, respectively; see Table 1), 
evolved in two flows with $M=0.9$ and $M=6.2$. The pollutant injection 
scale $L_{\rm p}$ of  those fields was the box size, or about twice the flow driving scale $L_{\rm f}$. 
sIn this section, we consider scalar fields in category \rmnum{3}\ in the $M=0.9$ and 6.2 
flows as primary examples. The injection scale of these fields is smaller, 
with $L_{\rm p} \simeq L_{\rm f}$. 
In the subsequent subsections, we will discuss in details 
the dependence of the pristine fraction decay on various parameters.     
 
Fig.\  (\ref{Mach0.9Four}) shows the mass fraction $P({10^{-8}}, t)$ of the 
flow with concentration level below $10^{-8}$ for scalar fields \rmnum{3}A, \rmnum{3}B, \rmnum{3}C, and \rmnum{3}D 
in the $M=0.9$ flow. The data points are 
simulation results, and the lines are fitting functions based on the predictions 
of the self-convolution PDF models discussed in \S 3.2. The 
left and right panels are the same figure on linear-linear 
and linear-log scales, respectively. 
The linear-linear scale shows the early evolution more clearly, 
while with a linear-log plot  one can see the late-time 
behavior better.  The initial pollutant fraction, $H_0$, of  
the four cases in this figure ranges from 0.5 (\rmnum{3}A) to $10^{-3}$ (\rmnum{3}D).  
As shown in PSS, the prediction, eq.\ (\ref{pfnconvsolution}), of 
the self-convolution models can successfully fit the simulation 
results for scalar fields with $H_0 \ge 0.1$. The fitting lines in Fig.\ (\ref{Mach0.9Four}) 
for the two cases with $H_0 =0.5$ and $0.1$ are the predictions of the convolution models 
with $n=10$. The initial pristine fraction $P_0$ in 
eq.\ (\ref{pfnconvsolution}) is set to 0.5 and 0.9, and the timescale $\tau_{\rm con}$ is taken 
to be $0.27$ and $0.25  \tau_{\rm dyn}$, respectively. Both the 
linear-linear and the linear-log plots show that 
the model prediction matches the simulation data well, 
suggesting that 
the pollution process in turbulent flows may be 
adequately described as a self-convolution process.

If $H_0$ is smaller than $\simeq 0.1$,  the evolution of the pristine 
fraction is more complicated, and one cannot satisfactorily 
fit the entire evolution of $P(Z_{\rm c}, t)$ with  the convolution 
model, eq.\ (\ref{pfnconvsolution}), by properly choosing the 
parameters $n$ and $\tau_{\rm con}$. In this case, the pollution 
process shows different behaviors at early and late evolution 
phases.  A two-phase behavior for scalar fields with small $H_0$ was 
actually seen earlier in the scalar variance decay (see the 
bottom panel of Fig.\ (\ref{var}) for scalars
\rmnum{3}C and \rmnum{3}D in the Mach 6.2 flow).
For these cases, only a small fraction of the flow material,
near the pollutant blobs, experiences PDF convolution 
at early times, because the amount of pollutants 
available for mixing is limited. This suggests that the convolution of the concentration 
PDF is local in space in the early phase, and, based on the physical 
discussion in \S 3.1,  the mixing process in this phase would 
be better described by a ``discrete" version of the convolution model 
(with $n$=1). Consistent with this picture, we find that
the pristine fraction in the early phase is in good agreement 
with the prediction, eq.\ (\ref{pfintegralsolution}), of the ``discrete" convolution model, 
or equivalently eq.\ (\ref{pfnconvsolution}) with $n$=1. With time, more and more flow 
is polluted, and the mixed flow material then acts as sources for further pollution. 
The PDF convolution would thus becomes more global in spatial 
space and hence more continuous in Laplace space, leading to an increase in $n$. 
As described in \S 3.1, $n$ essentially corresponds to the 
degree of spatial locality for the PDF convolution. Recognizing 
the different mixing behaviors at early and late times, we attempted to 
apply a two-phase fitting procedure for scalar fields  with 
$H_0\le 0.01$ (see PSS).  

For a two-phase fit, we need to determine the transition time at 
which the two behaviors connect. Since the generalized 
convolution model with a single phase provides perfect fits to scalar fields 
with $H_0 \ge 0.1$, one may expect that the second phase with a more global 
PDF convolution starts when the pristine fraction, $P(Z_{\rm c}, t)$, 
decreases to 0.9. We thus first  tried to obtain a fitting function that connects 
the two phases at the time $t_{0.9}$ when $P(Z_{\rm c}, t) = 0.9$. 
The results are shown as dashed lines in Fig.\  (\ref{Mach0.9Four}). In 
these lines, the early phases are fit by the ``discrete" model, eq.\ (\ref{pfintegralsolution}), 
with $\tau_{\rm con} = 0.17 \tau_{\rm dyn}$ for both case \rmnum{3}C ($H_0= 0.01$) 
and case \rmnum{3}D ($H_0=0.001$). Once $P(Z_{\rm c}, t)$ decreases to 0.9, we 
use the generalized model prediction, $P(Z_{\rm c}, t) = 0.9/[0.9^{1/n} + (1-0.9^{1/n}) \exp((t-t_{0.9})/\tau_{\rm con})]^n$
(c.f. eq.\ (\ref{pfnconvsolution})) with  $n=10$. The timescale $\tau_{\rm con} $ 
for the late phase is set to $0.23$ and $0.25 \tau_{\rm dyn}$ 
for  case  \rmnum{3}C and case \rmnum{3}D,  respectively. The fitting values adopted for 
$n$ and $\tau_{\rm con}$ in the late phase are close to those used for the scalar fields 
with $H_0 \ge 0.1$. This means that, once the polluted fraction becomes larger than $\simeq 0.1$,  
the pristine fraction decays in a similar way as the $H_0 \gsim 0.1$ fields.  
The fitting quality of the dashed lines appears to be acceptable.  
To distinguish the two convolution timescales in the early and late 
phases, we denote them as $\tau_{\rm con1}$ and $\tau_{\rm con2}$, 
respectively. We will also use $\tau_{\rm con2}$ to denote the 
convolution timescale for scalar fields with $H_0 \ge 0.1$ because the decay 
of the pristine fraction for those fields is similar to the later-phase evolution of  
the $H_0 \le 0.01$ cases.    

We find that one can obtain better fits  for scalar fields with $H_0 \le 0.01$ by 
connecting the two phases at later times. As shown in PSS, for these fields  
the ``discrete" model well matches the simulation data in an extended time range 
until $P(Z_{\rm c}, t)$ drops to 0.2-0.3. This allows us to connect the early and late 
behaviors at a time significantly larger than $t_{0.9}$. 
It turns out that the fitting quality is actually significantly 
improved if we start to use the generalized model with $n=10$ at 
times when $P(Z_{\rm c}, t)$ is smaller than $\simeq0.7$. The 
solid lines in Fig.\  (\ref{Mach0.9Four}) for cases \rmnum{3}C and \rmnum{3}D 
show the fitting functions  that connect the ``discrete" model and the later phase 
at  $t_{0.5}$ when $P(Z_{\rm c}, t)$ decreases to 0.5.   
In the fitting curves,  $\tau_{\rm con1}$ for 
the ``discrete" phase is set to $0.18 \tau_{\rm dyn}$ for both case \rmnum{3}C and case \rmnum{3}D.   
Starting from $t_{\rm 0.5}$, we use the generalized model $P(Z_{\rm c}, t) = 0.5/[0.5^{1/n} + (1-0.5^{1/n}) \exp((t-t_{0.5})/\tau_{\rm con})]^n$ 
with $n=10$. The timescale $\tau_{\rm con2} $ is set to $0.25$ and 
$0.27 \tau_{\rm dyn}$ for  case \rmnum{3}C and case \rmnum{3}D, respectively.  
From Fig.\  (\ref{Mach0.9Four}), the two-phase fitting lines connecting at $t_{0.5}$ 
agree with the  data considerably better than the dashed lines that connect at $t_{0.9}$.   
Our choice here to connect the two phases at $t_{\rm 0.5}$ is somewhat 
arbitrary because there is an extended time range where both the ``discrete" model and the $n=10$ 
model can match the simulation data (PSS).  In fact, combining the two 
models at any time with $0.2 \lsim P(Z_{\rm c}, t) \lsim 0.7$ would give fitting 
curves of similar quality. The parameter $n$ adopted in both the 
dashed lines and the solid lines is 10, i.e.,  the same as used for the scalar fields 
with $H_0 \ge 0.1$. This is also the case for the convolution timescale $\tau_{\rm con2}$ 
in the second phase. The values of $\tau_{\rm con2}$ used in the solid 
lines almost coincide with those adopted in the fitting lines for scalar fields 
with $H_0 \ge 0.1$,  and in the dashed lines the $\tau_{\rm con2}$ values are only 
slightly smaller (by $\lsim 10\%$).  For  the early phase,  the adopted values for 
the timescale $\tau_{\rm con1}$  in the solid and dashed lines are very close too.    

Our result that connecting the two phases at $t_{0.5}$ yields better 
fits than at $t_{0.9}$ seems to suggest that, for scalar fields with $H_0 \lsim 0.01$, 
the pollution process does not make an immediate transition from the ``discrete" to the 
generalized convolution model with larger $n$, when the pristine fraction decreases 
to ${0.9}$. The transition tends to occur later. Considering that the generalized model 
with a single phase works perfectly for scalar fields with $H_0 \gsim 0.1$, this implies 
that the time at which the generalized convolution phase starts is not simply 
controlled by the value of the pristine or polluted fraction: It 
appears to have some dependence on whether the initial 
pristine fraction is larger or smaller than $\simeq 0.9$. This  does not cause 
any problems in a practical application if the exact value of the 
initial pollutant fraction, $H_0$, is known. One can use the generalized 
convolution model with a single phase if $H_0 \ge 0.1$, or adopt a two-phase 
model connecting at, say, $t_{0.5}$ if  $H_0 \lsim 0.1$. 

However, there is some complication when applying this procedure to the subgrid 
model we will construct in \S 7 for large-eddy simulations for the pollution of primordial 
gas in early galaxies. For example, if at a given time the pristine fraction in a 
computational cell is, say, in between 0.9 and 0.5,  then the choice of 
using the ``discrete" model or the generalized model at that 
moment depends on whether the pristine fraction in that cell was larger 
or smaller than 0.9 when it was first polluted. This would make the 
implementation of our subgrid model complicated, as it requires 
keeping some information on the pollution history in each cell. 
We advocate simply using the generalized 
convolution model for any cells with a 
pristine fraction smaller than 0.9, as it gives acceptable, if not 
perfect, fits to our simulation data for $H_0 \le 0.01$ scalars at any time 
after $ t_{0.9}$.  
In the following subsections, we will only consider fitting functions that 
connect at $t_{0.5}$ for scalar fields with $H_0 \le 0.01$, as they are in 
better agreement with the simulation data. We will 
tabulate the fitting parameters obtained from such fits in \S 5.4.6.  
If in a particular application connecting the early and later 
phases at $t_{0.9}$ is preferred rather than at $t_{0.5}$, our 
tabulated parameters would still be applicable, as the best-fit  
parameters used in the fitting curves that connect 
at $t_{0.9}$ and $t_{0.5}$ are very close.

In Fig.\ (\ref{Mach6.2Four}), we show the simulation 
results and the fitting curves for scalar fields from the same category (\rmnum{3}), but in the 
Mach 6.2 flow. For the fields with $H_0 =0.5$ and $0.1$, 
the data points are fit by the convolution model, eq.\ (\ref{pfnconvsolution}), 
with $n=3$. The timescale $\tau_{\rm con2}$ for the two cases  is set to 
$0.30$ and $0.31 \tau_{\rm dyn}$, respectively.  Similar to the $M=0.9$ case, 
two-phase models connecting  at $t_{0.9}$ (dashed lines) and $t_{0.5}$ 
(solid lines) are used for the rest two cases with $H_0 \le 0.01$.  
For the dashed lines,  the early phase is fit by the ``discrete"  
convolution model with $\tau_{\rm con1} = 0.22\tau_{\rm dyn}$ for scalar field \rmnum{3}C and 
$\tau_{\rm con1} = 0.24\tau_{\rm dyn}$ for case \rmnum{3}D, and for the late phase  we used 
the $n=3$ convolution model with $\tau_{\rm con2} = 0.31 \tau_{\rm dyn}$ and $\tau_{\rm con2} = 0.33 \tau_{\rm dyn}$ for the two cases, 
respectively. For the solid lines that connect at $t_{0.5}$,  
the fitting parameters for  the  ``discrete"  phase are $\tau_{\rm con1} = 0.23\tau_{\rm dyn}$ 
and $\tau_{\rm con1} = 0.25\tau_{\rm dyn}$ for case \rmnum{3}C for case \rmnum{3}D, 
respectively, and the late evolution stage is fit with $n=3$ and $\tau_{\rm con2} = 0.34 \tau_{\rm dyn}$ 
for both cases. Again, the fitting quality is better with a connection at $P(Z_{\rm c}, t) =0.5$.  
In all cases, the fitting parameters, $n$ and $\tau_{\rm con2}$, adopted for the 
scalar fields with $H_0 \ge 0.1$ and for the late phases of the $H_0\le 0.01$ fields are very close, 
suggesting a universal decay behavior of the pristine fraction 
once the polluted fraction exceeds $\simeq 0.3$.

We find that,  for scalar fields with  $H_0 \le 0.01$, it is more difficult to fit the 
early phases as $H_0$ decreases, and the fitting quality 
becomes poorer with decreasing $H_0$ (see Fig.\ (\ref{Mach6.2Four} and \ref{Mach0.9Four})). 
The pollutant size is smaller for smaller values of $H_0,$ and this may 
cause some complexities for the prediction of the pristine fraction. For example,  
as $H_0$ decreases to 0.001, the blob diameter is about the size of 30 computational 
cells, which is close to the scale where the flow inertial range ends. The 
first effect is that, with time, the size of the polluted region around each pollutant blob 
increases, and the turbulent stretching timescale in the polluted regions 
may increase with time. This is not accounted for in the convolution 
models since the convolution timescale is set to be constant.   
Another effect arises from the fact that the turbulent stretching 
rate has larger spatial variations at smaller scales. The turbulent eddy ``seen" 
by a small pollutant blob may have a stretching rate different from the 
average value at the pollutant size. The increase in the  
amplitude of the stretching rate fluctuations with decreasing length scale 
indicates the turbulent intensity around smaller blobs is more ``random". 
In the early phase, the flow mass polluted by a single blob is expected to 
increase exponentially with the stretching rate. Therefore, using an 
average stretching rate for all the pollutant blobs may not give a precise prediction. 
The overall pollution rate depends on the turbulent stretching rates  ``seen" 
by all the pollutants at early times. The blobs encountering more intense 
eddies provide a larger contribution to the pollution process, and vice versa.   
The effect is further amplified by the phenomenon of intermittency:  the PDF 
of the stretching intensity exhibits fatter tails toward smaller scales. Therefore, 
the small blobs have large chance to encounter extreme stretching events. 
Clearly, the effect of intermittency makes it more difficult to predict the 
pristine fraction for scalar fields with smaller $H_0$.  

We point out that, for a given scalar field, the parameters, $n$ and $\tau_{\rm con2}$, that can fit  
scalar fields with $H_0 \ge 0.1$ or the late phases of $H_0 < 0.1$ cases are not unique. 
In fact, a (small) range of parameter pairs ($n$,  $\tau_{\rm con2}$) can give acceptable fits to 
the simulation data. For example, if a somewhat smaller (larger) $n$ is used, one could also 
have a similar fit with correspondingly smaller (larger) value of $\tau_{\rm con2}$.  
When obtaining the best-fit parameters, we attempted to select a 
single value of $n$ that provides good fits to all scalar fields in each 
category. With the chosen $n$, we then determine the best-fit value 
of $\tau_{\rm con2}$ for each scalar field in the category. As discussed above, 
the timescale turns out to be similar for all cases in a given category.  

\begin{figure*}[ht]
\centerline{\includegraphics[width=2.\columnwidth]{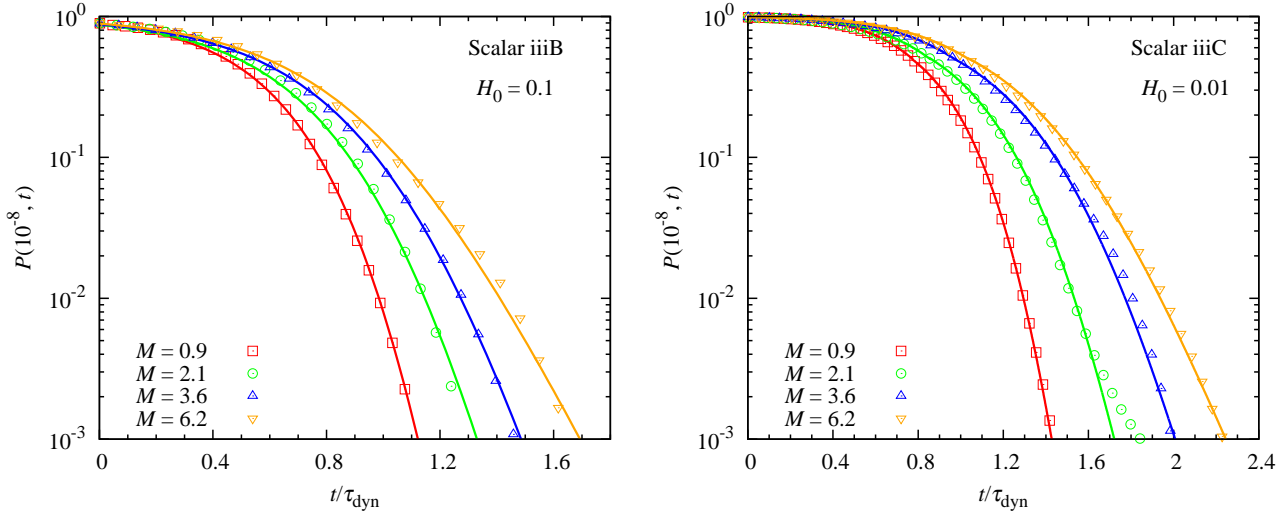}}
\caption{Mach number dependence of the pristine fraction. 
The two panels show $P(10^{-8}, t)$ as a function of time for 
scalar fields \rmnum{3}B (left) and \rmnum{3}C (right) in four simulated flows at Mach 0.9, 2.1, 3.5 and 6.2. 
Lines are fitting functions based on the predictions of self-convolution 
models. See text for fitting parameters.} 
\label{Machdependence} 
\end{figure*}

A comparison of Figs.\ (\ref{Mach0.9Four}) and (\ref{Mach6.2Four}) shows that, when the time is 
normalized to the flow dynamical timescale, the pristine fraction in the Mach 6.2 flow survives 
for significantly longer than in the Mach 0.9 case (see also PSS).  We discuss this Mach number dependence in the following subsection.  

\subsubsection{Dependence on the Flow Mach Number}

In Fig.\ (\ref{Machdependence}), we show the evolution of the pristine fraction for scalar 
fields \rmnum{3}B (left) and \rmnum{3}C (right) in our four simulated flows. As observed earlier, 
with $t$ normalized to the flow dynamical time, $\tau_{\rm dyn}$, the decrease of the 
pristine fraction becomes slower with increasing Mach number.  In the fitting lines for 
case  \rmnum{3}B (left panel), the parameter pair ($n$, $\tau_{\rm con2}$) is set to (10, 0.26$\tau_{\rm dyn}$),  (6, 0.29 $\tau_{\rm dyn}$),  
(5, 0.31$\tau_{\rm dyn}$), and (3, 0.31$\tau_{\rm dyn}$) for the four flows with $M=0.9$, 2.1, 3.5, and 6.2, 
respectively. 
Again, we see that $\tau_{\rm con2}$ first increases with $M$ and then saturates for $M\gsim 3$. 
This trend is similar to that of the variance decay timescale, $\tau_{\rm m}$, as a 
function of $M$ (see the top panel of Fig.\ (\ref{var})). As explained in \S 5.3, $\tau_{\rm m}$ 
increases as the energy fraction in compressible modes increases and then becomes roughly 
constant when compressible energy fraction saturates at $M \gsim 3$. The same reasoning 
also applies here for the trend of $\tau_{\rm con2}$ with $M$.   
Similar to the discussion in \S 5.3 on $\tau_{\rm m}$, the convolution timescale may 
have a different behavior with $M$ in compressively driven flows.

In \S 5.2, we showed that, at a similar concentration variance, the PDF tail becomes 
broader as $M$ increases,  
most likely because of a the increase in turbulent intermittency.
Because the pristine fraction corresponds to 
the far left tail of the concentration PDF, the effect also 
slows the pristine gas pollution at larger $M$. Another effect of the 
broadening of the PDF tail with increasing $M$ is that it changes the shape 
of the pristine fraction vs. time curve, as seen in Fig.\  (\ref{Machdependence}). 
To fit the pristine fraction  in flows at different $M$, we varied the parameter, $n$, in 
the self-convolution model, which controls the shape of the fitting function, 
eq.\ (\ref{pfnconvsolution}). The best-fit value of $n$ decreases from 10 to 3 as $M$ 
increases from 0.9 to 6.2. This decrease of $n$ is expected from the fact that the 
self-convolution model with smaller $n$ would predict broader PDF tails (see \S 3.1). 
The convolution model was originally proposed for mixing in incompressible 
turbulence, where $n$ was a pure parameter without a clear connection 
to the mixing physics. Our finding that the convolution model 
with properly chosen $n$ can well describe the pristine fraction evolution in 
compressible turbulent flows motivates a physical interpretation of  the parameter.

A possible intuitive reason for why $n$ decreases with the flow Mach 
number is that $n$ reflects the degree of spatial 
locality of the PDF convolution, with more local mixing events 
implying a smaller $n$. In a highly supersonic turbulence, the majority of the flow 
mass resides in a small fraction of volume, i.e., in the dense postshock regions. 
Therefore, mixing of the pollutants into and within local regions of 
high densities is crucial toward the final homogenization. The dense 
postshock regions are persistent with a lifetime on the order of the flow 
dynamical time, and the time scale for the homogenization between 
different postshock regions is  expected to be on the same order. This 
suggests that the presence of local dense regions may suppress the 
possibility of a global PDF convolution. In that case, as the flow Mach number 
increases, the convolution would become more local, leading to a decrease in $n$.  
Based on this argument, we speculate that, in compressively driven flows 
at similar $M$, the parameter $n$ would be smaller than  in our simulated 
flows. The convolution in a compressively driven flow is expected to be more local 
because the density fluctuations are stronger (e.g., Federrath et al.\ 2010).

\begin{figure*}[ht]
\centerline{\includegraphics[width=2\columnwidth]{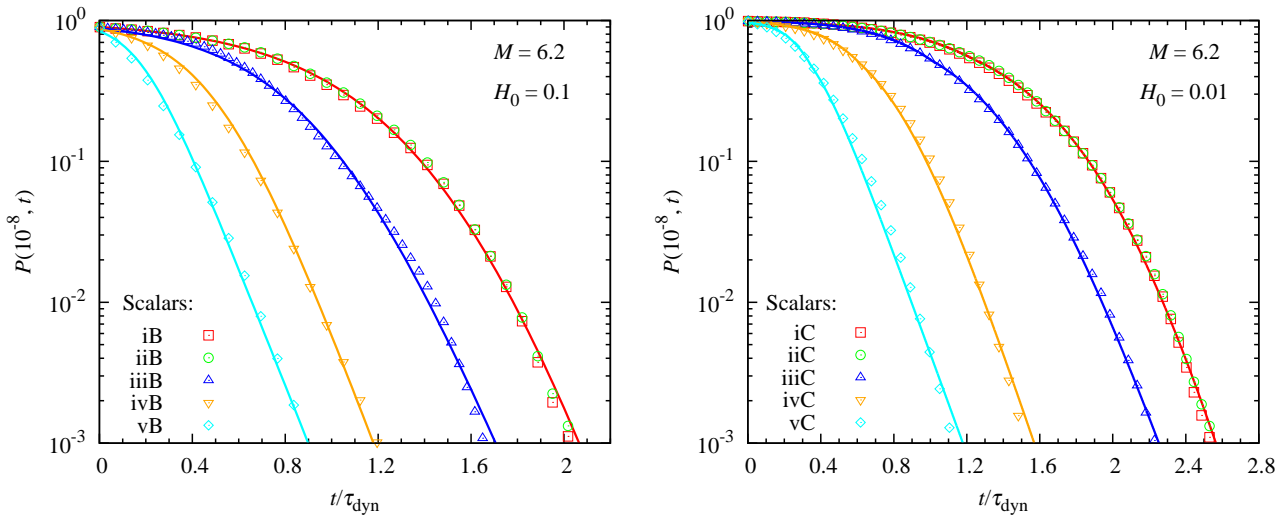}}
\caption{Dependence of the pristine fraction on the pollutant injection scale, $L_{\rm p}$. 
The figure plots $P(10^{-8}, t)$ as a function of time for scalar fields  with 
different pollutant  shapes and spatial distributions in the Mach 6.2 flow.  
The left and right panels show B and C fields with  $H_0 =0.1$ and 0.01, 
respectively. The five cases in each panel are from the five categories listed in 
Table 1. The simulation data for scalar fields in categories \rmnum{1}\ (single cubic pollutant) 
and \rmnum{2}\ (single spherical pollutant) almost coincide. The pollution of 
pristine gas becomes progressively faster as the pollutant injection length scale 
decreases. See text for description of the fitting lines.}
\label{lengthscale} 
\end{figure*}

The right panel of Fig.\ (\ref{Machdependence}) plots the results for case \rmnum{3} C
 with $H_0 =0.01$ in the four simulated flows. Again the decrease of the pristine fraction 
is slower in flows with larger $M$. As discussed earlier, a two-phase fitting scenario is 
needed for scalar fields
with $H_0 \le 0.01$. Using the discrete convolution model to fit the early-phase evolution, 
we find that the timescale $\tau_{\rm con1}$ is 0.17, 0.19, 0.22 and 0.23 $\tau_{\rm dyn}$ for $M=0.9$, 
2.1, 3.5 and 6.2, respectively. The two phases are connected at $t_{0.5}$.
Note that the timescale $\tau_{\rm con1}$ also increases with $M$ 
at first and then saturates for $M\gsim 3$.  For the late phase of scalar case \rmnum{3} C, we adopted the 
same values of $n$ (i.e., 10, 6, 5, and 3) for the 
flows as used in the case of \rmnum{3} B. To match the simulation data, $\tau_{\rm con2}$ in 
the late phase is set to 0.25, 0.30, 0.34 and 0.34 $\tau_{\rm dyn}$ 
for $M=0.9$, 2.1, 3.5 and 6.2, respectively. Again, these numbers are close to the 
best-fit values for case \rmnum{3} B shown in the left panel.

In summary, we found that the  pollution of the pristine gas is slower in flows at 
higher $M$. Two reasons are responsible for this behavior. First, the mixing 
(or variance decay) timescale $\tau_{\rm m}$ becomes larger as $M$ 
increases. Second, at the same concentration variance, the left PDF tail broadens  with 
$M,$  and this corresponds to a larger pristine fraction.

\subsubsection{Dependence on the Pollutant Injection Length Scale}

Next, we study the dependence of the pristine fraction evolution on 
the initial spatial configuration of  the pollutants, i.e., on how the pollutants 
are released into the flow. Each category in Table 1 represents a 
different pollutant shape or distribution at the initial time. In Fig.\ (\ref{lengthscale}), 
we compare the simulation results for scalar fields from different categories 
in the Mach 6.2 flow. The left panel shows five B fields with $H_0= 0.1$, and the 
right panel is for C cases with $H_0 =0.01$. The initial condition for the 
scalar fields in categories \rmnum{1}\ and \rmnum{2}\ is a single pollutant 
cube and a single spherical  blob, respectively,  
and the pristine fraction evolution for scalar fields in these two 
categories is almost the same, suggesting the geometric shape of 
the pollutant blob does not affect the pollution rate.

On the other hand, the pollution process has a sensitive dependence on 
on the injection length scale, $L_{\rm p}$. For scalar fields in first two 
categories (\rmnum{1}\ and \rmnum{2}), $L_{\rm p}$ is 
about equal to the box size, or twice the flow driving scale, $L_{\rm f}$.
For categories \rmnum{3},  \rmnum{4}\ and \rmnum{5},  
$L_{\rm p} \simeq L_{\rm f}$, $ L_{\rm f}/2$ and $L_{\rm f}/4$, respectively,  
As expected in \S 4,  
the decay of the pristine fraction becomes progressively faster with 
decreasing $L_{\rm p}$. The four lines in each panel of Fig.\ (\ref{lengthscale}) 
are fitting functions based on the self-convolution models. Since the data 
points almost coincide for scalar fields in categories \rmnum{1}\ and \rmnum{2}, 
a single fitting curve (the line on the right) works for both cases. The other three fitting lines, 
from the right to the left,  correspond to scalar fields in categories \rmnum{3}, \rmnum{4}, and \rmnum{5}\, 
respectively. The fitting parameters used for the B fields in the left panel 
are  $n=5, 3, 2,$ and 1, and $\tau_{\rm con2} = 0.42, 0.32, 0.19$ and 
$0.11\tau_{\rm dyn}$, respectively, for the four lines from the right to the left.
  
The timescale $\tau_{\rm con2}$ decreases by $\simeq 20 \%$ as $L_{\rm p}$ 
changes from $2L_{\rm f}$ to $L_{\rm f}$, 
and, as$L_{\rm p}$ decreases further 
below $L_{\rm f}$,  the decrease of $\tau_{\rm con2}$ is faster,
dropping by $\simeq 40\%$ for each factor of 2 in  $L_{\rm p}$.
This trend is similar to the dependence of the variance decay 
timescale, $\tau_{\rm m}$, on $L_{\rm p},$ 
which is controlled by the eddy turnover time at $\simeq L_{\rm p}$
and decreases with decreasing $L_{\rm p}$. This also explains the 
faster pollution of the pristine gas if the pollutants are injected at smaller scales. 
Recalling that $\tau_{\rm m}$ was measured to be $0.72, 0.57, 0.48$ 
and 0.34 $\tau_{\rm dyn}$ for the same B fields in the Mach 6.2 flow 
with $L_{\rm p}  \simeq 2, 1, 0.5$ and 0.25$L_{\rm f}$, respectively 
(see the mid panel of Fig.\ \ref{var}), we see that $\tau_{\rm con2}$ 
has a more sensitive dependence on $L_{\rm p}$ 
than $\tau_{\rm m}$.  A possible reason for 
this is that the exposure of the pollutants to the pristine flow may be 
an important factor for the pollution efficiency, and, with decreasing $L_{\rm p}$, 
the number of pollutant blobs increases rapidly, leading to enhanced pollutant exposure.  

The trend that  $n$ becomes smaller for scalars injected at smaller 
scales corresponds to broadening of the PDF tails with decreasing 
$L_{\rm p}$ found in \S 5.2.  
The PDF tails become broader because the flow structures 
``seen"  by the scalars with smaller $L_{\rm p}$ are more 
intermittent, and thus the dependence of $n$ on $L_{\rm p}$ 
is related to the higher degree of turbulent intermittency 
at smaller scales.  
If the turbulent flow is driven compressively, the decrease 
of $n$ with decreasing  $L_{\rm p}$ may be faster due to 
stronger intermittency of the flow. 
Note that broadening of the PDF tails 
makes
the pristine fraction larger, 
but this effect is minor in comparison to the faster 
decrease of the pristine fraction caused by the smaller 
mixing timescale at smaller $L_{\rm p}$. 
The decrease of $n$ 
with decreasing $L_{\rm p}$ may also be understood from a more intuitive 
argument.

For scalar fields with a small $L_{\rm p}$, each pollutant is 
stretched by a local velocity structure, and the mixing of each 
pollutant blob with the surrounding flow proceeds largely
independently at early times. The pollution process is almost 
complete when the mixed areas by the pollutant blobs 
start to overlap (see bottom panels of Fig.\ 1), meaning that 
the PDF convolution occurs locally and independently in 
different regions of size $L_{\rm p}$ during most of the 
mixing process.  As the injection scale decreases,  
the PDF convolution becomes 
more local,
leading to a smaller value of $n$, which corresponds to a higher 
degree of spatial locality in the PDF convolution (see \S 3.1).

For C fields shown in the right panel, a two-phase scenario connecting at $t_{0.5}$ is used to obtain 
the fitting lines. In the early phase, the timescale, $\tau_{\rm con1}$, in the 
discrete convolution model is taken to be 0.30, 0.24, 0.17,  and 0.1 $\tau_{\rm dyn}$, 
respectively, for the four fitting lines from right to left.  The 
dependence of $\tau_{\rm con1}$ is similar to that of $\tau_{\rm con2}$ for the B 
cases. It is first reduced by 20\%, as $L_{\rm p}$ decreases to $L_{\rm f}$, 
and then decreases faster, by $\simeq 30-40\%$, as $L_{\rm p}$ 
decreases further by each factor of 2.  For the late phase, we adopted the 
same values (5, 3, 2 and 1) of $n$ as for the corresponding B cases
shown in the left panel, and $\tau_{\rm con2}$ is set  to $0.43, 0.34$, 0.22, and 0.12 
$\tau_{\rm dyn}$ for scalar fields with $L_{\rm p}  \simeq 2$, 1, 0.5 and $0.25 L_{\rm f}$, 
respectively.  Again, these values of $\tau_{\rm con2}$ are close to those 
used in the fitting lines for the corresponding B cases.  It is interesting to note that, 
for case \rmnum {5}C, $n=1$ is adopted in both the early and late phases, although 
the timescales $\tau_{\rm con1}= 0.1 \tau_{\rm dyn}$ and $\tau_{\rm con2}= 0.12 \tau_{\rm dyn}$ 
are slightly different.

We also examined the $L_{\rm p}$ dependence for all the other scalar fields including those in 
the other three flows. We found similar trends for the parameters, $n$, $\tau_{\rm con1}$ 
and $\tau_{\rm con2}$, with varying pollutant injection scale. The results
are tabulated and further discussed  in \S 5.4.6.

\begin{figure*}
\centerline{\includegraphics[width=2\columnwidth]{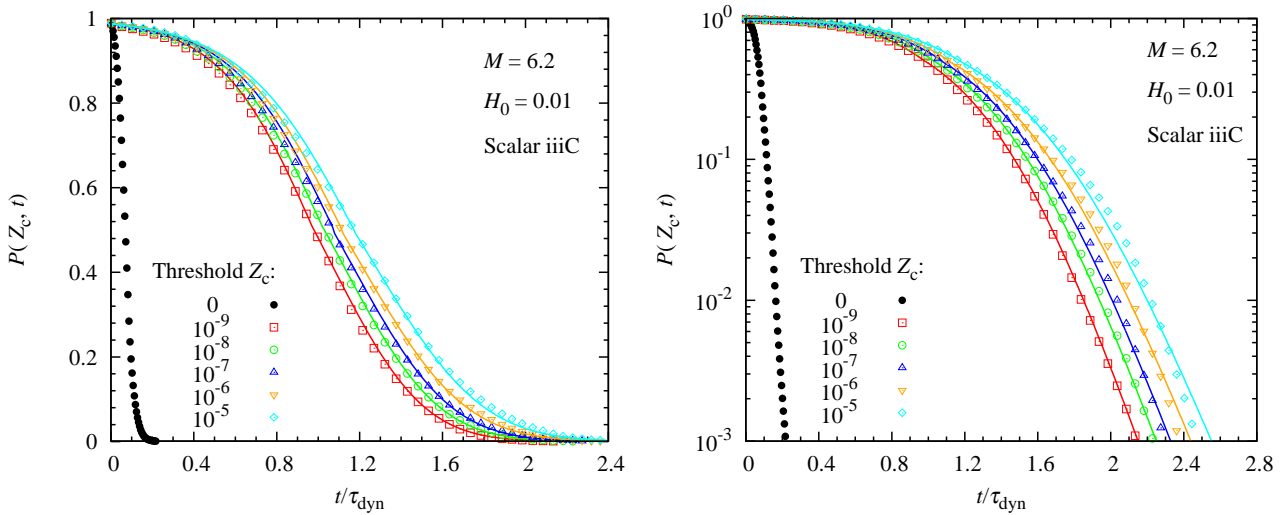}}
\caption{Dependence of the pristine fraction on the metallicity threshold,  $Z_{\rm c}$. 
The figure plots  $P(Z_{\rm c}, t)$, as a function of time for different values of $Z_{\rm c}$. 
The scalar field shown here is \rmnum{3}C in the Mach 6.2 flow. 
The left and right panels are the same figure plot on linear-linear 
and linear-log scales, respectively.}
\label{Threshdependence} 
\end{figure*}

\subsubsection{Dependence on the Threshold Metallicity}

When presenting simulation results in earlier subsections, we set the 
threshold metallicity to $Z_{\rm c} = 10^{-8}$ as a representative value,
but, as discussed in the Introduction, the threshold value for the 
transition to normal star formation is uncertain. We therefore need to 
study the dependence of the pristine fraction $P(Z_{\rm c}, t)$ on the 
$Z_{\rm c}$. In Fig.\ (\ref{Threshdependence}), we plot $P(Z_{\rm c}, t)$ at 
different threshold values for scalar \rmnum{3}C in the Mach 6.2 flow. 
The two panels show the same figure on linear-linear and linear-log scales, respectively.  
We consider the scalar case C as an example, with which we can 
examine the $Z_{\rm c}$ dependence of both convolution timescales, 
$\tau_{\rm con1}$ and $\tau_{\rm con2}$, for the early and late phases, respectively.  

The filled circles in Fig.\ (\ref{Threshdependence}) correspond to the fraction of 
exactly pristine flow material with $Z=0$.  This fraction decreases to zero 
almost instantaneously. This is caused by the effect of numerical diffusion. During each timestep, 
any computational cell adjacent to one that contains 
pollutants or has been polluted by earlier mixing events will obtain 
a finite, 
but often extremely small concentration. This means 
that the exactly-pristine flow material would be completely lost
in a number of steps $ \approx L_{\rm p}/\Delta$ with $L_{\rm p}$ 
and $\Delta$ the average pollutant separation 
and the computation cell size, respectively. The timestep in our 
simulation is approximately given by $\Delta/v_{\rm max}$, where 
$v_{\rm max}$ is the maximum flow velocity at a given time. Therefore, the survival time 
of exactly-pristine gas is $L_{\rm p}/v_{\rm max}$, which is much 
smaller than the flow dynamical time $L_{\rm f}/v_{\rm rms}$ because 
$v_{\rm max} \gg v_{\rm rms}$. The almost immediate removal of exactly 
metal-free gas by numerical diffusion is analogous to the expectation in \S 3.2 
that the molecular diffusivity alone tends to reduce the exactly-pristine 
fraction $P(t)$ to zero instantaneously (see also PSS), although the numerical diffusion in our simulation probably
has a different form and amplitude than the realistic molecular diffusivity.  

The open symbols in Fig.\ (\ref{Threshdependence}) show 
simulation data for finite, and more realistic,
threshold values in the range from  $10^{-9}$ 
to $10^{-5}$. For $Z_{\rm c}$ in this range, the simulation data for $P(Z_{\rm c}, t)$ 
can be fit by the self-convolution models. The fitting lines in 
Fig.\ (\ref{Threshdependence}) are obtained using a two-phase 
scheme which combines the early and late behaviors at $t_{\rm 0.5}$. 
Fitting the early phase with the discrete convolution model,  
we find that the dependence of the timescale $\tau_{\rm con1}$ on $Z_{\rm c}$ is very weak, with 
$\tau_{\rm con1} =$  0.226, 0.233, 0.244, 0.253 and 0.271 $\tau_{\rm dyn}$ for 
the five threshold values increasing from $10^{-9}$ to $10^{-5}$. 
If we express the dependence as a power-law, $\tau_{\rm con1} \propto Z_{\rm c}^{a_1}$, 
the exponent, $a_{1}$, would be very small, $\simeq 0.015$. 
Note that the increase seems to be faster (by about $7\%$), as the threshold increases 
from $10^{-6}$ to $10^{-5}$.  For the late phase, we fixed the parameter $n$ at 3 
(the same as used before for this scalar),  and adjusted the timescale $\tau_{\rm con2}$ 
to match the data points for different threshold values. The best-fit value for $\tau_{\rm con2}$ was 
found to be 0.33, 0.344, 0.355, 0.375 and 0.39 $\tau_{\rm dyn}$ for $Z_{\rm c} = 10^{-9}, 10^{-8}, 10^{-7}, 10^{-6}$, and $10^{-5}$, 
respectively.  On average, $\tau_{\rm con2}$ increases by 4\% as $Z_{\rm c}$ 
increases by each factor of 10, and the dependence can be roughly written 
as $\tau_{\rm con2} \propto Z_{\rm c}^{a_2}$ with $a_2 = 0.02$.  
A similar $Z_{\rm c}$ dependence of the convolution timescales 
was also found for other cases with different $L_{\rm p}$ 
and in flows at different $M$.  There is also a general trend that 
the increase of the convolution timescale with the threshold becomes 
faster as $Z_{\rm c}$ increases to the highest value, $10^{-5}$, considered in our study.

The weak power-law dependence of the convolution timescales,  $\tau_{\rm con1}$ 
and $\tau_{\rm con2}$, on $Z_{\rm c}$ may extend to a range of threshold 
values below $10^{-9}$, although as $Z_{\rm c} \to 0$, the numerical diffusion 
would finally be able to directly act to reduce $P(Z_{\rm c}, t)$ 
and the scaling of  the convolution  timescale with $Z_{\rm c}$ given earlier would fail.
The $Z_{\rm c} \to 0$ limit may not be of practical 
interest, as the critical metallicity is likely to be higher than $10^{-9}$ by 
mass. In the other limit, with increasing $Z_{\rm c}$, the weak 
power-law scaling will also break down eventually.  As pointed out 
above, the increase of the convolution timescales is already faster as 
$Z_{\rm c}$ increases to $10^{-5}$. In fact, if $Z_{\rm c}$ approaches the average 
concentration $\langle Z \rangle,$ (0.01 for the scalar field shown in Fig.\ \ref{Threshdependence}), 
eq.\ (\ref{pfnconvsolution}),
which was derived in the limit $Z_{\rm c} \to 0,$ will become invalid.
For illustration, let us consider 
the case in which $Z_{\rm c}$ is exactly equal to 
$\langle Z \rangle$. In this case, the fraction $P(Z_{\rm c},t)$ 
would not decrease to zero in the long time limit; instead it would 
approach 1/2. 
A more extreme example is that, if $Z_{\rm c}$ is larger than $\langle Z \rangle$, 
$P(Z_{\rm c}, t)$ would first decrease when the pollutants mix with a small 
amount of the flow material, and then increase and finally approach 
unity when the flow is completely homogenized. This situation may occur at 
very early times in the history of a galaxy, before heavy elements produced 
by supernova explosions increased the average metallicity to 
above the critical threshold. However, there is the possibility
metals from the explosion of a single massive Pop III star could make $\langle Z \rangle >Z_{\rm c}$ 
in a small high-redshift galaxy (Frebel et al.\ 2009). 
Even if  the  average metallicity in the entire interstellar medium is larger 
than $Z_{\rm c}$, there may exist local regions where the average metallicity 
is smaller than the threshold.  One would need to deal with this 
situation in a subgrid mode for the large -scale simulation for 
the primordial gas pollution in an early galaxy (see \S 7).  
In this work, we do not examine the evolution of $P(Z_{\rm c}, t)$ for $Z_{\rm c}$ 
close to or even larger than $\langle Z \rangle$. We defer it to a later work. 
In the case of $\langle Z \rangle< Z_{\rm c}$, a good approximation is perhaps 
to assume that $P(Z_{\rm c}, t)$ is a constant $\approx 1.$

Considering that $P(Z_{\rm c}, t)$ would show qualitatively 
different behaviors as the ratio, $Z_{\rm c}/\langle Z \rangle$, gets close to 
unity, it appears appropriate to take it as a function of  $Z_{\rm c}/\langle Z \rangle$, 
instead of the absolute value of $Z_{\rm c}$.  Another motivation is that, 
at a given ratio $Z_{\rm c}/\langle Z \rangle$, $P(Z_{\rm c}, t)$ samples 
a concentration range at a similar distance to the central 
part of the PDF.  Thus, in \S5.4.6, we 
tabulate the fitting parameters for the evolution of  $P(Z_{\rm c}, t)$ 
with $Z_{\rm c}/\langle Z \rangle = 10^{-7}$ as functions of the 
flow Mach number and the pollutant injection scale. 
For other values of  $Z_{\rm c}/\langle Z \rangle$, the 
timescales, $\tau_{\rm con1}$ and $\tau_{\rm con2}$ can be inferred using the 
weak power-law scaling given earlier, as long as $Z_{\rm c}/\langle Z \rangle \lsim 10^{-3}$.

\begin{figure*}[ht]
\centerline{\includegraphics[width=2\columnwidth]{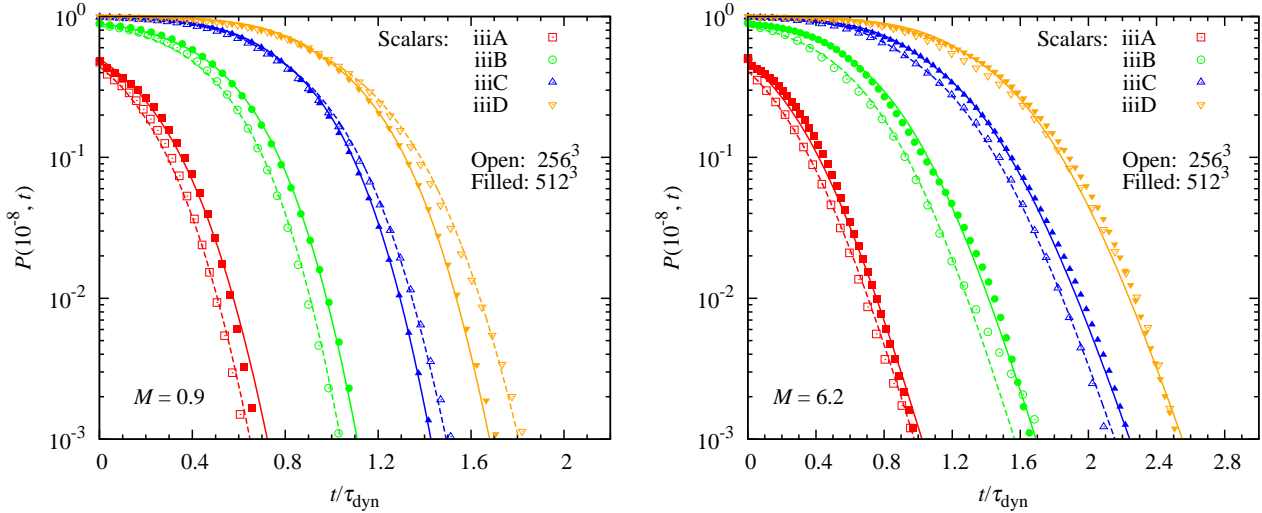}}
\caption{The pristine fraction, $P(Z_{\rm c}, t)$, as a function of time from 256$^3$ (open) and 512$^3$ (filled) 
simulations. The scalar fields shown here are from  category \rmnum{3}\  in the Mach 0.9 (left) and 6.2 (right) flows. The results from 
the 512$^3$ runs, already shown in Figs.\ (\ref{Mach0.9Four}) and (\ref{Mach6.2Four}), are replotted here for a comparison 
with the 256$^3$ data. No model fit is given to the 256$^3$ data for case \rmnum{3}D 
in the $M=6.2$ flow because the points are close to the 512$^3$ result.}
\label{resolutiondependence} 
\end{figure*}

\subsubsection{Dependence on the Numerical Resolution}

Finally, we examined the effect of numerical resolution. As discussed in \S 3.2,  
the timescale for the pollution of the pristine 
gas to a significant concentration level is mainly determined by the rate at 
which turbulence stretches the pollutants, and independent 
of the amplitude of the molecular or numerical diffusion 
if it is sufficiently small to allow a scale separation 
between the pollutant injection scale and the diffusion scale. 
To verify this expectation, we carried out $256^3$ 
simulations and compared them with the results from $512^3$ runs. 
The scale separation mentioned above exists at both resolutions, 
but the separation is limited for the $256^3$ runs. We drive the flows in 
exactly the same pattern in the $256^3$ and $512^3$ runs. 
However, due to the ``chaotic" nature of turbulence, 
the developed turbulent velocity field at given locations 
in the two runs are different. This means that, when 
released to the simulated flows at different resolutions, the 
pollutant blobs might encounter completely different velocity structures. 
In this sense, the comparison of our simulation results at two 
resolutions is somewhat different from the usual convergence check. 

In Fig.\ (\ref{resolutiondependence}), we plot $P(Z_{\rm c}, t)$ with $Z_{\rm c} =10^{-8}$ for 
scalar fields in category \rmnum{3}\ from the $256^3$ (open symbols) and $512^3$ 
(filled symbols) runs, with $M=0.9$ and $M=6.2.$
The filled data points and the solid fitting curves for the $512^3$ runs were already 
presented in  Figs.\ (\ref{Mach0.9Four}) and (\ref{Mach6.2Four}).  Here the early 
and late phases of cases (\rmnum{3}C) and (\rmnum{3}D) are connected 
at $t_{0.5}$. The fitting curves for the $256^3$ data are obtained with the 
same fitting scenario as in the $512^3$ case. In both the Mach 0.9 and 6.2 flows, the pristine 
fraction for the two scalar fields with $H_0=0.5$ (\rmnum{3}A) and $H_0=0.1$ (\rmnum{3}B) is 
smaller in the $256^3$ runs. In fact, the fraction becomes smaller than in the $512^3$ 
runs almost immediately after the pollutants are released into the flows.
This can be explained by considering the action of  numerical diffusion on the 
initial concentration field.
Since our initial concentration field consists of pure 
pollutants ($C=1$) and exactly pristine gas ($C=0$), 
there exist sharp edges between the pollutant blobs and 
the pristine flow. Numerical 
diffusion may operate on the large 
concentration gradient at the edges, and a fraction 
of the flow material surrounding the pollutants would be polluted 
immediately. This results in an instantaneous drop in the pristine fraction. 
In the $512^3$ runs, the effect was found to be weak, and the initial 
drop was slight.  The drop is significantly larger in the $256^3$ 
simulations due to the larger numerical diffusion, leading to smaller 
pristine fractions for scalars \rmnum{3}A and \rmnum{3}B in 
the $256^3$ runs than in the $512^3$ cases. 

Recognizing this effect of initial drop, we adjusted the 
initial pristine fraction, $P_0$, to smaller values when fitting the 
256$^3$ data. For scalar fields \rmnum{3}A and \rmnum{3}B,  
we used the same values of $n$ (i.e., $n=10$ and 3 for $M=0.9$ and 6.2, respectively) 
as in the corresponding $512^3$ cases.  With the adjusted 
values of $P_0$, the best-fit timescales $\tau_{\rm con2}$ for these two 
cases in the $256^3$ runs differ slightly, only by $\lsim 2\%$, 
from those used to fit the $512^3$ data. This is the case for 
both Mach 0.9 and Mach 6.2 flows. 
Therefore, except for the initial drop, the numerical resolution does 
not affect how the pristine fraction evolves for the two fields 
with $H_0 \ge 0.1$, and one may claim a numerical convergence 
of the convolution timescale. 
Note that,
in realistic interstellar turbulence, the effect of the initial drop 
would be minimal because the molecular diffusivity is much smaller 
than the numerical diffusion in our simulations. Also the sharp pollutant-flow 
edges in the simulations are artificial and may not exist in reality.     

The dependence on numerical resolution is more complicated 
for cases \rmnum{3}C and \rmnum{3}D with $H_0 = 0.01$ and $H_0 = 0.001,$ 
respectively. As seen in Fig.\ (\ref{resolutiondependence}), the pristine fraction decay 
in the $256^3$ runs can be either faster or slower than in the 
$512^3$ simulations. The stronger initial drop in the $256^3$ runs 
still exists in the early evolution phases. However, unlike 
scalar fields with $P_0 \ge 0.1$, it is not the dominant effect.
The velocity field at a given location in the $256^3$ 
and $512^3$ simulations is different (see above), and thus 
the same pollutant blob may encounter very different velocity 
structures in the two runs. As discussed earlier, 
for cases with small $H_0$, the pollutant size is small, 
and the turbulent stretching rate around the blobs 
would show larger variations.  
Therefore, the stretching rate in the eddy across a small pollutant blob may deviate
significantly from the mean value at that scale.
Since the flow mass polluted by an individual blob scales
nonlinearly with the local stretching rate, the overall pollution rate 
for scalar fields with tiny $H_0$ cannot be predicted 
by an average stretching rate, instead it depends on the distribution 
of the stretching rates over all the pollutant blobs. This is different from 
the case of $H_0 \gsim 0.1$ fields with large pollutant sizes, 
where the amplitude of the stretching rate fluctuations is smaller and the 
stretching rate for each blob is similar and equal to the average value.
Thus, the pollution for scalar cases with small $H_0$ ($\lsim 0.01$) is more sensitive to the 
details of the stretching rates encountered by all the blobs.  If the overall 
stretching rates in the eddies ``seen" by the pollutant blobs in the $256^3$ run 
is relatively higher, the pollution would proceed relatively faster than 
the $512^3$ run and vice versa. 
It appears that the origin of the observed difference at late times
is stochastic and has nothing to do with the numerical diffusion/resolution.  
The above picture also suggests that the difference 
may become larger as $H_0$ decreases further below 0.001.

When fitting the early phases of  cases \rmnum{3}C and \rmnum{3}D in the $256^3$ flows, 
we decreased $P_0$ to account for the initial drop. 
In the $M=0.9$ flow, $\tau_{\rm con1}$ for the early phases of these two cases are 
close to those used to fit the $512^3$ data, and the difference is at the level 
of $\lsim 5 \%$. The best-fit timescales $\tau_{\rm con2}$ for the late phases of 
the two scalars are larger (by $\simeq 10\%$) than for the $512^3$ data.  
In the $M=6.2$ flow, the fitting parameters for scalar \rmnum{3}C in the $256^3$ 
run are the same as in the corresponding 512$^3$ case, once the initial drop is accounted for. 
For case \rmnum{3}D in the $M=6.2$ flow, the data points 
from the $256^3$ and $512^3$ runs almost coincide,  and we do not give a separate fit to the 
$256^3$ data. It appears that the resolution dependence of 
the best-fit parameters is quite weak.

In summary, we find that the larger numerical diffusion 
in the $256^3$ simulations causes a larger initial drop in the 
pristine fraction. This effect is successfully accounted 
by adjusting the value of $P_0$ in our model fits to 
the simulation results.
The effect is expected to be
negligibly weak in the interstellar gas where the molecular diffusivity 
is tiny.  
For scalar fields with small $H_0 \le 0.01$,  the timescales to fit the simulation 
data differ by $\lsim 10\%$ at the two resolutions,
and 
numerical convergence may be claimed. The origin of the 
``random" dependence of the pristine fraction on the resolution for these fields 
is related to
the larger fluctuations of turbulent stretching rate at smaller scales 
and suggests that a precise prediction of the pristine fraction in the case 
of  tiny $H_0$ may require the detailed eddy conditions at  the initial pollutant locations. 
We finally point out that numerical convergence would not exist at all if the resolution 
did not allow a separation between the pollution injection 
and the diffusion length scales. 

\subsubsection{Summary}

We summarize our simulation results in Tables 2, 3, and 4. The parameters listed in the 
tables are obtained by fitting the fraction $P(10^{-7} \langle Z \rangle, t)$ from simulation data 
for scalar fields from different categories and in different flows. Here, for each scalar, the threshold 
$Z_{\rm c}$ is set to $10^{-7}\langle Z \rangle$. The average concentration $\langle Z \rangle$ is 
equal to the initial pollutant fraction $H_0$ for a double-delta initial condition, 
eq.\ (\ref{initialpdf}). For scalar fields A, B, C and D, $H_0 =0.5, 0.1, 0.01,$ and 0.001, 
and we choose $Z_{\rm c}$ to be $5\times 10^{-8}$, $10^{-8}$, $10^{-9}$ and $10^{-10} $,  
respectively.  The choice of a fixed ratio $Z_{\rm c}/\langle Z \rangle$ is 
more convenient for practical applications. The timescales in Tables 2 and 4 are 
slightly different from those used in the figures in previous subsections, 
where (except in \S 5.4.4) the threshold was 
fixed at $Z_{\rm c} =10^{-8},$ for all values of $H_0$.  The numbers 
in these two tables are in units of the flow dynamical time, $\tau_{\rm dyn}$.  The first columns of 
Tables 2, 3 and 4 show results for scalar fields with the injection scale $L_{\rm p}$ 
close to the box size or $\simeq 2 L_{\rm f}$.  These parameters  
are measured from scalar cases in category  \rmnum{1}. 
Measuring the parameters using category  \rmnum{2}\ fields with the same $L_{\rm p}$ would 
give essentially the same results.

Table 2 lists the timescale, $\tau_{\rm con1},$ for the early 
phase of scalar fields with $H_{0} \le 0.01$. In this phase, the pristine fraction 
evolution is fit by the ``discrete" convolution model with $n=1$. For scalar 
cases in each category ($L_{\rm p}$) and each flow ($M$), we measured 
$\tau_{\rm con1}$ for the early phases of fields C (with $H_0 = 0.01$) and D (with $H_0=0.001$), 
and  the number given in Table 2 is  the average of the measured values 
for these two fields. As found in \S 5.4.2,  at a given injection scale, 
$\tau_{\rm con1}$ first increases as $M$ increases  from 0.9 to 2-3,  and then 
saturates for larger $M$. The overall increase in  $\tau_{\rm con1}$ is about 20\% for $M$ in 
the range from 0.9 to 6.2. This is in general agreement with the trend of the mixing timescale $\tau_{\rm m}$ 
with $M$ found in PS10. 
At a given Mach number, $\tau_{\rm con1}$ decreases with decreasing 
injection length scale $L_{\rm p}$.
As $L_{\rm p}$ decreases from $2 L_{\rm f}$ to $L_{\rm f}$, $\tau_{\rm con1}$ 
is smaller by $\sim 25\%$. The decrease is faster for smaller $L_{\rm p}$, 
and a further decrease of $L_{\rm p}$ by each factor of 2 
reduces  $\tau_{\rm con1}$ by $\simeq 35\%$. 
If we express the $L_{\rm p}$ dependence of $\tau_{\rm con1}$ roughly as a power law 
for $L_{\rm p} \lsim L_{\rm f}$, we have $\tau_{\rm con1} \propto L_{\rm p}^{0.62}$. 

\begin{table}[ht]
\begin{center}
\caption{The convolution timescale $\tau_{\rm con1}$ for the early-phase evolution of $P(10^{-7}\langle Z \rangle, t )$ for scalar fields 
with initial pollutant fraction $H_{0} \lsim0.1$}
\label{tbl-2}
\vspace{1mm}
\scalebox{1.1}{
\begin{tabular}{c@{\hspace{3mm}}|@{\hspace{3mm}}c@{\hspace{3mm}}c@{\hspace{3mm}}c@{\hspace{3mm}}c}
\hline\hline
$M$  &  $L_{\rm p}  = 2 L_{\rm f} $ & $ L_{\rm p} = L_{\rm f}$ & $L_{\rm p} = L_{\rm f}/2$  & $ L_{\rm p } = L_{\rm f}/4$  \\
\hline
0.9       &   0.25   &  0.18     &  0.12    &  0.075\\
2.1       &   0.30  &   0.20      &  0.14   & 0.085\\
3.5       &   0.29  &   0.22     &  0.14    &  0.09 \\
6.2       &   0.28   &  0.23     &  0.15    & 0.09\\
\hline
\end{tabular}}
\end{center}
\end{table}

\begin{table}[ht]
\begin{center}
\caption{The parameter $n$ for the pristine fraction evolution of scalar fields with $H_{0} \gsim0.1$ and  for the later-phase evolution of  $H_{0} \lsim 0.1$ fields}
\label{tbl-3}
\vspace{1mm}
\scalebox{1.1}{
\begin{tabular}{c@{\hspace{3mm}}|@{\hspace{2mm}}c@{\hspace{3mm}}c@{\hspace{3mm}}c@{\hspace{3mm}}c}
\hline\hline
$M$  &  $L_{\rm p}  = 2 L_{\rm f} $ & $ L_{\rm p} = L_{\rm f}$ & $L_{\rm p} = L_{\rm f}/2$  & $ L_{\rm p } = L_{\rm f}/4$  \\
\hline
0.9       & $\infty$   &      10                  &  5        &  2.5   \\
2.1       &  10          &        6                   &  3        &  2   \\
3.5       &   8           &        5                   &  2.5      & 1.5 \\
6.2       &   5           &        3                   &  2         & 1 \\
\hline
\end{tabular}}
\end{center}
\end{table}

\begin{table}[ht]
\begin{center}
\caption{The convolution timescale $\tau_{\rm con2}$ for the evolution of  
the pristine fraction $P(10^{-7}\langle Z \rangle, t)$ for  scalar fields with $H_{0} \ge 0.1$ 
and for the later phases of scalar cases with $H_0 \lsim 0.1$}
\label{tbl-4}
\vspace{1mm}
\scalebox{1.1}{
\begin{tabular}{c@{\hspace{3mm}}|@{\hspace{3mm}}c@{\hspace{3mm}}c@{\hspace{3mm}}c@{\hspace{3mm}}c}
\hline\hline
$M$  &  $L_{\rm p}  = 2 L_{\rm f} $ & $ L_{\rm p} = L_{\rm f}$ & $L_{\rm p} = L_{\rm f}/2$  & $ L_{\rm p } = L_{\rm f}/4$  \\
\hline
0.9       &  0.35   &   0.26    & 0.17   &  0.10   \\
2.1       &  0.42   &   0.31    & 0.21   &  0.12   \\
3.5       &  0.42   &   0.33    & 0.22   &  0.13   \\
6.2       &  0.42   &   0.32    & 0.21   &  0.11   \\
\hline
\end{tabular}}
\end{center}
\end{table}

Tables 3 and 4 give the parameters $n$ and $\tau_{\rm con2}$ as 
functions of $M$ and $L_{\rm p}$. These  
are measured for the pristine fraction evolution of scalar fields with $H_0 \ge 0.1$ or 
the late-time behavior of scalars with smaller $H_0$.  
For  a given category ($L_{\rm p}$) and a given flow ($M$), we choose a 
single value of $n$, with which the  self-convolution model prediction 
can well match the simulation data simultaneously for both the two scalar 
cases with $H_0 \ge 0.1$ and the late phases of the other two cases 
with $H_0 \le 0.01$.  In Table 3,  the parameter $n$ is  taken to be 
$\infty$ for scalar fields with $L_{\rm p} \simeq 2 L_{\rm f}$ in the Mach 0.9 flow,
which corresponds to the continuous convolution model (eq.\ \ref{pfcconvsolution}). 
PSS showed that the continuous model can be used to obtain successful fits to category \rmnum{1}\ 
scalars in the  $M=0.9$ flow.  We find that, for a given $\tau_{\rm con2}$, 
the predicted pristine fraction by the convolution model barely changes with increasing 
$n$ once $n$ exceeds $\sim 20$. This means that replacing $\infty$ in 
Table 3 by  any number larger than $20$ would also work for 
category \rmnum{1}\ (or \rmnum{2}) fields in the Mach 0.9 flow. 

From Table 3, we see that $n$ decreases with increasing Mach number and decreasing $L_{\rm p}$. 
This is due to the higher degree of flow intermittency at larger $M$ (Pan \& Scannapieco 2011) or 
smaller $L_{\rm p}$, which causes broader concentration PDF tails.  
As described previously, 
a smaller $n$ indicates
a more local PDF convolution .

After fixing the parameter $n$ for each $L_{\rm p}$ and $M$, 
we measure the timescale, $\tau_{\rm con2}$, for scalar cases 
A and  B  and the late phases of cases C and D. The measured 
values for the four cases are not exactly the same, but show slight variations. 
The variations are stronger at larger $M$ or smaller $L_{\rm p}$. 
We found that the amplitude of the variations is smaller when using a  
fixed $Z_{\rm c}/\langle Z \rangle$ ratio rater than a fixed threshold $Z_{\rm c}$. 
This also justifies taking the pristine fraction as a function of $Z_{\rm c}/\langle Z \rangle$.  
The numbers given in Table 4 are the averages of the best-fit values 
for the four scalar cases in each category and each flow.  The dependence 
of $\tau_{\rm con2}$ on $M$ and $L_{\rm p}$ is very similar to that 
of  $\tau_{\rm con1}$ shown in Table 2.  Again, it 
increases  by about 20\% as $M$ increases from $0.9$  to $2-3$,  and then 
stays constant at larger $M$. Like $\tau_{\rm con1}$, the decrease 
of $\tau_{\rm con2}$ with decreasing $L_{\rm p}$ also appears to be faster
 at smaller $L_{\rm p}$. It is reduced by 25\%, 35\%, and 40\%, 
 respectively, as $L_{\rm p}$ decreases by each factor of 2 from 
 $2L_{\rm f}$ to $L_{\rm f}/4$. Roughly, $\tau_{\rm con2}$ scales 
 with the injection scale as $\tau_{\rm con2} \propto  L_{\rm p}^{0.65}$
 for $L_{\rm p} \lsim L_{\rm f}$.

We point out that, when measuring the model parameters from 
all the scalar fields with $H_{0} \le 0.01$, we connected the 
early and late phases at the time $t_{0.5}$ as the pristine fraction 
decreases to 0.5. However, as discussed in \S 5.4.1, one can still 
use the parameters given in Tables 2, 3, and 4 if a connection 
at an earlier time, $t_{0.9}$, is preferred in a particular application.     

Tables 2, 3 and 4 can be used for practical applications.  
One may first fix the three parameters, $\tau_{\rm con1}$, $n$ 
and $\tau_{\rm con2}$, by interpolating the tabulated values according to the flow Mach number, $M$, 
and the pollutant injection scale, $L_{\rm p}$.  
and for interpolation purposes, 
one can replace $n \to \infty$ by, say, $n=20$ for the case with $M=0.9$ 
and $L_{\rm p} = 2 L_{\rm f}.$ 
For subsonic flows with $M < 0.9$,  we expect the parameters 
to be very close to those measured here for the $M=0.9$ flow. 
As shown in PS10 and Pan and Scannapieco (2011), 
the velocity structures at all orders in the Mach 0.9 flow are essentially the same as 
in incompressible turbulence (corresponding to the limit $M \to 0$). In the other 
limit  of large $M$,  the timescales would not change with $M$ for $M\gsim 6$, 
since they already saturate at $M=2-3$. The parameter $n$ may keep 
decreasing as $M$ increases above 6.2, and in that case one may obtain $n$ by extrapolation,
with the expectation that $n$ has a minimum value of 1, corresponding to the 
highest degree of spatial locality in the PDF convolution.
For the dependence of the timescales on $L_{\rm p}$, 
we can use the approximate power-law scalings given above 
for $L_{\rm p} \lsim L_{\rm f}$. Next, depending on the initial pollutant fraction 
$H_0$,  one may decide whether to start with an early phase using the 
discrete convolution model. For different values of the 
ratio $Z_{\rm c}/\langle Z \rangle$, $n$ does not change, and the 
timescales $\tau_{\rm con1}$ and $\tau_{\rm con2}$ may be obtained 
from the weak power-law scaling with $Z_{\rm c}$ given in \S 5.4.4. 
The scaling applies for $Z_{\rm c}/\langle Z \rangle \lsim 10^{-3}$. 

For convenience, we have computed fits to $\tau_{\rm con1}$, $n$ and $\tau_{\rm con2},$ which can be used
in place of interpolating along the table.  
Because the regime in which $L_{\rm p} \leq L_{\rm f}$ is the most important one for 
most astrophysical systems, we have focused on this case
when computing our $L_{\rm p}$ dependence, and furthermore, because of
the statistical noise in our measurements, we have taken an average 
scaling of  $L_{\rm p}^{0.63}$  for both $\tau_1$ and $\tau_2.$  Imposing a strict floor of
$n \geq 1$ and the $Z_c$ scaling measured above we find
\begin{gather}
\tau_{\rm con1} =   \left[0.225 - 0.055 \exp(-M^{3/2}/4) \right] \left(\frac{L_{\rm p}}{L_{\rm f}} \right)^{0.63} \times \notag\\ \left(\frac{Z_{\rm c}}{10^{-7} \langle Z \rangle} \right)^{0.015}, \notag \\
\tau_{\rm con2}  =   \left[0.335 - 0.095 \exp(-M^2/4) \right] \left(\frac{L_{\rm p}}{L_{\rm f}} \right)^{0.63}\hspace{.3cm} \times \notag \\ \left(\frac{Z_{\rm c}}{10^{-7} \langle Z \rangle} \right)^{0.02}, 
\notag\\
n      =   1 + 11 \, \exp(-M/3.5) \left(\frac{L_{\rm p}}{L_{\rm f}}\right)^{1.3}, \hspace{1.7cm}
\label{eq:taunfit}
\end{gather}
which provides good fits for all Mach numbers and pollution properties, as 
long as $Z_{\rm c}/\langle Z \rangle \lsim 10^{-3},$ and $L_{\rm p} \leq L_{\rm f}.$ 
We finally point out that  the parameters may have a dependence on how the turbulent flow is driven.  For example, 
 in a compressively driven flow at the same Mach number, $n$ may be smaller  than measured from our simulations (see \S 5.4.2).

\section{Application to the Pollution of Primordial Gas in Early Galaxies}

\subsection{The Global Pristine Fraction}

In previous sections, we have focused on understanding the fundamental physics 
of the pollution of pristine flow material by turbulent mixing.  We now describe how our
results can be applied to investigate the pollution 
of primordial gas in the interstellar media of high-redshift galaxies. In this section, 
we discuss using our results to obtain a qualitative estimate 
of the pollution timescale in early galaxies, similar to the formalism of Tinsley (1980) in 
which the evolution within a galaxy is reduced to a few general parameters. 
A more accurate approach based on large-eddy simulations and subgrid modeling 
will be presented in the next section.   

To study the mixing of heavy elements in interstellar turbulence, we need to specify 
the source term in the PDF equation (\ref{pdfeq}), which can be evaluated by 
considering how the pollutants, including fresh metals from supernova 
explosions and low-metallicity or pristine infall gas, affect the metallicity PDF 
(see \S 2.2).  If the supernova rate per unit volume in 
a given region of a galaxy is $\dot{n}_{\rm SN}({\bf x}, t)$, the source 
term due to new metals from supernovae would be ${\dot{n}_{\rm SN}  m_{\rm ej}} \left[\delta(Z-Z_{\rm ej}) -\Phi(Z; {\bf x}, t)\right] $ 
where it is assumed that, on average, each supernova produces an ejecta 
mass of $m_{\rm ej}$ with a mass fraction of metals $Z_{\rm ej}$, 
and that Rayleigh-Taylor and Kelvin-Helmholtz instabilities arising during the explosion mix the fresh 
metals with the envelope material. In reality, the source term by supernovae may have a 
finite width instead of being a delta function, because the ejecta mass and the heavy element yield vary with the 
mass of the progenitor star. One can refine the form of the source term by using 
nucleosynthesis results for the ejecta mass and metal yield as functions 
of the progenitor mass (\eg Maeder 1992; Woosley \& Weaver 1995; Heger \& Woosley 2002)
and accounting for the initial stellar mass function. 
The $-\Phi$ term corresponds to the replacement of the existing 
PDF in a fraction of the interstellar gas by $\delta(Z-Z_{\rm ej})$ due to the 
release of new metals from supernovae, and it guarantees the source term conserves the total 
probability. 

During the formation of an early galaxy, there may exist an 
infall of primordial gas that continuously flows from the halo into the galaxy.
This provides another source
term, $\dot{m}_{\rm I} [\delta (Z) -\Phi(Z; {\bf x}, t)]$, 
where $\dot{m}_{\rm I} ({\bf x}, t)$ denotes the local  infall rate. The infall rate should be take to be 
zero except at the boundary, where the pristine gas enters 
the galaxy. Again the $-\Phi$ term is to ensure the conservation 
of the total probability. Clearly, new metals from 
supernovae and the pristine infall gas force spikes at 
high and low concentration levels in the PDF, respectively. 

We define a global pristine fraction as $P_{\rm g} (Z_{\rm c}, t) =
 \int_0^{Z_{\rm c}} dZ \int_{V} dx^3 \langle \rho({\bf x}, t) \rangle \Phi(Z; {\bf x}, t)/M_{\rm g}$ 
where $V$ is the total volume of the galaxy and 
$M_{\rm g} = \int_{V} \langle \rho \rangle dx^3$ is the total mass of the interstellar gas. 
An equation for $P_{\rm g}$ can be derived by performing a double integration of eq.\ (\ref{pdfeq}) 
over space and concentration.
The advection term vanishes when integrated over 
space,
and the double integral of the supernova source term 
gives  $- \frac{\dot{N}_{\rm SN} m_{\rm ej} }{M_{\rm g}} P_{\rm g}(Z_{\rm c}, t)$, 
where $\dot{N}_{\rm SN}$ is the total supernova rate in the 
galaxy. Clearly, the contribution from supernovae is always 
negative. On the other hand, the infall of primordial 
gas contributes a positive term $\frac {\dot{M}_{\rm I} }{M_{\rm g}} [1 -P_{\rm g}(Z_{\rm c}, t )]$, 
where $ {\dot{M}_{\rm I} }$ is the global  infall rate.  
Using  the self-convolution model for the diffusivity term in the $P_{\rm g}$ equation, we obtain,  
\begin{equation}
\frac{dP_{\rm g}}{dt} = -\frac{n}{\tau_{\rm con}} P_{\rm g}(1-P_{\rm g}^{1/n})-  \frac{\dot{N}_{\rm SN}m_{\rm ej}}{M_{\rm g}} P_{\rm g} 
+  \frac {\dot{M}_{\rm I}} {M_{\rm g}} (1-P_{\rm g}). 
\label{eq:pfgalaxy}
\end{equation}
A similar equation with $n=1,$
and without the infall term was 
first given in Pan and Scalo (2007). 

When writing eq.\ (\ref{eq:pfgalaxy}), we have made an implicit assumption 
of statistical homogeneity, which may break down for several reasons.
First, the prediction of the self-convolution model for the pristine fraction 
evolution is tested and verified only in statistically homogeneous flows, and it may not 
be valid for a system with large-scale inhomogeneities. 
Second, eq.\ (\ref{eq:pfgalaxy}) adopts single values for the parameters $n$ and $\tau_{\rm con}$,  
equivalent to assuming similar turbulence conditions everywhere in the interstellar 
gas. Finally, the first (mixing) term on the r.h.s. is nonlinear with $P_{\rm g}$, 
and this nonlinearity would affect the prediction accuracy, if, for example, the star formation and hence the 
metallicity have a large-scale gradient. The parameters in eq.\ (\ref{eq:pfgalaxy}) 
should be viewed as the effective averages over the turbulence and 
metallicity conditions of the entire galaxy.  These suggest 
the solution of eq.\ (\ref{eq:pfgalaxy}) only provides a rough estimate 
for the global pristine fraction, which can be improved by accounting for 
realistic complexities.
Nevertheless, the equation is a useful guideline for the study of the primordial 
gas pollution in early galaxies. 

The turbulence conditions in the interstellar media of early galaxies are essentially 
unknown, and thus the parameters in eq.\ (\ref{eq:pfgalaxy}) cannot be estimated with 
certainty. Here we will make various assumptions for the turbulence parameters, 
and discuss how the pollution of the pristine gas proceeds under different 
conditions. Future observations will help constrain the parameter space and give a 
clearer picture of the mixing process in high-redshift galaxies. 

\subsection{The Pollution Timescale}   

A crucial parameter for mixing in the interstellar gas is the driving length scale
of   the interstellar turbulence, $L_{\rm f}.$
If the turbulence is driven at the largest scales, 
e.g., by the collapse of the baryonic matter into the potential well of the dark matter halo,
then $L_{\rm f}$ is close to the size  of the galaxy, $L_{\rm G}.$ In this case, the primary 
energy source for turbulence is the gravitational energy, and the driving force of the 
interstellar turbulence is associated with the source term in eq.\ (\ref{eq:pfgalaxy}) for the pristine infall.  
The driving scale may also remain close to $L_{\rm G}$ at late times if the infall 
from the halo is persistent during the galaxy evolution. On the other hand, if the primary 
energy source for interstellar turbulence is the explosion energy of supernovae, 
then $L_{\rm f}$ is likely on the order of the typical size of supernova remnants,  
$L_{\rm SNR}.$  In general, we expect $L_{\rm SNR} \lsim L_{\rm f} \lsim L_{\rm G},$ and, 
depending on how $L_{\rm f}$ compares with $L_{\rm G}$, the pollution will proceed in 
qualitatively different ways. 

We first consider the case where the turbulent driving scale, $L_{\rm f}$, is close to the 
galaxy size, $L_{\rm G}$.  With $L_{\rm f} \simeq L_{\rm G}$, we may roughly think of  the 
entire interstellar medium as corresponding to our simulation box, and the 
dynamical time, $\tau_{\rm dyn}$, may be calculated by dividing 
$L_{\rm G}$ by the rms turbulent velocity.  
As new metals from supernovae are released to the interstellar 
turbulence, the supernova source term in eq.\ (\ref{eq:pfgalaxy}) reduces the pristine 
fraction, $P_{\rm g}$. We can start applying the self-convolution 
model in eq.\ (\ref{eq:pfgalaxy}) to calculate $P_{\rm g}$, once the average metallicity 
exceeds the threshold value, $Z_{\rm c}$. The parameters $n$ and $\tau_{\rm con}$ 
can be estimated based on  our simulation results  tabulated in \S 5.4.6. 
With more supernovae exploding, the pollution process would become 
faster due to the increased amount of pollutants.  
Also, the average pollutant separation and hence the injection scale, $L_{\rm p}$, 
will decrease with the total number of supernovae, $N_{\rm SN}(t).$
Assuming a random supernova distribution, $L_{\rm p}$ scales 
like $N_{\rm SN}^{-1/3}$. The convolution timescale $\tau_{\rm con}$ 
would thus decrease with time, 
according to the power-law scaling of $\tau_{\rm con}$ with $L_{\rm p}$ 
resulting in a faster pollution rate. 
A subtle and minor effect is that  the increase of the average 
metallicity reduces the threshold to average ratio, $Z_{\rm c}/\langle Z \rangle$, 
leading to a slight additional
decreases in $\tau_{\rm con}$. This t 
may be accounted for using the $Z_{\rm c}$-dependence of $\tau_{\rm con}$
given in \S 5.4.4. If the infall of pristine gas is persistent,  the infall term in eq.\ (\ref{eq:pfgalaxy}) 
provides a continuous source for the pristine fraction, and there may exist  a quasi-steady-state for $P_{\rm g}$ (see Pan and Scalo 2007), 
which is controlled by three timescales, the convolution timescale, $\tau_{\rm con}$, 
the timescale for supernova sources, $M_{\rm g}/(\dot{N}_{\rm SN} m_{\rm ej})$, 
and the timescale for the mass accretion by infall, $M_{\rm g}/\dot{M}_{\rm I}$.   
  
The estimate of $P_{\rm g}$ is more complicated if the driving scale, $L_{\rm f}$, 
is much smaller than $L_{\rm G}$.  If $L_{\rm f} \ll L_{\rm G}$, the correlation 
length scale of the turbulent velocity field is much smaller than the size 
of the entire interstellar medium, and one may view the interstellar medium 
as a collection of many ``independent" turbulent regions of size $\sim L_{\rm f}$.  
The pollution process in each region would be similar to that in our simulation box, with 
a timescale determined by the local stretching/convolution timescale, 
$\sim \tau_{\rm dyn}$ ($\equiv L_{\rm f}/v_{\rm rms}$). 
However, the pollution in the entire interstellar medium may not be simply described 
by a self-convolution model or eq.\ (\ref{eq:pfgalaxy}) with a local convolution timescale. 
This is because the situation in individual regions of size $L_{\rm f}$ may 
be completely different.  For example, the regions that had supernova explosions at early times
may have already been significantly polluted,  while the pollution process may have 
not yet started in the regions that had not experienced supernovae or received any heavy elements.  
Thus the mixing/pollution timescale over the entire galaxy may depend on the 
large-scale turbulent transport of pollutants between the ``independent" regions. 
Assuming a random walk model for turbulent transport at scales $\gg L_{\rm f}$, the transport timescale 
at the galactic scale may be roughly estimated as $\tau_{\rm trans} \equiv L_{\rm G}^2/(L_{\rm f} v_{\rm rms})$, 
which is much larger than  the local stretching timescale $L_{\rm f}/v_{\rm rms}$ and the timescale $L_{\rm G}/v_{\rm rms}$.  

If $L_{\rm f} \ll L_{\rm G}$, another timescale of interest is  $\tau_{\rm SN}$, 
defined as the time needed for the average separation between 
the supernova remnant locations to decrease below $\simeq L_{\rm f}$. 
In other words,  $\tau_{\rm SN}$ represents the time for supernovae to populate the 
interstellar medium at a level of about one per region of size $L_{\rm f}$. If the 
supernovae are randomly distributed, $\tau_{\rm SN}$ can be estimated from $N_{\rm SN}(\tau_{\rm SN}) \simeq (L_{\rm G}/L_{\rm f})^3$, 
where $N_{\rm SN}(t)$ is the total number of  supernovae 
exploded before time $t$.  At $t \ll \tau_{\rm SN}$, only a smaller number of supernovae occur, 
and the supernova sources would be statistically inhomogeneous at the 
scale $L_{\rm f}$. In that case,  eq.\ (\ref{eq:pfgalaxy}) is not directly applicable 
as it implicitly assumes statistical homogeneity (see above).

Thus, the pristine fraction evolution in the $L_{\rm f} \ll L_{\rm G}$ case 
depends on a comparison of three timescales, $\tau_{\rm dyn}$,  $\tau_{\rm SN}$ 
and $\tau_{\rm trans}$. From their definitions, 
$\tau_{\rm dyn} \equiv L_{\rm f}/v_{\rm rms} \ll \tau_{\rm trans} \equiv L_{\rm G}^2/(L_{\rm f} v_{\rm rms}),$ and the amplitude 
of $\tau_{\rm SN}$ relative to these two timescales is crucial for how the pollution proceeds. If the star formation or 
supernova rate is so high that $\tau_{\rm SN} \ll \tau_{\rm dyn}$, 
the supernovae fill the interstellar medium quickly, and its spatial distribution 
would appear more or less homogeneous at the scale $L_{\rm f}$
before each region of size $L_{\rm f}$ is significantly polluted. 
This suggests that the pollution in all the ``independent" 
regions would roughly proceed at a  similar pace, and the 
pristine fraction evolution in each region may approximately reflect 
the global pristine fraction.
Therefore, at $t  \gsim \tau_{\rm SN}$, one may apply  eq.\ (\ref{eq:pfgalaxy}) 
to estimate 
the global pristine fraction
using $n$ and $\tau_{\rm con}$ corresponding to the 
physical conditions at the scale $L_{\rm f}$. In this case, the timescale for 
the decay of $P_{\rm g}$ would be $\simeq \tau_{\rm dyn}$.  

If  $\tau_{\rm dyn} \ll \tau_{\rm SN} \ll \tau_{\rm trans}$, the mixing of fresh metals from 
a supernova with the surrounding region of size $L_{\rm f}$ is fast with a relatively short timescale 
($\sim \tau_{\rm dyn}$), and the interstellar medium would have 
been completely polluted if, on average, each region of size
$L_{\rm f}$ in the galaxy had one supernova explosion.  This is expected to occur 
at time $t \simeq \tau_{\rm SN}$, and thus the pollution timescale in the entire galaxy 
is on the order $\sim \tau_{\rm SN}$. 

Finally, if the star formation rate is very low and $\tau_{\rm SN}$ is 
significantly larger than
$\tau_{\rm trans}$, then the turbulent transport at 
large scales ($\gg L_{\rm f}$) plays a crucial role in the pollution process. 
The delivery of heavy elements by the large-scale transport provides the entire galaxy 
with pollutants before the metal deposit by supernova events 
covers most of the interstellar medium. The pollution in the galaxy would
be completed at $\tau_{\rm trans}$. In this case, modeling the effect of the 
large-scale turbulent transport is essential.

One interesting limiting case is when the interstellar turbulence is completely 
driven by supernova explosions, and turbulent motions are weak outside the influence radius $L_{\rm SNR}$ of each supernova. 
In that case, we have $L_{\rm f} \simeq L_{\rm SNR}$ in regions affected by supernovae, and the transport of metals in 
between supernova locations would be slow with a large timescale $\tau_{\rm trans}$. 
From the discussion above, the pollution timescale would be determined by the 
maximum of the two timescales, $\tau_{\rm dyn}$ and $\tau_{\rm SN}$.  
We note that a quantitatively accurate prediction for this case may need to carefully 
account for the correlation between  the metal injection and the turbulence driving force.

While eq.\ (\ref{eq:pfgalaxy}) provides a rough estimate for  the pollution of  primordial gas 
in early galaxies, perhaps the best tool for a quantitative prediction is a large-scale 
numerical simulation that can include complexities such as  
large-scale velocity and metallicity
inhomogeneities of  the interstellar medium,  and the effect of large-scale transport.
So far we have ignored the advection term in the PDF 
equation (\ref{pdfeq}), which is responsible for the transport of the local 
PDF by the velocity field. The transport effect on the primordial gas pollution is substantial 
under certain circumstances, as seen earlier in the pollution timescale estimate. 
In the next section, we will establish a formulation for  large-scale simulations of 
the pristine gas pollution in early galaxies. In the context of large-eddy simulations, 
the advection term corresponds to the local PDF or pristine fraction 
exchange between neighboring computational cells due to both the large-scale 
velocity and the subgrid turbulent motions. Modeling the advection 
term in the PDF equation is crucial in these simulations, and we will adopt a commonly-used 
subgrid closure for the transport effect by subgrid turbulent motions.
 
\section{Large-Eddy Simulations and Subgrid Modeling}
 
The complexities present in a realistic high-redshift galaxy can only be dealt with 
in detail through direct numerical simulations of the pollution of pristine gas in the 
interstellar medium of a high-redshift galaxy. 
However, it 
is prohibitively expensive for such simulations to resolve the scale 
at which
homogenization by molecular diffusivity occurs in interstellar turbulence. 
Limited resolution implies significant numerical diffusion, 
which causes artificial mixing, erasing any 
metallicity fluctuations that would exist below the size of a computational cell. 
In fact, due to the vast range of scales existing in the problem, resolving any inertial-range scales 
at all is extremely challenging (\eg Scannapieco \& Br\" uggen 2010).
This results in an underestimate in the degree of metallicity fluctuations/inhomogeneity 
in the interstellar gas. Nevertheless, a large-scale
simulations still provide useful estimates for the low-order metallicity statistics, such as the metallicity variance, if 
they manage to resolve a small portion of inertial range, since the majority of the scalar fluctuation 
power is at large scales. 
  
On the other hand, the problem is much more severe for 
the pollution of the primordial gas,  
which corresponds to high-order statistics of the metallicity fluctuations.   
Since the threshold metallicity $Z_{\rm c}$ for the transition to Pop II 
star formation is  tiny, a computational cell would essentially lose all 
the pristine gas once it is subject to the pollution by even a small amount of 
heavy elements. A significant underestimate in the pristine 
mass fraction is therefore expected in simulations that do 
not resolve a considerable inertial range.    

Here we propose to approach the problem using large 
eddy simulations (LES) that keep track of the 
concentration fluctuations at subgrid scales. In   
such simulations, the flow at large 
scales is directly computed, while the effects of 
turbulent motions at subgrid scales are modeled.  
The existence of scale invariance in the inertial range of 
turbulent flows is crucial for subgrid modeling (Meneveau \& Katz 2000), 
which justifies using the resolved flow structures to infer the feedback effect of  
small-scale fluctuations. 

In this section, we outline an LES scenario for simulating the pollution 
of pristine gas in  early galaxies.  In \S 7.1, we first 
derive the LES equations for the interstellar turbulent flow and 
introduce subgrid models to close the equations,   
taking
so-called one-equation subgrid model (e.g., Lilly 1966), which 
evolves the turbulent kinetic energy at subgrid scales, as an illustrative 
example.
In \S 7.2, we develop an LES formulation for the pristine 
fraction based on an equation for the local concentration PDF filtered at the resolution scale, 
using the self-convolution PDF models. The model parameters can be
determined  
with the simulation results summarized in \S 5.4.6.       
By retaining the subgrid concentration fluctuations, 
the model provides a remedy to the over-pollution 
by numerical diffusion, and is expected to significantly improve 
the predicting power of large-scale simulations for the primordial 
gas pollution in high-redshift galaxies.

\subsection{Subgrid Modeling of the Interstellar Turbulent Flow}

We start by introducing the basic filtering procedure used 
to derive the governing flow equations at resolved scales in LES. 
The procedure employs  a low-pass filter function, $G ({\bf x}-{\bf x}')$, 
which eliminates fluctuations below the resolution scale
of the simulation grid, $\Delta$. 
Examples of the filtering function are a window function of width ${\Delta}$ 
or a Gaussian function with variance $\Delta^2$.  For any flow variable, $A({\bf x}, t)$, the filtered quantity $\overline{A}({\bf x}, t)$ 
is defined as, 
\begin{equation}
\overline{A}({\bf x}, t) = \int_V A({\bf x}', t) G ({\bf x}-{\bf x}') d {x'}^3,   
\label{filter} 
\end{equation}
and it represents the variable at the resolved scales. From eq.\ (\ref{filter}),  we have,  
\begin{equation}
\overline{\left(\frac {\partial A} {\partial t}\right)}  = \frac {\partial \overline{A} } {\partial t}, \hspace{5mm}  \overline{\left(\frac {\partial A} {\partial x_i}\right)} = \frac {\partial \overline{A}} {\partial x_i}
\label{filteredvariables} 
\end{equation}  
where integration by parts is used to obtain the second equality.  
For compressible flows, it is more convenient to use the Favre filtering (e.g., Speziale et al.\ 1988, Moin et al.\ 1991, Erlebacher et al.\ 1992), 
defined as, 
\begin{equation}
\widetilde{A}({\bf x}, t) =  \frac { \overline{\rho A}}{\overline{\rho}},      
\label{Favrefilter} 
\end{equation}
where a density-weighted factor is included.

Applying the filtering procedure to the continuity and momentum equations gives, 
\begin{equation}
\frac{ \partial \overline{\rho}} {\partial t}  + \frac{\partial}{\partial x_i}  (\overline{\rho} \hspace{0.5mm} \widetilde{v}_i)  = 0,       
\label{FiilteredContinuity} 
\end{equation}
and,  
\begin{equation}
\frac{ \partial (\overline{\rho}   \hspace{0.5mm} \widetilde{v_i}) } {\partial t}  + \frac {\partial ( \overline{\rho}   \hspace{0.5mm}  \widetilde{v_i}  \hspace{0.5mm} \widetilde{v_j} ) } {\partial x_j}  =  
-  \frac {\partial (\overline{\rho}  \hspace{0.5mm} \tau_{ij})} {\partial x_j}  
-  \frac {\partial \overline{p} }  {\partial x_i} +  \frac {\partial \overline{\sigma}_{ij}  } {\partial x_j}  + \overline{\rho} \hspace{0.5mm} {\widetilde f_i}, 
\label{FilteredMomentum} 
\end{equation}
where $\tau_{ij}$, called the subgrid-scale stress tensor, is defined as 
\begin{equation}
\tau_{ij} = \widetilde{v_i v_j} - \widetilde{v_i}  \hspace{0.5mm} \widetilde{v_j}. 
\label{SGSstress} 
\end{equation}
This tensor cannot be evaluated exactly because of the closure problem, and developing an adequate 
model for it is essential for large-eddy simulations. The filtered pressure can be written as $\overline {p} =  \overline{\rho} R  \widetilde{T} $, where $T$ is the gas temperature, 
and $R =  k_{\rm B}/(\mu_{\rm H} m_{\rm H})$ is the ideal gas constant  with  $k_{\rm B}$, $\mu_{\rm H}$ and 
$m_{\rm H}$ the Boltzmann constant, the molecular weight and the atomic mass unit, respectively. 
The viscous stress tensor, $\sigma_{ij}$, in eq.\ (\ref{FilteredMomentum}) is given 
by $\sigma_{ij} = 2 \rho \nu (S_{ij} - \frac{1}{3} \delta_{ij}  S_{kk})$ where $\nu$ 
is the kinematic viscosity and $S_{ij} = \frac{1}{2}(\partial_i v_j +  \partial_j v_i)$ is the rate of  
strain tensor. We approximate the filtered viscous stress by $\overline{\sigma}_{ij} = 2 \overline{\rho} \nu (\widetilde{S}_{ij} - \frac{1}{3} \widetilde{S}_{kk} \delta_{ij} )$, 
where $\widetilde{S}_{ij} \equiv \frac{1}{2}(\partial_i \widetilde{v}_j + \partial_j \widetilde{v_i})$ is the strain tensor at the 
resolution scale.  For interstellar  turbulence,  various sources contribute to 
the driving force, $f_i$, in the momentum equation, including, e.g., gravity and the acceleration by supernovae.    

To evaluate the pressure term in eq.\ (\ref{FilteredMomentum}),  one needs to consider the filtered 
energy or temperature equation, which
reads (e.g., Garnier et al.\ 2009), 
\begin{gather}
C_{\rm V}\frac{ \partial (\overline{\rho} \hspace{0.5mm}  \widetilde{T}) } {\partial t}  + 
C_{\rm V} \frac {\partial ( \overline{\rho} \hspace{0.5mm}  \widetilde{T} \hspace{0.5mm}
\widetilde{v_i} ) } {\partial x_i}  =  - \overline{\rho} R \widetilde{T}  \hspace{0.5mm} \frac{\partial \widetilde{v}_i}{\partial x_i}   + \widetilde{S}_{ij}  \hspace{0.5mm} \overline{\sigma}_{ij}   + \overline{\rho} \hspace{0.5mm} \widetilde{\Gamma} - \overline{\rho} \hspace{0.5mm} \widetilde{\Lambda} 
 \hspace{2.5cm}   \notag\\ \hspace{2.8cm} 
- \left(\overline{p \frac {\partial v_i}  {\partial x_i} } - \overline{p}  \hspace{0.5mm} \frac{\partial \widetilde{v}_i}{\partial x_i}\right) +  
\left (\overline{S_{ij} \sigma_{ij}}  -   \widetilde{S}_{ij}  \hspace{0.5mm} \overline{\sigma}_{ij}  \right)
\hspace{2cm} \notag\\ \hspace{1.4cm}  + \frac {\partial} {\partial x_i} \left(\overline {\kappa} \frac{\partial \widetilde{T}} {\partial x_i}  \right)  
- C_{\rm V} \frac {\partial \left(  \overline{\rho} q_i \right) } {\partial x_i} ,   
\label{FilteredTemperature} 
\end{gather}
where $C_{\rm V}$ is the heat capacity of the flow material, 
equal to $3 R/ 2$ for a monoatomic gas.  The last two terms, $\Gamma$ and $\Lambda,$ on the first line are the heating rate by external sources  
and the cooling rate by radiation, respectively. 
The  two  terms in the second line of equation  (\ref{FilteredTemperature}) correspond to the effects of $pdV$ work and 
heating by viscous dissipation at subgrid scales, which we model later in this section.  
The first term on the third line represents thermal conduction with $\kappa$ the thermal conductivity, where we 
assumed $\overline{\kappa \hspace{0.5mm} \partial_i T} \simeq  \overline{\kappa} \hspace{0.5mm} \partial_i \widetilde{T}$. 
The last term in eq.\ (\ref{FilteredTemperature}) is  the heat  transport by subgrid turbulent motions, and the temperature flux $q_i$ is  defined as,  
 \begin{equation}
 q_i= \widetilde{T v_i} - \widetilde{T} \hspace{0.5mm} \widetilde{v}_i, 
 \end{equation} 
which will be modeled later. 

A variety of subgrid models have been developed to approximate $\tau_{ij}$ 
(see reviews by Lesieur and Metais (1996) and Meneveau and Katz (2000) for 
LES of incompressible flows). 
A major class of models adopt an eddy-viscosity assumption, relating the deviatoric part of $\tau_{ij}$ to the resolved strain tensor, 
 \begin{equation}
\tau_{ij}  - \frac{1}{3} \tau_{kk} \delta_{ij} =  - 2 \nu_{\rm t} \left(\widetilde{S}_{ij} - \frac{1}{3} \widetilde{S}_{kk} \delta_{ij}\right),
\label{eddyviscosity} 
\end{equation}
where the eddy viscosity, $\nu_{\rm t},$ is usually constructed as the product of a length scale ($\simeq \Delta$) 
and a velocity scale characteristic of the subgrid turbulent motions. In an LES for interstellar turbulence, $\nu_{\rm t}$ is 
typically much larger than the kinematic viscosity $\nu$, and the viscous stress term in eq.\ (\ref{FilteredMomentum}) 
may be neglected. For incompressible flows, $\widetilde{S}_{kk}$ in eq.\ (\ref{eddyviscosity}) vanishes because 
$\partial_i \widetilde{v}_i =0$, and the isotropic part, $\frac{1}{3} \tau_{kk} \delta_{ij}$, of the subgrid stress can 
be absorbed in the pressure term. Therefore, one obtains a complete subgrid model for LES of incompressible flows by setting $\tau_{ij} = -2 \nu_{\rm t} \widetilde{S}_{ij}$. 
On the other hand, in compressible flows, the isotropic part must be modeled explicitly. This part behaves like 
a pressure term, and is sometimes named ``turbulent pressure".  
Note that $\tau_{kk} = \widetilde{v_k v_k} - \widetilde{v}_k \widetilde{v}_k = 2 K$, with $K$ the turbulent 
kinetic energy per unit mass at subgrid scales.   Similar to the eddy-viscosity model for the subgrid stress, 
one may adopt an eddy-diffusivity assumption for  the temperature flux, $q_i$, caused by subgrid  turbulent 
motions,  
\begin{equation}
q_i = -  {\alpha_{\rm t}} \frac{\partial \widetilde{T}} {\partial x_i},  
\label{TFlux}
\end{equation}
where the ``eddy conductivity"  $\alpha_{\rm t}$ is of the same order as $\nu_{\rm t}$, and is 
usually parameterized by a subgrid Prandtl number, $\alpha_{\rm t} = \nu_{\rm t}/Pr_{\rm t},$
where $Pr_{\rm t}$ is typically  taken to be $\simeq 0.7$ (e.g., Edison 1985, Erlebacher et al.\ 1992, Jaberi et al.\ 1999).  

Eddy-viscosity models differ in how $\nu_{\rm t}$ is evaluated. In the Smagorinsky (1963) model,  
$\nu_{\rm t}$  is calculated by $(C_{\rm s} \Delta)^2 |\widetilde{S}|$ with $|\widetilde{S}| = (2\widetilde{S}_{ij} \widetilde{S}_{ij})^{1/2}$,  
which essentially assumes the amplitude of the subgrid velocity fluctuations 
goes like $\propto \Delta |\widetilde{S}|$. The Smagorinsky model has also been 
used in the LES of compressible flows (e.g., Moin et al.\ 1991,  Erlebacher et al.\ 1992, 
Vreman et al.\ 1997).  For compressible flows, Yoshizawa (1986) proposed to 
set $\tau_{kk}  \equiv 2K  =2 C_{\rm I} \Delta^2 |\widetilde{S}|^2$ for  the isotropic 
part of the subgrid stress, which appears to underestimate the subgrid kinetic energy 
(Park \& Mahesh 2007).  
  
A variant of the eddy-viscosity model is the so-called one-equation model, 
where an equation for the subgrid kinetic energy, $K$, is derived, modeled and solved 
(e.g., Lilly 1966, Schumann 1975, Moeng 1984,  Ghosal et al.\ 1995, Menon \& Kim 1996; 
for one-equation models of compressible flows, see, e.g., Schmidt et al.\ 2006, Park \& Mahesh 2007, 
Genin \& Menon 2010, and Chai \& Mahesh 2012). Using the solved 
subgrid kinetic energy, the eddy-viscosity is then estimated by, 
\begin{equation}
\nu_{\rm t}= C_{\nu} \Delta \sqrt{2K}.
\label{nut}
\end{equation} 
In this paper, we will consider the one-equation  model primarily  as an 
example to illustrate  the construction of an LES for the pollution of primordial gas in interstellar turbulence.  
For a compressible flow, the subgrid kinetic energy equation is given by, 
\begin{gather}
{\displaystyle \frac{ \partial (\overline{\rho} \hspace{0.5mm} K) } {\partial t}  + \frac {\partial ( \overline{\rho}   K \widetilde{v_i} ) } {\partial x_i}  
= - \overline{\rho} \widetilde{S}_{ij} \tau_{ij}  + \overline{\rho} \left(\widetilde{v_i f_i} - \widetilde{v}_i \hspace{0.5mm}  \widetilde{f}_i  \right)} \hspace{2cm} \notag \\ 
\hspace{0.5cm} + \left( \overline{p \frac {\partial v_i}{\partial x_i}} - \overline{p} \frac{ \partial \widetilde{v}_i} { \partial x_i}  \right) -  \left( \overline{S_{ij} \sigma_{ij}} - \widetilde{S}_{ij} \overline{\sigma}_{ij} \right) \notag \\ \hspace{1.4cm}
 + \frac{\partial}{\partial x_i} \Bigl[  \overline{\rho} \hspace{0.5mm}  \widetilde{v}_j \tau_{ij} +  (\overline{ v_j \sigma_{ij} } - \widetilde{v}_j \hspace{0.5mm}  \overline{\sigma}_{ij} ) 
\hspace{2.2cm}  \notag \\
\hspace{1.8cm} - \frac{1}{2} \overline{\rho} \hspace{0.5mm} ( \widetilde{v_jv_j v_i} - \widetilde{v_j v_j}  \hspace{0.5mm}  \widetilde{v}_i) 
   - \left(\overline{p v_i}  - \overline{p} \hspace{0.5mm}  \tilde{v}_i \right)
\Bigr] ,
\label{subgridkinetic}
\end{gather}
where the first  term on the r.h.s.\  represents the production of subgrid kinetic energy 
by the cascade from resolved scales.
The two terms on the second line appeared earlier 
in the filtered temperature equation,  corresponding to the $pdV$ work (or pressure-dilation)  
and the viscous dissipation at subgrid scales. The pressure-dilation 
term is sometimes neglected
for weakly compressible flows
because it is difficult to model (e.g., Moin et al.\ 1991, Erlebacher et al.\ 1992), 
but in highly compressible flows the $pdV$ work is not negligible and needs 
to be accounted for.
Using direct numerical simulations of supersonic turbulence, 
PS10 showed that, despite its reversible nature, the $pdV$ work tends to convert kinetic energy to thermal energy, 
and thus acts as a significant kinetic energy sink in addition to the viscous dissipation. 
Based on their results, one can model 
the pressure-dilation and viscous dissipation terms together as $C_{\rm diss} \overline{\rho}  K/ \tau_{\rm sdyn} = C_{\rm diss} \sqrt{2}\overline{\rho}K^{3/2}/\Delta $, 
where the subgrid dynamical time, $\tau_{\rm sdyn}$, is assumed to be $\Delta/\sqrt{2K}$.  
Here we have implicitly assumed that the filter size lies in the inertial range of the real flow and also 
assumed that the flow ``driving" length at subgrid scales is $\simeq \Delta$.  
This may not be true if, for example, the supernova explosion 
is the main energy source for turbulence and
the resolution scale is significantly larger than the size of supernova 
remnants. We do not consider this complexity here, as the one-equation subgrid 
model is used largely for an illustration purpose.  The dimensionless parameter $C_{\rm diss}$ 
is expected to be a function of the subgrid Mach number, $M_{\rm s} \equiv {(2K/ R \widetilde{T})}^{1/2}$,  
and the dependence of $C_{\rm diss}$ on  $M_{\rm s}$ may be determined 
using the simulation results of PS10.  
 
The last term on the first line of eq.\ (\ref{subgridkinetic}) corresponds to the addition of kinetic energy at subgrid scales by the 
driving force, $f_i$. If the characteristic length scale of  $f_i$  
is much larger the filter size, $f_i \simeq \widetilde{f}_i$, and $(\widetilde{v_i f_i} - \widetilde{v}_i \hspace{0.5mm}  \widetilde{f}_i)$ 
would be negligible, meaning that the driving force stores kinetic energy mainly at the 
resolved scales. On the other hand, a significant fraction of energy input from supernova explosions 
may be deposited primarily as subgrid kinetic energy, if the simulation does not resolve 
the typical size of supernova remnants (Scannapieco \& Bruggen 2010). In that case, 
$\widetilde{f}_i \simeq 0$ for  isotropically expanding supernova remnants, 
and the subgrid kinetic energy input can be estimated 
as the product of the supernova explosion energy and the local supernova rate per unit volume. 
 
The transport (or flux)  terms in the last two lines of eq.\ (\ref{subgridkinetic}) are usually grouped 
and modeled together as a diffusion of the subgrid kinetic energy (Lilly 1966, Schumann 
1975, Moeng 1984,  Ghosal et al.\ 1995, Schmidt et al.\ 2006, Genin \& Menon 2010, see, however, Chai \& Mahesh 2012 
for separate treatment of each individual term).  Here we adopt  an eddy-diffusion assumption 
for the first  three flux terms and approximate them together by  $(\nu +  \nu_{\rm k}) \partial_i K $, where  $ \nu_{\rm k}  =  C_{\rm k} \Delta \sqrt{2K} $.  
The parameter $C_{\rm k}$ is sometimes set to be equal to 
$C_{\rm \nu}$ in eq.\ (\ref{nut}) for the subgrid stress tensor (e.g., Kim \& Menon 1999). 
In general, they may be different and need to be treated separately (e.g., Schmidt et al.\ 2006). 
The last flux term in eq.\ (\ref{subgridkinetic}) can be 
written as  $-\overline{\rho} R (\widetilde{Tv}_i -\widetilde{T} \hspace{0.5mm}\widetilde{v}_i) =  -\overline{\rho} R q_i$ 
where the temperature flux, $q_i$, by subgrid motions is modeled by  eq.\ (\ref{TFlux}).        
With these assumptions,  we have (see Genin \& Menon 2010),  
\begin{gather}
\frac{ \partial (\overline{\rho} \hspace{0.5mm} K) } {\partial t}  + \frac {\partial ( \overline{\rho}   K \widetilde{v_i} ) } {\partial x_i}  
= - \overline{\rho} \widetilde{S}_{ij} \tau_{ij}  -  C_{\rm diss} \frac {\sqrt{2} \overline{\rho}K^{3/2} } {\Delta} \hspace{2.3cm} \notag \\  \hspace{2.3cm}
+ \frac{\partial}{\partial x_i} \left[ \overline{\rho} (\nu  + \nu_{\rm k} ) \frac{\partial K}{\partial x_i}  + 
\overline{\rho} R \frac{\nu_{\rm t}}{Pr_{\rm t}} \left( \frac{\partial \widetilde{T} } {\partial x_i}  \right) \right] \notag\\  \hspace{-1.1cm}
+ \overline{\rho} \left(\widetilde{v_i f_i} - \widetilde{v}_i \hspace{0.5mm}  \widetilde{f}_i  \right),
\label{subgridkineticmodel}
\end{gather} 
which is in a closed form and can be evolved to obtain the subgrid turbulent energy. 

We next consider the filtered temperature equation (\ref{FilteredTemperature}). 
We use the eddy-diffusivity model, eq.\ (\ref{TFlux}), for the temperature flux, $q_i$, in 
the last term of  eq.\ (\ref{FilteredTemperature}), and model
the pressure-dilation and viscous dissipation terms 
as in eq.\ (\ref{subgridkineticmodel}).
With these  assumptions, we obtain,  
\begin{gather}
C_{\rm V}\frac{ \partial (\overline{\rho} \hspace{0.5mm}  \widetilde{T}) } {\partial t}  + 
C_{\rm V} \frac {\partial ( \overline{\rho} \hspace{0.5mm}  \widetilde{T} \hspace{0.5mm}
\widetilde{v_i} ) } {\partial x_i}  =  - \overline{\rho} R \widetilde{T}  \hspace{0.5mm} \frac{\partial \widetilde{v}_i}{\partial x_i}   + \widetilde{S}_{ij}  \hspace{0.5mm} \overline{\sigma}_{ij}  \hspace{2cm}  \notag \\ \hspace{2.3cm} + \frac {\partial} {\partial x_i} \left(\overline {\kappa} \frac{\partial \widetilde{T}} {\partial x_i}  \right)
 +  C_{\rm V} \frac {\partial } {\partial x_i} \left(\overline{\rho} \frac{\nu_{\rm t} } {Pr_{\rm t}} \frac{\partial \widetilde{T}} {\partial x_i}\right)   \notag \\\hspace{0.8cm} 
+ C_{\rm diss} \frac {\sqrt{2} \overline{\rho} K^{3/2} } {\Delta} 
 + \overline{\rho} \hspace{0.5mm} \widetilde{\Gamma} - \overline{\rho} \hspace{0.5mm} \widetilde{\Lambda}  , 
\label{FilteredTemperaturemodel} 
\end{gather}
where the thermal conductivity term can be neglected if $\overline{\kappa} \ll \overline{\rho} C_{\rm V} \alpha_{\rm t}$.    
An alternative approach to obtain $\widetilde{T}$ is to model and evolve the equation of 
the filtered total energy, $\widetilde{E}$($\equiv \frac{1}{2} \widetilde{v_i}  \widetilde{v_i} + K  + C_{\rm V}  \widetilde{T} $), per unit mass (e.g., Vreman et al.\ 1997, Kosovic et al.\ 2002, Schmidt et al.\ 2006, Park \& Mahesh 2007, 
Genin \& Menon 2010,  Scannapieco \& Bruggen 2010, Chai \& Mahesh 2012).    

To solve the LES and $K$ equations, one needs to determine four parameters, 
$C_{\rm \nu}$, $Pr_{\rm t}$, $C_{\rm diss}$, and $C_{\rm k}$. Traditionally, these are assumed to 
be positive constants and specified {\it a priori} and then tuned by testing against experiments or  numerical simulations. 
On the other hand, this approach has the weaknesses of failing to fully 
account for the flow-dependence of these parameters, as well as 
not allowing the backscatter of the subgrid kinetic energy to the resolved scales, 
which does occur in some local regions of a turbulent flow (Piomelli et al.\ 1991).
These limitations motivated a dynamic procedure for subgrid 
modeling where the model coefficients are computed 
in a localized and adaptive way using the flow structures at resolved scales 
and the assumption of scale invariance
(e.g.\ Germano et al.\ 1991;  Moin et al.\ 1991, Germano 1992; Lilly 1992; Ghosal et al.\ 1995, Kim \& Menon 
1999; Schmidt et al.\ 2006; Park \& Mahesh 2007; Genin \& Menon 2010; Chai \& Mahesh 2012).

Here we have restricted our attention to the eddy-viscosity models and focused particularly 
on the one-equation model. The interested reader is referred to, e.g., Vreman et al.\ (1997), 
Meneveau \& Katz (2000), and De Stefano et al.\ (2008), for non-eddy-viscosity subgrid models 
and their dynamic versions.  Two-equation subgrid models have also been developed, 
which, in addition to the subgrid kinetic energy, evolve another subgrid quantity, 
such as the dissipation rate (e.g., Gallerano et al.\ 2005) or the characteristic length 
scale of subgrid turbulent motions (e.g., Fang \& Menon 2006; Dimonte \& Tipton 2006;  Scannapieco \& Bruggen 2010).  

\subsection{Subgrid Model for Turbulent Mixing and the Pollution of Pristine Gas}

In this subsection, we construct a subgrid model for the pollution of primordial 
gas in early galaxies. We first consider the equation for the filtered concentration field, 
which provides a general illustration for subgrid modeling of turbulent mixing. 
Applying the filtering procedure 
to the advection-diffusion equation (\ref{advection}) gives,   
\begin{equation}
\frac{ \partial (\overline{\rho} \hspace{0.5mm} \widetilde{C} )} {\partial t}  +  \frac {\partial (\overline{\rho} \hspace{0.5mm} \widetilde{v}_i \widetilde{C})}{\partial x_i}  
=  - \frac{\partial (\overline{\rho} g_i )}{\partial x_i}   +  \frac{\partial}{\partial x_i}  \left( \overline{\rho \gamma  \frac{\partial C}{\partial x_i}}  \right)  + 
\overline{\rho} \widetilde{S}, 
\label{FiilteredConcentration} 
\end{equation}
where $g_i  = \widetilde{v_iC} - \widetilde{v_i}\widetilde{C} $ is the concentration flux 
caused by subgrid turbulent motions. The equation is similar to the temperature equation 
(\ref{FilteredTemperature}) except for the pressure-dilation and viscous dissipation terms. 
In analogy to the subgrid temperature flux,  $q_{i}$, one may adopt an eddy-diffusivity 
assumption for the concentration flux, $g_i = - \gamma_{\rm t} \partial_i \widetilde{C}$, yielding,  
\begin{equation}
\frac{ \partial (\overline{\rho} \hspace{0.5mm} \widetilde{C})} {\partial t}  +  \frac {\partial (\overline{\rho} \hspace{0.5mm} \widetilde{v}_i \widetilde{C})}{\partial x_i}  
=  \frac{\partial}{\partial x_i} \left(  \overline{\rho} (\gamma+ \gamma_{\rm t} )  \frac {\partial \widetilde{C}}{\partial x_i}  \right)  + \overline{\rho} \widetilde{S}, 
\label{FilteredConcentrationAdvection} 
\end{equation}
where we also assumed $\overline{\rho \gamma \partial_i C} \simeq \overline{\rho} \gamma \partial_i \widetilde{C}$. 
The eddy diffusivity, $\gamma_{\rm t}$, is of the same order as the eddy viscosity, and the subgrid 
Schmidt  number $Sc_{\rm t}$($\equiv \nu_{\rm t}/\gamma_{\rm t}$) is sometimes set to 
be the same as the subgrid Prandtl number $Sc_{\rm t} = Pr_{\rm t} \approx 0.7$ (e.g., Jaberi et al.\ 1999). 
Somewhat smaller values, $Sc_{\rm t} \simeq 0.3-0.4$, have also been proposed (e.g., Pitsch \& Steiner 2000 and 
Jimenez et al.\ 2001). $Sc_{\rm t}$ can also be computed from the local flow structures using 
the dynamic procedure discussed above (see e.g., Moin et al.\ 1991, Pierce \& Moin 1998). For the LES of interstellar turbulence,  
$\gamma_{\rm t}$ is expected to be much larger
 than the molecular diffusivity  $\gamma$.  

Similar to the subgrid kinetic energy, we can derive an 
equation for the subgrid concentration variance, defined as $\widetilde{(\delta C)^2} = \widetilde{C^2}  - (\widetilde{C})^2$. 
The equation reads, 
\begin{gather}
 \frac{ \partial \left(\overline{\rho} \hspace{0.5mm} \widetilde{(\delta C)^2}\right)} {\partial t}  +  \frac {\partial \left(\overline{\rho} \hspace{0.5mm} \widetilde{v}_i \hspace{0.5mm} \widetilde{ (\delta C)^2 }\right)}{\partial x_i}  
=  - 2 \overline{\rho}  \hspace{0.5mm} g_i  \hspace{0.5mm} \frac{\partial \widetilde{C} }{\partial x_i}  
 + \frac{\partial}{\partial x_i} \bigg\{ 2 \overline{\rho} \hspace{0.5mm} \widetilde{C}  \hspace{0.5mm} g_i 
 \hspace{3cm} 
\notag\\ \hspace{1.6cm}
 +  \left( \overline{\rho \gamma  \frac {\partial C^2} {\partial x_i}}  - 2 \widetilde{C} \hspace{0.5mm} \overline{ \rho \gamma  \frac {\partial C} {\partial x_i} }\right)   -  \overline{\rho} \left(\widetilde{C^2 v_i} - \widetilde{C^2} \hspace{0.5mm} \widetilde{v}_i\right)   \bigg\}\notag \\ \hspace{-0.8cm}
 - 2 \left[\overline{ \rho \gamma \left(\frac{\partial C}{\partial x_i} \right)^2 } -  \overline{\rho \gamma  \left( \frac{\partial C} {\partial x_i}\right)}
\frac{\partial \widetilde{C} }{\partial x_i} \right]  \notag\\ \hspace{-3.5cm}
+ 2 \overline{\rho} ( \widetilde{SC} - \widetilde{S}  \hspace{0.5mm} \widetilde{C})  ,   
\label{FilteredConcentrationVariance} 
\end{gather}
which is in close analogy to eq.\ (\ref{subgridkinetic}).
The first term, $- 2 \overline{\rho}  g_i \partial_i \widetilde{C}$, on the r.h.s.\ represents the production 
of the concentration variance by the scalar cascade from the resolved scales. 
The term on the third line corresponds to the subgrid scalar dissipation 
by molecular diffusivity. We model the dissipation as $\overline{\rho} \widetilde{ (\delta C)^2}/\tau_{\rm sm}$ 
(see eq.\ (\ref{ensemblevariance})), where the subgrid mixing timescale $\tau_{\rm  sm}$ is expected to scale with 
the subgrid dynamical time, $\tau_{\rm sdyn} \equiv  \Delta/\sqrt{2K}$. Parametrizing 
$\tau_{\rm sm}$ with respect to $\tau_{\rm sdyn}$, we set the dissipation term 
to $C_{\rm m}   \overline{\rho} \sqrt{2K} \widetilde{ (\delta C)^2}/\Delta$, where $C_{\rm m} = \tau_{\rm sdyn}/\tau_{\rm m}$. 
The parameter $C_{\rm m}$ depends on the local subgrid Mach 
number, $M_{\rm s} = (2K/RT)^{1/2}$, and also on the subgrid length 
scale $L_{\rm sp}$ at which the pollutants are injected, 
and it can be  calibrated using the simulation results of PS10, 
who tabulated  the mixing timescale of passive scalars forced 
at different length scales in turbulent flows at a range of Mach numbers.
If the pollutants are forced at large scales and 
the subgrid fluctuations are contributed primarily by the cascade 
from resolved scales,  
and
it is appropriate to set $L_{\rm sp}  = \Delta $.  However, 
$L_{\rm sp}$ could be smaller than the resolution scale, $\Delta$, if, for 
example, multiple supernovae explode in a single computational cell (\eg Scannapieco \& Br\" uggen 2010). 
Finally, we model the transport (flux) terms (the last term in the first line and all the terms in the second line) together as 
a diffusion term,  $\partial_i \left( (\gamma+ \gamma_{\rm t2}) \overline \rho \partial_i \widetilde{\delta C^2}\right)$.
The eddy-diffusivity,  $\gamma_{\rm t2}$, here is likely to be close to $\gamma_{\rm t}$ in the 
$\widetilde{C}$ equation, although it is not clear if they are exactly equal (see below). 
These assumptions result in a closed variance equation (c.f. Jimenez et al.\ 2001), 
\begin{gather}
\frac{ \partial (\overline{\rho} \hspace{0.5mm} \widetilde{(\delta C)^2})} {\partial t}  +  \frac {\partial (\overline{\rho} \hspace{0.5mm} \widetilde{v}_i \widetilde{ (\delta C)^2 })}{\partial x_i}  
=  2 \overline{\rho} \gamma_{\rm t} \left( \frac{\partial \widetilde{C} }{\partial x_i} \right)^2  \notag \hspace{2.cm} \\ \vspace{2mm}\hspace{1.5cm}
-  C_{\rm m} \overline{ \rho} \frac{\sqrt{2K} \widetilde{(\delta C)^2}}{\Delta} 
+ \frac{\partial}{\partial x_i} \left[ \overline{\rho}(\gamma + \gamma_{\rm t2})  \frac {\partial \widetilde{ (\delta C)^2 }}{\partial x_i}  \right]  \notag \\
+2 \overline{\rho} ( \widetilde{SC} - \widetilde{S} \widetilde{C}), \hspace{2.9cm}
\label{ModeledConcentrationVariance} 
\end{gather}
which illustrates the basic picture for modeling the subgrid concentration 
fluctuations, and provides a useful guideline for formulating a 
subgid model for the pollution of pristine gas. A dynamic 
procedure for the subgrid scalar variance and dissipation was developed in Pierce and Moin (1998).

The subgrid model we construct for the pollution of pristine gas is based on the
the PDF formulation in the 
context of LES. Applying a Favre filter, eq.\ (\ref{Favrefilter}), to the fine-grained 
concentration PDF  $\phi = \delta (Z-C({\bf x},t))$, we define a density-weighted PDF at the 
resolution scale,    
\begin{equation}
\widetilde {\phi}(Z; {\bf x},t)  =  \frac{\overline {\rho \phi(Z; {\bf x},t)}} {\overline{\rho}}. 
\end{equation}
An exact equation for the filtered PDF, $\widetilde{\phi}$, is derived in Appendix A,  
\begin{gather}
\frac {\partial (\overline {\rho} \hspace{0.5mm}\widetilde {\phi})} {\partial t}  +  \frac{\partial}{\partial x_i} \left( \overline{\rho} \hspace{0.5mm} \widetilde{\phi} \hspace{0.5mm}\overline{[v_i|C=Z]}_{\rho}  \right) = \frac{\partial}{\partial x_i} \left( \overline{\rho \gamma \frac{\partial \phi}{\partial x_i}} \right)  \hspace {3cm}\notag  \\ \hspace{2.2cm}  
-  \frac{\partial^2}{\partial Z^2} \left(\overline{\rho} \hspace{0.5mm} \widetilde{\phi} \overline{\left[\gamma \left(\frac{\partial C}{\partial  x_i} \right)^2 \left\vert  \vphantom{\frac{1}{1}} \right. C=Z \right]}_\rho \right)\notag  \\ \hspace{.3cm}  
  -  \frac{\partial}{\partial Z}  \left( \overline{\rho}  \hspace{0.5mm}\widetilde{\phi}  \hspace{0.5mm} \overline{ [S|C=Z]} _{\rho}  \right), 
\label{rhofilteredeqtext}
\end{gather}
where $\overline{[\cdot \cdot \cdot|C=Z]}_\rho$ denotes density-weighted filtering 
conditioned on the local concentration value. The definition of the conditional filtering is given in Appendix A.

Eq.\ (\ref{rhofilteredeqtext}) is essentially identical to eqs.\ (\ref{pdfeq}) and 
(\ref{ensemblediffusivityterm}) for the ensemble-defined PDF, $\Phi$. This implies 
that, first,  the same closure problem exists for the advection and diffusivity terms in 
eq.\ (\ref{rhofilteredeqtext}), and, second, the PDF closure models in the ensemble 
context can be applied to the filtered PDF equation. Although our primary goal is not to 
solve the equation for the entire filtered PDF, we give an outline for modeling 
the PDF equation, which is helpful for understanding our LES approach for the pristine fraction. 
We first consider the advection term, which is responsible for the transport of the 
PDF between different regions by the turbulent velocity. We write it in two terms,  
$\widetilde{\phi} \hspace{0.5mm}\overline{[v_i|C=Z]}_{\rho}  =  \widetilde{\phi} \hspace{0.5mm}  \widetilde{v}_i  +    \widetilde{\phi} \left( \overline{[v_i|C=Z]}_{\rho} -  \widetilde{v}_i\right)$, 
and then modeling the second term with an eddy-diffusivity assumption gives $\widetilde{\phi} \hspace{0.5mm}\overline{[v_i|C=Z]}_{\rho}  =  \widetilde{\phi} \hspace{0.5mm}  \widetilde{v_i}  -  \gamma_{\rm t \phi} { \partial_i \widetilde{\phi}} $ where $\gamma_{\rm t \phi}$ is the eddy-diffusivity for the PDF flux by the subgrid motions (see, e.g., Gao \& O'Brien (1993), Colucci et al.\ (1998),  Jaberi et al.\ (1999)).    
The filtered PDF equation then becomes, 
\begin{gather}
\frac {\partial (\overline {\rho} \hspace{0.5mm}\widetilde {\phi})} {\partial t}  +  \frac{\partial}{\partial x_i} \left( \overline{\rho} \hspace{0.5mm} \widetilde{\phi} \hspace{0.5mm} \widetilde{v}_i  \right) 
= \frac{\partial}{\partial x_i} \left( \overline{\rho} ( \gamma_{\rm t \phi} + \gamma) \frac{\partial \widetilde{\phi} }{\partial x_i}  \right) 
\hspace{3cm}\notag\\ \hspace{2cm}
- \frac{\partial^2}{\partial Z^2} \left(\overline{\rho} \hspace{0.5mm} \widetilde{\phi} \overline{\left[\gamma \left(\frac{\partial C}{\partial  x_i} \right)^2 \left\vert  \vphantom{\frac{1}{1}} \right. C=Z   \right]}_\rho \right) \notag\\
\hspace{0.1cm}
  -  \frac{\partial}{\partial Z}  \left( \overline{\rho}  \hspace{0.5mm}\widetilde{\phi}  \hspace{0.5mm} \overline{ [S \vert C=Z]} _{\rho}  \right),
\label{rhofilteredeqadvection}
\end{gather}
where we also assumed that $\overline{\rho \gamma \partial_i \phi} \simeq \overline{\rho} \gamma \partial_i \widetilde{\phi}$ 
(see, e.g., Jaberi et al.\ 1999). Taking the first-order moment of eq.\ (\ref{rhofilteredeqadvection}) gives an equation for 
the filtered concentration $\widetilde{C}$,
which is the same as eq.\ (\ref{FilteredConcentrationAdvection}) 
except that  $\gamma_{\rm t}$ is replaced by $\gamma_{\rm t \phi}$. This suggests that $\gamma_{\rm t \phi} \simeq \gamma_{\rm t}$ (Gao \& O'Brien 1993). 
Also, using the second-order moment of eq.\ (\ref{rhofilteredeqadvection}), we can derive an equation for the subgrid variance, $\widetilde{(\delta C)^2}$,  
which is the same as
eq.\ (\ref{ModeledConcentrationVariance}) 
except that 
$\gamma_{\rm t \phi}$ replaces both $\gamma_{\rm t}$ and $\gamma_{\rm t2}.$ 
This indicates
that the eddy diffusivities for the concentration 
mean ($\gamma_{\rm t}$) and variance ($\gamma_{\rm t2}$) are automatically set to 
be equal if one models the advection term in the PDF equation with an eddy-diffusivity closure.   

The term on the second line of eq.\ (\ref{rhofilteredeqadvection}) represents homogenization by molecular diffusivity, 
and can be modeled using established PDF closure approximations for turbulent 
mixing,  such as those discussed in \S 3.1.  In \S 6.1, we derived 
an expression for the source term in the ensemble PDF equation. 
Using the same method,  we estimate the source term in the last line of  eq.\ (\ref{rhofilteredeqadvection}) for the filtered PDF.  The source term for new metals from supernovae is 
$\overline{\dot{n}}_{\rm SN} m_{\rm ej} \left(\delta(Z-Z_{\rm ej}) -\widetilde{\phi}\right)$, where $\overline{\dot{n}}_{\rm SN}({\bf x}, t)$ 
is the filtered number rate of supernova explosions per unit volume, and ejecta from each supernova 
is assumed to have the same mass $m_{\rm ej},$ with metallicity $Z_{\rm ej}$.
Again the $-\widetilde{\phi}$ term ensures the conservation of  the total probability. 
With the supernova source term in the filtered PDF equation, it is straightforward to 
calculate the source terms in the filtered concentration and variance equations (\ref{FilteredConcentrationAdvection} and \ref{ModeledConcentrationVariance}). 
We find  that the source terms are $\overline{\dot{n}}_{\rm SN} m_{\rm ej} (Z_{\rm ej} -\widetilde{C})$ and 
$\overline{\dot{n}}_{\rm SN} m_{\rm ej} [(Z_{\rm ej} -\widetilde{C})^2-\widetilde{\delta C^2}]$ in the $\widetilde{C}$ and $\widetilde{\delta C^2}$ equations, respectively.   
If a continuous infall of pristine gas from the halo exists 
during the formation and evolution of a galaxy, one can maintain a mass flux 
at the boundary of the simulation box  and set $\widetilde{\phi} = \delta(Z) $ 
as the boundary condition for the (filtered) concentration PDF. 

We finally consider modeling the pollution of the primordial gas in an LES. Clearly, the fine-grained 
pristine fraction, $P(Z_{\rm c}; {\bf x},t)$, at a given point is an integral of the fine-grained PDF, ${\phi}(Z; {\bf x},t)$, from $Z=0$ 
to the threshold, $Z_{\rm c}$,  and, similarly, the filtered pristine fraction, $\widetilde{P}$, 
at the resolution scale is given by, 
\begin{equation}
\widetilde{P} (Z_{\rm c}; {\bf x},t)= \int_{0}^{Z_{\rm c}} \widetilde{\phi} (Z; {\bf x},t) dZ. 
\label{filteredpristine}
\end{equation}
We can therefore derive an equation for $\widetilde{P}$ by integrating the filtered 
PDF equation (\ref{rhofilteredeqtext}) from $0$ to $Z_{\rm c}$. Performing such an
integration for the advection term in eq.\ (\ref{rhofilteredeqtext}) yields $\widetilde{P v_i}$, 
which corresponds to the flux of the pristine fraction into and out of a computational cell 
due to the transport/advection of the turbulent velocity (see \S 2.1). The term can be rewritten 
as $\widetilde{P} \hspace{0.5mm} \widetilde{v}_i + (\widetilde{P v_i} - \widetilde{P} \hspace{0.5mm} \widetilde{v}_i)$ 
where the term in the brackets is the pristine fraction flux caused by the subgrid turbulent motions.  
We model this subgrid flux with an eddy-diffusion 
assumption, 
\begin{equation}
\widetilde{P v_i} - \widetilde{P} \widetilde{v}_i = -\gamma_{\rm P} \partial_i \widetilde{P},  
\end{equation}
where $\gamma_{\rm P}$ is the eddy diffusivity for the pristine fraction. We  
use the self-convolution models (\S 3.2) for the effect of the diffusivity term on the pristine fraction.  
Integrating the supernova source term from 0 to  $Z_{\rm c}$ 
gives $- \overline{\dot{n}}_{\rm SN} m_{\rm ej} \widetilde{P}$. With these models and assumptions, we obtain, 
\begin{gather}
 \frac {\partial (\overline {\rho} \hspace{0.5mm}\widetilde {P})} {\partial t}  +  \frac{\partial}{\partial x_i} \left( \overline{\rho} \hspace{0.5mm} 
\widetilde{P} \hspace{0.5mm}\widetilde{v}_i  \right) = \frac{\partial}{\partial x_i} \left( \overline{\rho} (\gamma + \gamma_{\rm P})  \frac{\partial \widetilde{P}}{\partial x_i} \right)
 \hspace{2cm} \notag\\ \hspace{2cm}
- \frac{n_{\rm s}}{\tau_{\rm scon}} \widetilde{P}\left(1- \widetilde{P}^{1/n_{\rm s}} \right) - \overline{\dot{n}}_{\rm SN} m_{\rm ej} \widetilde{P} ,
\label{filteredprimordial}
\end{gather}
where it is also assumed $\overline{\rho \gamma \partial_i P} = \overline{\rho} \gamma \widetilde{P}$, 
and $n_{\rm s}$ and $\tau_{\rm scon}$ correspond to the parameters, $n$ and $\tau_{\rm con}$, in the 
self-convolution models discussed in \S 3. As $\gamma$ is likely much smaller than $\gamma_{\rm P}$, 
the pristine fraction flux due to the molecular diffusivity can be neglected. 
The choice for $n_{\rm s}$ and $\tau_{\rm scon}$ according to the turbulence and pollutant 
conditions at subgrid scales will be described and discussed below.  If a pristine mass 
flux is enforced at the boundary of the simulation box to imitate the primordial infall gas, 
one should set  $\widetilde{P} =1$ as a boundary condition. 

A comparison of eq.\ (\ref{filteredprimordial}) with eq.\ (\ref{ModeledConcentrationVariance}) shows that 
both equations have transport or flux terms, a mixing/homogenization term by the molecular diffusivity, 
and a source term. A similar analogy also exists with the equation for the subgrid kinetic energy, eq.\ (\ref{subgridkineticmodel}). 
There is, however, an interesting difference. The concentration variance equation has a term that tends to increase the 
subgrid variance, representing the scalar cascade from the resolved scales. On the other hand, there is no such 
production term in the $\widetilde{P}$ equation, 
because no mechanism exists in the mixing process that 
can produce pristine gas at subgrid scales.  

To derive the equation for the filtered pristine fraction, we 
could also have started from eq.\ (\ref{rhofilteredeqadvection}) 
for the filtered PDF, where the advection term is already modeled by an eddy-diffusion 
assumption.  In that case, we
would have found that the primordial flux due to subgrid 
turbulent motions is given by $-\gamma_{\rm t \phi} \partial_i \widetilde{P}$. This suggests that, 
when applying an eddy-diffusivity closure to the advection term in the filtered PDF equation, it is 
implicitly assumed that the three eddy diffusivities, $\gamma_{\rm t}$, $\gamma_{\rm t2}$,  and $\gamma_{\rm P}$, 
respectively for the mean, variance and the pristine fraction, are the same and all equal to  $\gamma_{\rm t \phi}$. 
The quantitative accuracy of this assumption is not clear, although all the eddy-diffusivities are expected 
to be of the same order. A simple estimate for $\gamma_{\rm P}$ is to scale 
it with the eddy viscosity as, $\gamma_{\rm P} = \nu_{\rm t}/Sc_{\rm P}$, with the 
Schmidt  number $Sc_{\rm P}$ for the pristine fraction in the range from $0.3-0.7$,  
as in the case of $\gamma_{\rm t}$ discussed earlier (see text below eq.\ (\ref{FilteredConcentrationAdvection})).    

To implement eq.\ (\ref{filteredprimordial}) in an LES for the pollution of 
primordial gas in early galaxies,  we now only need to specify the two parameters, 
$n_{\rm s}$ and $\tau_{\rm scon}$, from the self-convolution model. 
When using the convolution model in eq.\ (\ref{filteredprimordial}), we have 
implicitly assumed that statistical homogeneity is restored at the resolution scale, $\Delta$, 
because the applicability of the model is tested and confirmed only in statistically homogeneous 
turbulent flows. With this assumption, $n_{\rm s}$ and $\tau_{\rm scon}$ can be determined using our 
simulation results. These parameters are functions of the flow Mach number and the pollutant 
injection scale relative to the flow driving scale. The subgrid Mach number, $M_{\rm s}$, 
can be easily computed by $(2K/R\widetilde{T})^{1/2}$ in the one-equation model,  
where $K$ and $\widetilde{T}$ are, respectively,  the subgrid kinetic energy and the filtered gas 
temperature.  The subgrid source injection scale, $L_{\rm sp}$, in a computational 
cell would be close to the cell size, $\Delta$, if the pollutant source was  transported into the cell by advection,  
or if only one supernova exploded in the cell. In that case, it is appropriate to set $L_{\rm sp} = \Delta$. On the 
other hand, if multiple supernova explosions occurred in a single cell, 
then $L_{\rm sp}$ would roughly go like the number of supernovae to -1/3 power, 
assuming a random distribution. We assume the subgrid flow ``driving" scale, $L_{\rm sf}$, is roughly 
given by the cell size, $\Delta$. 

With $M_{\rm s}$ and the ratio $L_{\rm sp}/L_{\rm sf}$,  one can fix the parameters, 
$n_{\rm s}$  and $\tau_{\rm scon}$, by interpolating Tables 2, 3 and 4 in \S 5.4.6, or using the fits given 
in eqs.(\ref{eq:taunfit})). The timescales $\tau_{\rm con1}$ and $\tau_{\rm con2}$ given in 
Tables 2 and 4 are normalized to the flow dynamical time, therefore the values for $\tau_{\rm scon}$ 
are in units of the subgrid dynamical time, $\tau_{\rm sdyn}$$(\equiv \Delta/\sqrt{2K})$. 
We point out there is an uncertainty in the applicability of our tabulated parameters to computation cells with 
supernova explosions. In these cells, the effective driving is likely better described by a pure compressive 
force rather than solenoidal. As discussed earlier, this may affect the parameters in the convolution model.  
Future simulations are needed to investigate the potential dependence of the parameters with the compressibility of the driving force.
The convolution timescale also has a dependence  on the threshold metallicity, 
$Z_{\rm c}$, relative to the mean concentration (see \S 5.4.4). Thus, to determine 
$\tau_{\rm scon}$, we need to compute the ratio, $Z_{\rm c}/\widetilde{C}$, 
of the threshold to the mean, $\widetilde{C}$, in a cell. 
For that purpose, it is necessary to solve the filtered concentration 
equation (\ref{FilteredConcentrationAdvection}) to keep track of  
$\widetilde{C}$ in all computational cells. 
As mentioned earlier, the source term in this equation 
is given by $\overline{\dot{n}}_{\rm SN} m_{\rm ej}(Z_{\rm ej} -\widetilde{C})$.
Based our results In \S 5.4.4, for small values of $Z_{\rm c}/\widetilde{C}$ ($\lsim 10^{-3}$),
one can use a weak power-law scaling (see \S 5.4.4) to 
rescale the convolution timescales listed in Tables 2 and 4.
However, it is possible that  $\widetilde{C}$ in 
a computational cell is close to or even smaller than $Z_{\rm c}$. As discussed in \S5.4.4, 
in the extreme case with $\widetilde{C} \lsim Z_{\rm c}$, the evolution of $\widetilde{P}$ 
in a cell would be qualitatively different from the prediction of  the self convolution models.  
A careful treatment is thus needed for cells  with $\widetilde{C}$ close or smaller than $Z_{\rm c}$.  
How the fraction, $\widetilde{P}$, evolves under this situation is not explored in the current work, and we defer it  to a future paper.

In the subgrid model outlined above for the pristine fraction, 
we adopted a simple approach to fix the model parameters,
prescribing them based on our simulation results and 
previous work  on LES.  An interesting question is whether  
the parameters can be determined dynamically using the resolved 
local flow and scalar structures. It seems highly uncertain whether the 
dynamic procedure is applicable at all to the problem of how the pristine 
gas is polluted in a turbulent flow. As mentioned earlier, the validity of 
the dynamical procedure relies on the existence of scale invariance. 
This can be justified, e.g., for the cascade of kinetic energy or the 
concentration variance, based on Kolmogorov's similarity theory of 
turbulence. However, unlike kinetic energy or the scalar variance, which 
are 2nd-order statistical measures, the pristine fraction 
corresponds to the extreme PDF tail, and it is unknown whether scale 
invariance exists for such a high-order quantity.  Exploring the possibility 
of developing a dynamic subgrid model for the pristine gas 
fraction would be an interesting topic for a future study.

\section{Conclusions}

The shift from Population III 
to normal star formation is a global transition of the 
universe that is dependent on mixing on scales 
smaller than a parsec (Pan \& Scalo 2007).  
This means 
that numerical simulations of this process will only be possible if we first develop a deep understanding of
the fundamental physics of how the 
pristine material is polluted in turbulent flows.
In an earlier paper (PSS), we developed a theoretical approach 
to modeling this process based on the PDF method for passive scalar mixing in statistically homogeneous turbulence,
and we explored the evolution of the pristine fraction, $P(Z_{\rm c}, t)$, defined as the mass 
fraction of the flow with pollutant concentration below a tiny threshold $Z_{\rm c}$. 
Then we used numerical simulations to show that
a class of PDF models, called self-convolution models, 
provide successful fitting functions to the solution of $P(Z_{\rm c}, t),$ which corresponds to the far left tail of the concentration PDF.

The convolution models are based on the physical picture of
turbulence stretching pollutants and causing a cascade 
of concentration structures toward small scales. Mixing then occurs as the 
scale of the structures becomes sufficiently small for molecular 
diffusivity to 
operate efficiently, 
and the homogenization between 
neighboring structures corresponds to a convolution of the concentration PDF.  
The picture suggests that the mixing/pollution timescale is determined by the 
turbulent stretching rate at the scale where the pollutants are injected, and 
the main result of PSS was the prediction for the pristine fraction evolution, 
i.e., eq.\ (\ref{pfnconv}), by the generalized self-convolution model.  For convenience, 
we repeat  eq.\ (\ref{pfnconv}) here, 
\begin{equation}
\frac{dP(Z_{\rm c}, t)}{dt} = - \frac{n}{\tau_{\rm con}} P(1-P^{1/n}),  
\label{pfnconvconc}
\end{equation}
where  $\tau_{\rm con}$ is the timescale for the PDF convolution, and 
the parameter $n$ is interpreted as an indicator of the degree of spatial 
locality of the PDF convolution process. A smaller $n$ corresponds to 
more local convolution and broader PDF tails. 

In the present work, we briefly reviewed the formulation of PSS, and conducted 
a systematic numerical study of the turbulent pollution process, exploring an 
extended parameter space. We simulated four statistically homogeneous 
turbulent flows with rms Mach number $M$ ranging from $0.9$ to $6.2$. In each flow, we evolved 20 
decaying scalars with different initial pollutant 
fractions, $H_0$, and different pollutant injection scales, $L_{\rm s}$. 
The simulation data further confirmed the validity of the convolution model and allowed us to 
measure the model parameters, $n$ and $\tau_{\rm con}$, 
in eq.\ (\ref{pfnconvconc}) over a wide range of turbulence and pollutant conditions.
Consistent with PSS, we find that, if the initial pollutant fraction $H_0 \gsim 0.1$, 
the simulation results for the pristine fraction can be well fit by 
the convolution model prediction, eq.\ (\ref{pfnconvconc}), with properly 
chosen parameters.   Eq.\ (\ref{pfnconvconc}) is solved by
\begin{equation}
P(Z_{\rm c}, t) = \frac{P_0}{\left[P_0^{1/n} + (1-P_0^{1/n} ) \exp\left( t /\tau_{\rm con2} \right)  \right]^n},  
\label{pfnconvsolutionconc}
\end{equation}
where $P_0$ is the initial pristine fraction, and  
we have denoted the convolution timescale as $\tau_{\rm con2}$ for 
these scalar fields. Using  eq.\ (\ref{pfnconvsolutionconc}) to fit the simulation 
data yielded best-fit parameters $n$ and $\tau_{\rm con2}$. 
On the other hand, if $H_0 \lsim 0.1$, the evolution of $P(Z_{\rm c}, t)$ shows 
different behaviors at early and late times. In the early phase, the PDF convolution occurs locally 
in space due to the limited amount of pollutants, 
and the pristine fraction evolution follows the prediction of the ``discrete" convolution 
model with $n=1$, i.e., 
\begin{equation}
P(Z_{\rm c}, t) = \frac{P_0}{P_0 + (1-P_0) \exp\left( t /\tau_{\rm con1} \right)},  
\label{pfintegralsolutionconc}
\end{equation}
where the convolution timescale for the early phase is denoted as $\tau_{\rm con1}$. 
Once a significant fraction (0.2-0.3) of flow is polluted, the pristine fraction evolves in 
the same way as the scalar fields with $H_{0} \gsim 0.1$.  We therefore named the convolution 
timescale as $\tau_{\rm con2}$ for both scalars with $H_0 \gsim 0.1$ and the late phases 
of  $H_{0} \lsim 0.1$ scalars (see \S 5.4). A successful two-phase fitting scenario was adopted for  scalars fields
with $H_{0} \lsim 0.1$, which connects eqs.\ (\ref{pfintegralsolutionconc}) and (\ref{pfnconvsolutionconc})
for early and late times.

We examined the dependence of the model parameters on the flow 
Mach number, $M$. We found that the convolution timescales, $\tau_{\rm con1}$ and $\tau_{\rm con2}$, 
normalized to the flow dynamical time increase
by $\simeq 20\%$ as $M$ goes from 0.9 
to 2.1 and then saturates at $M \gsim 2$. This is similar to the behavior 
of the variance decay timescale, $\tau_{\rm m}$, as a function of $M$. 
For $H_0 \gsim 0.1$ scalars 
or the late phase of  scalars with $H_0 \lsim 0.1$, the 
parameter $n$ decreases with increasing $M$, indicating that the PDF convolution 
proceeds more locally in supersonic turbulence 
with larger $M$. The decrease of $n$ is related to broader  
concentration PDF tails at higher $M$, corresponding to 
a larger pristine fraction at the same concentration variance.

The pristine fraction evolution also depends on the pollutant injection scale $L_{\rm s}$. 
As $L_{\rm s}$ deceases, the pollution of the pristine gas is faster and 
the timescales, $\tau_{\rm con1}$ and $\tau_{\rm con2}$, decrease. This is 
expected as the mixing timescale scales with the eddy turnover time at $L_{\rm s}$.   
For  scalars with $H_0 \gsim 0.1$ or the late phase of  $H_0 \lsim 0.1$ 
scalars, the parameter $n$ becomes smaller as $L_{\rm s}$ decreases
because, intuitively, the convolution is more local if 
the pollutants are injected at smaller scales. 

The dependence of  the model parameters, $n$, $\tau_{\rm con1}$ and $\tau_{\rm con2}$, on 
the turbulence and pollutant properties is summarized in Tables 2, 3, and 4, and for 
convenience we have fit these results with simple functions as,
\begin{gather}
\tau_{\rm con1} =   \left[0.225 - 0.055 \exp(-M^{3/2}/4) \right] \left(\frac{L_{\rm p}}{L_{\rm f}} \right)^{0.63} \times \notag\\ \left(\frac{Z_{\rm c}}{10^{-7} \langle Z \rangle} \right)^{0.015}, \notag \\
\tau_{\rm con2}  =   \left[0.335 - 0.095 \exp(-M^2/4) \right] \left(\frac{L_{\rm p}}{L_{\rm f}} \right)^{0.63}\hspace{.3cm} \times \notag \\ \left(\frac{Z_{\rm c}}{10^{-7} \langle Z \rangle} \right)^{0.02}, 
\notag\\
n           =   1 + 11 \, \exp(-M/3.5) \left(\frac{L_{\rm p}}{L_{\rm f}}\right)^{1.3}, \hspace{1.7cm}
\end{gather}
which are applicable for all Mach numbers and pollution properties, as long as $L_{\rm p} \leq L_{\rm f}.$
Note that unlike eqs.\ (\ref{pfnconvconc}) and (\ref{pfnconvsolutionconc}), these fits are for convenience only and not based 
on an underlying physical picture.
We showed that the model is valid for  $Z_{\rm c} \lsim 10^{-3} \langle Z \rangle$, where 
$\tau_{\rm con1}$ and $\tau_{\rm con2}$ only have a weak dependence on $Z_{\rm c}$ (\S 5.4.4).   
If $Z_{\rm c}$ is close to or larger than $\langle Z \rangle$, the model is no  
longer applicable, and we defer a study of this situation to a later work. 
We also tested the convergence of the model parameters with the numerical 
resolution (\S5.4.5). 
The parameters $n$ and $\tau_{\rm con}$ may have a dependence 
on the compressivity of the driving force, which will be systematically examined in a future work.

To apply our  model and simulation results to the mixing of heavy 
elements in the interstellar medium, we specified the source term in the 
concentration PDF equation, accounting for the effects of new metals 
from supernova explosions and the possible infall of pristine gas from 
the halo or the intergalactic medium. These two sources force spikes in the 
PDF at high and low concentration values, respectively. With the source term, 
we derived an equation (eq.\ (\ref{eq:pfgalaxy})) for the global 
pristine fraction in early galaxies. A description for how to use the equation 
and our simulation results to estimate the primordial gas fraction
was given in \S 6.    
We discussed the timescales of relevant processes that control how 
and how fast the pollution process proceeds.   
In particular, the spatial transport by turbulent motions over 
galactic scales may play an important role if the interstellar turbulence 
is driven at small scales, e.g., by supernova explosions.    

Numerical simulations accounting for the interstellar environment at galactic scales are 
a valuable tool to study the pollution of primordial gas from a less idealized point of 
view. In fact, recent efforts to track metal mixing in the context of the formation 
of protogalaxies have made significant improvements in tracking the spatial 
evolution of the metallicity, averaged over relatively large scales (e.g., Wise \& Abel 2008).
Greif \etal (2009, 2010) employed a turbulent diffusion formalism to mimic mixing by 
smoothing over the SPH kernel, and a similar approach was used in the 10 Mpc SPH 
simulations by Maio \etal (2010) and Campisi \etal (2011), who assumed the initial metal pollution 
was spread over $\approx$ kpc scales by cluster winds.  Another  recent simulation by Ritter \etal (2012) 
used a finite-difference code with adaptive mesh refinement coupled to Lagrangian tracer particles to keep track of  
the metals produced in an initially metal-free galaxy.  Interestingly, they found that a cold supersonically turbulent 
core developed because of the fallback of metal-enhanced ejecta.  However,  because the 
resolution scale was much larger than the scales on which turbulence-enhanced molecular 
diffusivity operates (Pan \& Scalo 2007), they could 
not resolve a sufficiently large range of scales to track the unmixed, primordial  fraction. 

To overcome limitations such as these,  we have developed a large-eddy simulation 
(LES) approach based on our model and simulation results. In LESs, 
the flow quantities at resolved scales are directly computed, 
while the feedback effect of subgrid turbulent motions is modeled. 
To overcome the over-pollution by numerical diffusion, a subgrid model was 
constructed to track the evolution of the concentration fluctuations below 
the resolution scale. Using the standard filtering procedure for the LES formulation, 
we derived 
an equation for the filtered concentration PDF representing metallicity 
fluctuations at subgrid scales, and discussed the treatment of each 
term in the equation. 
The core of our subgrid model is equation (\ref{filteredprimordial}) 
for the filtered pristine fraction (i.e., the pristine fraction in each 
computational cell), which was derived from the filtered PDF 
equation.   
Again, we repeat it here for convenience:
\begin{gather}
\frac {\partial (\overline {\rho} \hspace{0.5mm}\widetilde {P})} {\partial t}  +  \frac{\partial}{\partial x_i} \left( \overline{\rho} \hspace{0.5mm} 
\widetilde{P} \hspace{0.5mm}\widetilde{v}_i  \right) = \frac{\partial}{\partial x_i} \left( \overline{\rho} (\gamma + \gamma_{\rm P})  \frac{\partial \widetilde{P}}{\partial x_i} \right) \hspace{3cm}\notag \\ \hspace{2cm}
- \frac{n_{\rm s}}{\tau_{\rm scon}} \widetilde{P}\left(1- \widetilde{P}^{1/n_{\rm s}} \right) - \overline{\dot{n}}_{\rm SN} m_{\rm ej} \widetilde{P} ,   
\end{gather}
where $n_{\rm s}$ and $\tau_{\rm scon}$ are parameters for the subgrid pollution, corresponding to $n$ and $\tau$ in 
our convolution model, $\gamma_{\rm P}$ is the eddy diffusivity for the pristine fraction,   $\overline{\dot{n}}_{\rm SN}({\bf x}, t)$ is the 
filtered number rate of supernova explosions per unit volume and each supernova is assumed to have an ejecta 
mass $m_{\rm ej}$. 
This equation adopts the commonly-used eddy-diffusivity model 
for the transport effect of subgrid turbulent motions, and employs 
the convolution model for the pollution of the pristine gas within 
each cell. The implementation of our subgrid model was illustrated 
in the context of a one-equation LES model for the interstellar 
turbulence, which evolves the kinetic energy of subgrid turbulent 
motions. Together with the resolved temperature field, the subgrid 
kinetic energy specifies the turbulence properties in each cell, which are 
needed to calculate the eddy-diffusivity in the transport 
term and to determine the parameters in the convolution model 
for the subgrid pollution.  
The convolution model parameters depend on the metal/supernova 
sources in each cell, and can be evaluated using our simulation results.

The resulting physically-realistic model for the evolution of the unresolved primordial fraction serves as a prototype for future simulations aimed at interpreting many observations  currently probing the nature of early galaxies.   The continuing discovery of star-forming galaxies at $z \approx 7-10$ in broad-band photometric searches, for example, (as in the Hubble UDF12 survey, Ellis \etal 2013)  suggests that observations of galaxies with significant primordial fractions should soon become available.  The situation for galaxies selected  on the basis of strong Ly$\alpha$ emission (\eg Cowie \& Hu 1998; Rhoads \etal 2000) is even more promising, and it may only be a matter of time before several such galaxies are clearly identified as containing primordial stars
recognizable by their large Ly$\alpha$ equivalent width and weak He II emission  (Scannapieco \etal 2003; Jimenez \& Haiman 2006). In fact, a recent detailed analysis of deep Subaru images by Inoue \etal (2011) strongly supports the interpretation that the mass fraction of stellar populations with extremely small metal abundances in $z \approx 3$  Lyman alpha emitters may be 1-10 \% by mass,  based  on their very strong rest frame Lyman continua.   Based on similar diagnostics, Kashikawa et al. (2012) recently proposed a $z = 6.5$ Ly$\alpha$ emitter as a Pop III candidate although enhanced Ly$\alpha$ emission from a clumpy, dusty medium (Neufeld 1991; Hansen \& Oh 2006) cannot be ruled out conclusively in this case. If any of these galaxies are convincingly demonstrated to contain primordial stars, their evolution could only be simulated using an approach such as the one outlined here.

Currently,  a more direct constraint on the evolution of primordial gas is based on the absence of metal lines in absorption line systems in the intergalactic medium. Fumagalli \etal (2011) used this approach to obtain upper limits of $Z< 10^{-6}$ by mass in two Lyman limit systems associated with quasars at $z \sim 3.1$ and 3.4.  Simcoe \etal (2012) used the lack of metal lines in a $z \approx 7$ quasar spectrum that shows a large neutral hydrogen column density  to obtain an upper limit of $Z \approx 10^{-5} -10^{-6},$ depending on the whether the gas is bound in a galaxy or is diffuse intergalactic gas at that redshift.   The implications of these measurements can only be fully explored through models such as ours, which capture the unresolved, unmixed fraction.

Finally, at least four examples of Galactic stars with [Fe/H] $<$ -4.5 ($Z \lsim 10^{-6.5}$) are known (\eg Christlieb et al.\ 2002; Frebel et al.\ 2008;  Norris et al.\ 2007; Caffau et al.\ 2011), although only one is not enhanced in carbon.  Recently, Yong et al.\ (2012) have shown convincingly that the Milky Way metallicity pdf is still decreasing smoothly down to at least [Fe/H] = -4.1 ($Z \approx 10^{-6}$), without the sudden cutoff claimed in earlier work.   Once the rather severe selection effects are understood, this measurements could also be directly compared with our models, and even allow their two main parameters to be calibrated outside of numerical simulations. In fact, our proposed large-eddy simulation is expected to give reliable predictions for any physical problem, astrophysical our otherwise, in which the unresolved, low concentration tail of the pdf needs to be tracked.  Its numerical implementation and targeted application represents an extremely promising avenue for future studies.

\acknowledgements

We acknowledge  support from NASA under theory Grant No. NNX09AD106 and astrobiology institute grant 08-NAI5-0018 and from the National Science Foundation under grant AST 11-03608.  All simulations were conducted at the Arizona State University Advanced Computing Center and the Texas Advanced Computing Center, using the FLASH code, a product of the DOE ASC/Alliances funded Center for Astrophysical Thermonuclear Flashes at the University of Chicago.

\appendix
\section{Filtered PDF Equation}
 
We formulate a PDF approach for large-eddy simulations of turbulent mixing. LESs based on the PDF method have
been applied to study reacting turbulent flows (e.g., Gao \& O'brein 1993, Colucci et al.\ 1998, Jaberi et al.\ 1999, Pitsch 2006). 
We first derive an equation for the local fine-grained concentration PDF, and then apply the 
filtering procedure to obtain an exact equation for the filtered PDF at the resolution scale. 
The derivation is similar to that in Appendix A of PSS for the equation of the concentration 
PDF defined in a statistical ensemble. 

We start with the definition of the fine-grained concentration PDF as a delta function, 
\begin{equation}
\phi(Z;{\bf x},t)= \delta (Z-C({\bf x},t)), 
\label{volumefinepdf}
\end{equation}
because the concentration field in a given turbulent flow 
is single-valued at given position and time (PSS). Here $Z$ is 
the sampling variable. Since $\phi(Z;{\bf x},t)$ depends on $t$ only through 
the variable $Z-C({\bf x},t)$, the time-derivative $\phi(Z;{\bf x},t)$ can 
be written as, 
\begin{equation}
\frac {\partial \phi(Z;{\bf x},t)}{\partial t}  = -\frac {\partial \phi(Z;{\bf x},t)} 
{\partial Z} \frac {\partial C({\bf x},t) } {\partial t}.
\label{pdftimederivative} 
\end{equation}
Similarly, the spatial gradient of $\phi$ is given by 
\begin{equation}
\frac{\partial \phi(Z;{\bf x},t)}{\partial x_i}  = -\frac {\partial \phi(Z;{\bf x},t)} {\partial Z}   \frac{\partial C({\bf x},t)}  {\partial x_i}.
\label{pdfgradient} 
\end{equation}
Using eqs.\ (\ref{pdftimederivative}) and (\ref{pdfgradient}) and the advection-diffusion equation (\ref {advection}), we have  
\begin{equation}
\frac {\partial \phi(Z;{\bf x},t)}{\partial t} + v_i \frac{\partial \phi(Z;{\bf x},t)} { \partial x_i}  = - \frac{\partial}{\partial Z} \left[\phi(Z;{\bf x},t) \left(\frac{1}{\rho} \frac{\partial} {\partial x_i} \left(\rho \gamma \frac{\partial C}{\partial x_i}\right)+S\right) \right],
\label{finegraineq}
\end{equation}
where we used the fact that, except $\phi(Z;{\bf x},t)$, all the quantities on the r.h.s.\ are independent of $Z$. 

Combining eq.\ (\ref{finegraineq}) with the continuity equation, we obtain 
\begin{equation}
\frac {\partial  (\rho \phi)} {\partial t} + 
\frac{\partial (\rho \phi  v_i )}{\partial x_i} =  - \frac{\partial}{\partial Z} \left[ \rho \phi \left(\frac{1}{\rho} \frac{\partial} {\partial x_i} \left(\rho \gamma \frac{ \partial C } {\partial x_i} \right)+S\right)\right], 
\label{rhofinegraineq}
\end{equation}
which was also derived in Appendix A of PSS. The diffusivity term in eq.\ (\ref{rhofinegraineq}) can be rewritten as,  
\begin{equation}
 - \frac{\partial}{\partial Z} \left[\rho \phi \left(\frac{1}{\rho} \frac{\partial}{\partial x_i} 
 \left(\rho \gamma \frac{\partial C}{\partial x_i} \right)\right)\right] =  \frac{\partial}{\partial x_i} \left(\rho \gamma \frac{\partial \phi}{\partial x_i} \right)- \frac{\partial^2}{\partial Z^2} \left(\rho \phi \gamma \left(\frac{\partial C}{\partial x_i}\right)^2 \right),
\label{diffusivity}
\end{equation}
where the first term on the r.h.s. is  a spatial diffusion of the local PDF.  Note 
that eq.\ (\ref{ensemblediffusivityterm}) in \S 2.2 is derived by taking the ensemble 
average of this equation and using the definition and properties of the conditional ensemble average (see Appendix A of PSS).  

We next apply a filtering procedure to eq.\ (\ref{finegraineq}). A convolution of $\phi(Z; {\bf x}, t)$ 
with the filter function, $G$, gives a filtered PDF, $\overline{\phi} (Z; {\bf x}, t) = \int_V \phi(Z; {\bf x}', t) G( {\bf x}-  {\bf x}') dx'^3$, 
characterizing the concentration fluctuations within regions of the 
filter size (or resolution scale).  For compressible turbulence, we define a filtered PDF with 
density weighting (Jaberi  et al.\ 1999),  
\begin{equation}
\widetilde{\phi} (Z; {\bf x}, t) \equiv  \frac { \overline{\rho \phi} } {\overline{\rho}},     
\label{densweightedPDF} 
\end{equation}
which is a specific example of eq.\ (\ref{Favrefilter}).  Using eqs.\ (\ref{filteredvariables}) and 
(\ref{densweightedPDF}) in eq.\ (\ref{rhofinegraineq}), we obtain the filtered PDF equation, 
\begin{equation}
\frac {\partial \left( \overline{\rho} \hspace{0.5mm} \widetilde{\phi} \right)  } {\partial t}  +  \frac{\partial  \left(  \overline{ \rho  v_i \phi }  \right)}{\partial x_i} =  
- \frac{\partial}{\partial Z}  \left( \overline { \phi \frac{\partial}{\partial x_i} \left(\rho \gamma \frac{ \partial C}{\partial x_i} \right)} \right)  -  
\frac{\partial \left( \overline{\rho S \phi } \right)}{\partial Z} .
\label{rhofiltered}
\end{equation}
To write the equation in a more convenient form, we introduce conditional filtering based on the local 
concentration values.    

For any variable $A$ in the flow, we define a conditionally filtered quantity,       
\begin{equation}
\overline{A|C=Z}  =  \frac {\overline{A \phi (Z;{\bf x},t) } } {\overline{\phi}}.     
\label{conditionalvolume} 
\end{equation}
Since the fine-grained  PDF, $\phi$, is a delta function, the definition is straightforward 
to understand: the conditionally filtered variable is the average over the set 
of points within a filter size satisfying $C ({\bf x}, t) = Z$. It is analogous to the conditional average 
defined in 
the context of a statistical ensemble (see \S 2 \& PSS).  
We further introduce a density-weighted conditional filtering,    
\begin{equation}
\overline{[A|C=Z]}_{\rho}  = \frac{ \overline{\rho A|C=Z} } {\overline{\rho|C=Z} } = \frac { \overline { \rho \phi A} } {\overline {\rho \phi} }, 
\label{conditionalmass} 
\end{equation}
where the last step follows from eq.\ (\ref{conditionalvolume}). This definition is similar to eq.\ (\ref{densweightconditionalave}) 
for the density-weighted conditional average over an ensemble. 

Combining eqs.\ (\ref{densweightedPDF}) and (\ref{conditionalmass}), we have 
$\overline{\rho \phi  A } = \overline{\rho} \hspace{0.5mm} \widetilde{\phi} \hspace{0.5mm}\overline{[A|C=Z]}_\rho$.
Applying this relation to the last three terms in eqs.\ (\ref{rhofiltered}), we obtain,    
\begin{equation}
\frac {\partial (\overline {\rho} \hspace{0.5mm}\widetilde {\phi})} {\partial t}  +  \frac{\partial}{\partial x_i} \left( \overline{\rho} \hspace{0.5mm} \widetilde{\phi} \hspace{0.5mm}\overline{[v_i|C=Z]}_{\rho}  \right)
= - \frac{\partial}{\partial Z} \left(  \overline{\rho} \hspace{0.5mm} \widetilde{\phi} \hspace{0.5mm} \overline { \left[\frac {1}{\rho}  \frac{\partial} {\partial x_i} \left(\rho \gamma  \frac{\partial C} {\partial x_i} \right)  
\left\vert  \vphantom{\frac{1}{1}} \right. C=Z \right]}_{\rho}   \right) 
 -  \frac{\partial}{\partial Z}  \left( \overline{\rho}  \hspace{0.5mm}\widetilde{\phi}  \hspace{0.5mm} \overline{ [S|C=Z]} _{\rho}  \right), 
\label{rhofilteredeq}
\end{equation}
which is equivalent to eq.\ (22) in Jaberi et al.\ (1999). Using eq.\ (\ref{diffusivity}) for the diffusivity term gives,
\begin{equation}
\begin{array}{lll}
{\displaystyle \frac {\partial (\overline {\rho} \hspace{0.5mm}\widetilde {\phi})} {\partial t}  +  \frac{\partial}{\partial x_i} \left( \overline{\rho} \hspace{0.5mm} \widetilde{\phi} \hspace{0.5mm}\overline{[v_i|C=Z]}_{\rho}  \right)  =  \frac{\partial}{\partial x_i} \left( \overline{\rho \gamma \frac{\partial \phi}{\partial x_i}} \right) - 
 \frac{\partial^2}{\partial Z^2} \left(\overline{\rho} \hspace{0.5mm} \widetilde{\phi} \overline{\left[\gamma \left(\frac{\partial C}{\partial  x_i} \right)^2 \left\vert\vphantom{\frac{1}{1}}\right. C=Z \right]}_\rho \right)} \\ \hspace{5.5cm}
{\displaystyle -  \frac{\partial}{\partial Z}  \left( \overline{\rho}  \hspace{0.5mm}\widetilde{\phi}  \hspace{0.5mm} \overline{ [S|C=Z]} _{\rho}  \right)}.  
\end{array}   
\label{rhofilteredeqdiffusivity}
\end{equation}
Note that eq.\ (\ref{rhofilteredeq}) becomes identical to 
eq.\ (\ref{pdfeq}) for the ensemble-defined PDF, if  we replace $\overline{\rho}$, $\widetilde{\phi}$, and $\overline{[\cdot \cdot \cdot |C=Z]}_{\rho}$ by $\langle \rho \rangle$,  $\Phi$, 
and $\langle \cdot \cdot \cdot |C=Z \rangle_{\rho}$, respectively. 
The equivalence between the filtered PDF and the ensemble-defined PDF has 
been discussed in \S 2, based on the ergodic theorem and the assumption that 
statistical homogeneity is restored at the filter scale.   
Similar to the case of the ensemble PDF equation,  
the advection and diffusivity terms in eqs.\ (\ref{rhofilteredeq}) and (\ref{rhofilteredeqdiffusivity}) 
need to be modeled due to the closure problem. In \S 7, we adopted an eddy-diffusivity 
assumption for the advection term, and the self-convolution models 
discussed in \S 3 may be applied to approximate the diffusivity term. 
Using the convolution PDF models and the results of our simulations,  a subgrid 
model is constructed in \S 7 to investigate the pollution of primordial gas in  early galaxies.

\end{document}